\documentclass[aar]{svjour}
\usepackage{amsmath}
\usepackage{graphicx}
\usepackage{color,hyperref}
\definecolor{linkcolor}{rgb}{0,0,0.25}
\hypersetup{
  colorlinks=true,        
  linkcolor=linkcolor,    
  citecolor=linkcolor,    
  filecolor=linkcolor,    
  urlcolor=linkcolor     
}
\newcounter{address}
\usepackage{natbib}
\bibpunct{(}{)}{;}{a}{}{,}
\usepackage{aas_macros}
\usepackage{deluxetable}
\newcommand{\DM}{\mathcal{DM}}
\newcommand{\ldm}{\mathcal{L}\bigl(\DM\bigr)}
\newcommand{\sigmaz}{\sigma_z}
\newcommand{\sigmar}{\sigma_R}

\newcommand{\etal}{et~al.}

\newcommand{\vecx}{$\vec{x}$}
\newcommand{\vecv}{$\vec{v}$}
\newcommand{\vx}{\vec{x}}
\newcommand{\vv}{\vec{v}}
\newcommand{\vlos}{\ensuremath{v_{\mathrm{los}}}}
\newcommand{\vphi}{v_\phi}
\newcommand{\vcirc}{v_{\ensuremath{\mathrm{circ}}}}
\newcommand{\eg}{e.g.}
\newcommand{\abundances}{\ensuremath{[\overrightarrow{\mathrm{X/H}}]}}

\newcommand{\tage}{\ensuremath{t_{\mathrm{age}}}}
\newcommand{\age}{\tage}
\newcommand{\eqnname}{equation}

\newcommand{\feh}{\ensuremath{[\mathrm{Fe/H}]}}
\newcommand{\afe}{\ensuremath{[\alpha\mathrm{/Fe}]}}
\newcommand{\teff}{\ensuremath{T_{\mathrm{eff}}}}
\newcommand{\Teff}{\teff}
\newcommand{\logg}{\ensuremath{\mathrm{log~}g}}
\newcommand{\Ro}{\ensuremath{R_0}}
\newcommand{\hz}{\ensuremath{h_z}}
\newcommand{\hR}{\ensuremath{R_e}}
\newcommand{\re}{\hR}
\newcommand{\AV}{$A_V$}
\newcommand{\X}{\abundances}
\newcommand{\Pot}{\ensuremath{\Phi (\vx )}}
\newcommand{\Potlambda}{\Phi (\vx|\lambda_\Phi)}
\newcommand{\Like}{\mathcal{L}}
\newcommand{\Kz}{K_z(z)}
\newcommand{\tf}{\tilde{f}}
\newcommand{\LCDM}{$\Lambda$CDM}
\newcommand{\lesssim}{\le}
\newcommand{\gtrsim}{\ge}
\newcommand{\pdf}{PDF}

\newcommand{\selparams}{\ensuremath{\vec{r}}}
\newcommand{\distparams}{\ensuremath{\vec{f}}}

\newcommand{\ssf}{SSF}
\newcommand{\DF}{distribution function}
\newcommand{\df}{DF}
\newcommand{\dff}{\ensuremath{\mathbf{f}}}
\newcommand{\map}{MAP}
\newcommand{\dex}{\ensuremath{\,\mathrm{dex}}}
\newcommand{\Gyr}{\ensuremath{\,\mathrm{Gyr}}}
\newcommand{\kpc}{\ensuremath{\,\mathrm{kpc}}}
\newcommand{\pc}{\ensuremath{\,\mathrm{pc}}}
\newcommand{\kms}{\ensuremath{\,\mathrm{km\ s}^{-1}}}
\newcommand{\msun}{\ensuremath{\,\mathrm{M}_{\odot}}}

\usepackage{color}

\begin{document}
\authorrunning{Hans-Walter Rix \& Jo Bovy}
\title{The Milky Way's Stellar Disk}
\subtitle{Mapping and Modeling the Galactic Disk}
\author{Hans-Walter Rix\inst{1} and Jo Bovy\inst{2}
\thanks{Hubble fellow}
}                     
\institute{Max-Planck-Institut f\"ur Astronomie, K\"onigstuhl 17, D-69117
  Heidelberg, Germany \and Institute for Advanced Study, Einstein Drive, Princeton, NJ 08540, USA}

\date{Received: date / Accepted: date}

\maketitle

\begin{abstract}
A suite of vast stellar surveys mapping the Milky Way, culminating in the Gaia mission,
is revolutionizing the empirical information about the distribution and properties of
stars in the Galactic stellar disk. We review and lay out what analysis and modeling machinery
needs to be in place to test mechanism of disk galaxy evolution 
and to stringently constrain the Galactic gravitational potential, using such Galactic star-by-star
measurements. We stress the crucial role of stellar survey selection functions in any such
modeling; and we advocate the utility of viewing the Galactic stellar disk as made up from
`mono-abundance populations' (\map s), both for dynamical modeling and for constraining 
the Milky Way's evolutionary processes. We review recent work on
the spatial and kinematical distribution of \map s, and lay out how further
study of \map s in the Gaia era should lead to a decisively clearer picture of 
the Milky Way's dark matter distribution and formation history.

\end{abstract}

\section{Introduction}\label{intro}
In the world of galaxies, the Milky Way is in many ways as
typical as it gets: half of the Universe's present-day stars live in
galaxies that match our Milky Way in stellar mass, size, chemical
abundance, etc. within factors of a few \citep[e.g.,][]{Mo10a}. But
for us, it is the only galaxy whose stellar distribution we can see in
its full dimensionality: star-by-star we can obtain 3D positions and
3D velocities ($v_{\mathrm{los}},\mu_\ell,\mu_b$), coupled with the stars'
photospheric element abundances and constraints on their ages. We know
that in principle this enormous wealth of information about the
stellar body of our Galaxy holds a key to recognizing and
understanding some of the mechanisms that create and evolve disk
galaxies. And it holds a key to mapping the three-dimensional
gravitational potential and by implication the dark matter
distribution in the Milky Way.  A sequence of ongoing photometry and
spectroscopy surveys has recently hundred-folded the number of stars
with good distances, radial and transverse velocities, and abundance
estimates; this only forebodes the data wealth expected from ESA's
flagship science Mission Gaia to be launched next year.

\begin{figure}
  \centering
  \includegraphics[width=0.85\textwidth]{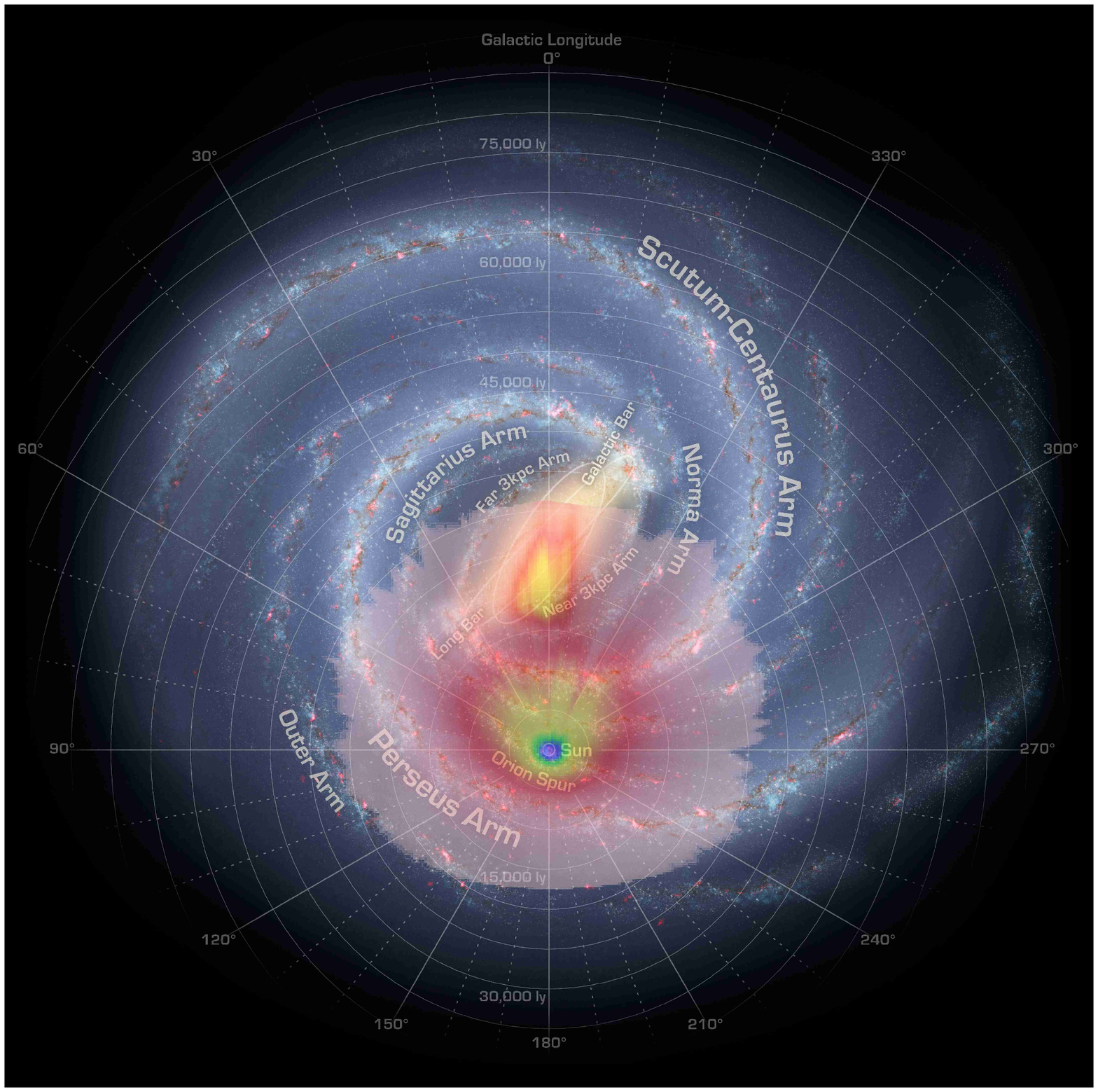}
  \includegraphics[width=0.85\textwidth]{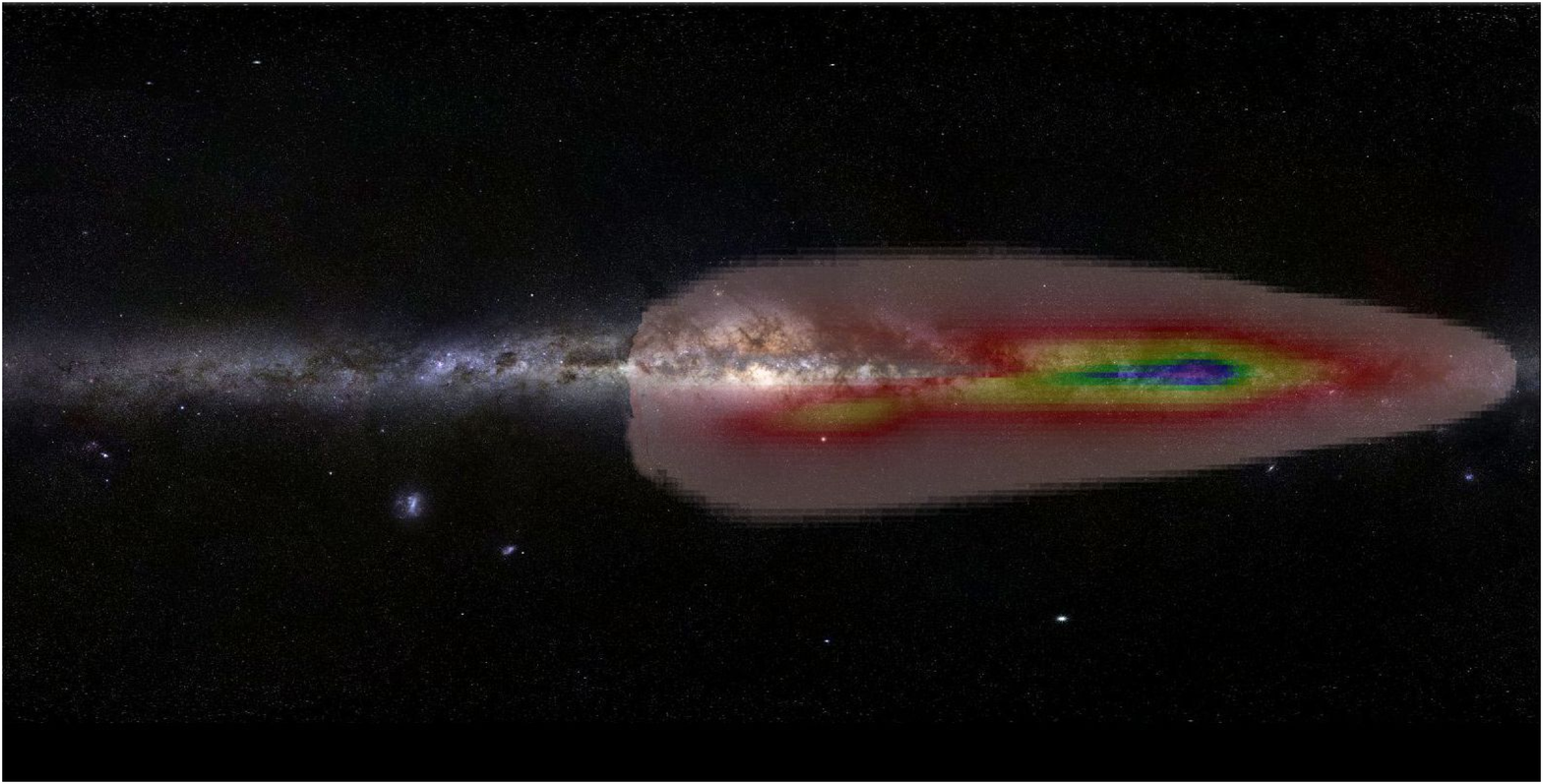}
  \caption{A view of our Galaxy and the effective volume that Gaia
    will survey (courtesy X. Luri \& A. Robin), based on current
    simulations of Gaia mock catalogs. Even in the age of Gaia, dust
    extinction and image crowding will limit the exploration of the
    Disk to only a quadrant with optical surveys.}
  \label{fig:Gaia-Survey-Volume}
\end{figure}

Yet, practical approaches to extract the enormous astrophysical
information content of these data remain sorely underdeveloped. It
is not even qualitatively clear at this point what will limit the
accuracy of any galaxy-formation or dark-matter inferences: the sample
sizes, the fraction of the Milky Way's stellar body covered
(cf.~\figurename~\ref{fig:Gaia-Survey-Volume}), the precision of the
\vecx - \vecv\ phase-space measurements, the quality and detail of the
abundance information, or the (lack of) stellar age estimates. Are
dynamical analyses limited by the precision with which sample
selection functions can be specified, or by the fact that dust
obscuration and crowding will leave the majority (by
stellar-mass-weighted volume) of Milky Way stars unobserved even if
all currently planned experiments worked our perfectly? Or are
dynamical inferences limited by the fact that the symmetry and
equilibrium assumptions, which underlie most dynamical modeling, are
only approximations?

The ongoing data deluge and the continuing recognition how much our
Milky Way, despite being `just one galaxy', may serve as a Rosetta
Stone for galaxy studies, have triggered a great deal of preparatory
work on how to analyze and model these data. Many of the scientific
and practical issues have been laid out, e.g., in \citet{Turon08a} or
\citet{Binney11b}. Yet, it appears (at least to the authors) that the
existing survey interpretation and modeling approaches are still woefully
inadequate to exploit the full information content of even the
existing Galactic stellar surveys, let alone the expected information
content of Gaia data.

And while the science theme of `understanding the current structure of
the Galaxy and reconstructing its formation history' remains of
central interest to astrophysics, the specific questions that should
be asked of the data have evolved since the original Gaia mission
science case was laid out, through advances in our understanding of
galaxy formation in the cosmological framework and recent work on
secular galaxy evolution. The aim of simply `reconstructing' the
formation history of the Galaxy in light of even idealized Gaia data,
now seems naive.

The review here will restrict itself to the Galactic stars (as opposed
to the Galactic interstellar medium), and 
in particular to the Milky Way's dominant stellar component,
its {\sl Disk}, which contains about three quarters of all Galactic
stars.  As a shorthand, we will use the term {\sl Disk} (capitalized)
to refer to the `Milky Way's stellar disk'. All other disks will be
labeled by qualifying adjectives. The discussion will focus on the
{\it stellar} disk of our Galaxy and dark matter in the {\it central}
parts of the overall Milky Way halo,\ ($\le 0.05\times R_{\mathrm{virial}}
\sim 12$kpc), with only cursory treatment of the other stellar
components, the gas and dust in the Milky Way, and the overall halo
structure. Perhaps our neglect of the Galactic bar is most problematic in
drawing scope boundaries for this review, as the interface between
disk and bar is both interesting and unclear, and because it is manifest 
that the present-day Galactic bar has an important dynamical impact on the
dynamics and evolution of the Disk; yet, the Galactic bar's role and complexity warrants
a separate treatment.

The questions about galaxy disk formation that a detailed
analysis of the Milky Way may help answer are manyfold.  What
processes might determine galaxy disk structure? In particular, what
processes set the exponential radial and vertical profiles seen in the
stellar distributions of galaxy disks? Were all or most stars born
from a well-settled gas disk near the disk plane and acquired their
vertical motions subsequently? Or was some fraction of disk stars
formed from very turbulent gas early on \citep[e.g.,][]{Bournaud09a,Ceverino12a},
forming a primordial thick disk?  Are there discernible signatures of
the stellar energy feedback to the interstellar medium that global
models of galaxy formation have identified as a crucial `ingredient'
of (disk) galaxy formation \citep{Nath09,Hopkins12}?  What was the
role of internal heating in shaping galaxy disks? What has been the
role of radial migration
\citep{SellwoodBinney,Roskar08a,Roskar08b,Schoenrich08a,Minchev11}, i.e., the
substantive changes in the stars' mean orbital radii that are expected
to occur without boosting the orbits' eccentricity? What was the
disk-shaping role of minor mergers \citep[e.g.,][]{Abadi03a}, which
are deemed an integral part of the \LCDM\ cosmogony? How much did
in-falling satellites impulsively heat the Milky Way's disk,
potentially leading to a distinct thick disk \citep{Villalobos08a}?
How much stellar debris did they deposit in this process? 
Is the radial orbit migration induced by satellite infall \citep{Bird12a}
distinguishable from purely internal processes? All of these
questions are not only relevant for the Milky Way in particular, but lead
generically to the question of how resilient stellar disks are to
tidal interactions; it has been claimed
\citep[e.g.,][]{Kormendy10a,Shen10a} that the existence of large, thin
stellar disks poses a challenge to the merger-driven \LCDM\ picture.

In the end, answers to these questions require a multi-faceted
comparison of the Disk's observable status quo to the expectations
from {\sl ab initio} formation models, in practice for the most part
hydrodynamical simulations.  However, the current generation of
`cosmological' disk galaxy formation models is more illustrative than
exhaustive in their representation of possible disk galaxy formation
histories; therefore the question of how to test for the importance of
galaxy formation {\sl ingredients} through comparison with
observational data is actively underway.

The maximal amount of empirical information about the Disk that one
can gather from data is a joint constraint on the gravitational
potential in which stars orbit $\Phi (\vx,t)$ and on the {\it
  chemo-orbital distribution function} of the Disk's stars. How to
best obtain such a joint constraint is a problem solved in principle,
but not in practice (see Sections~\ref{sec:DiskGravPot} \&
\ref{sec:DynMod} ).

It is in this context that this review sets out to work towards three
broad goals:
\begin{itemize}
\item synthesize what the currently most pertinent questions about
  dark matter, disk galaxy formation and evolution are that may
  actually be addressed with stellar surveys of the Galactic disk.
\item lay out what `modeling' of large stellar samples means and
  emphasize some of the practical challenges that we see ahead.
\item describe how recent work may change our thinking about how to
  best address these questions.
\end{itemize}

Compared to \citet{Turon08a} and \citet{IvezicBeersJuric} for an empirical description of the Milky Way
disk, and compared to a series of papers by Binney and collaborators
on dynamical modeling of the Galaxy
\citep{Binney10a,Binney11a,Binney12a,McMillan12a}, we place
more emphasis two aspects that we deem crucial in 
Galactic disk modeling:
\begin{itemize}
\item the consideration of `mono-abundance' stellar sub-populations (\map s),
  asking the question of `what our Milky Way disk would look like if
  we had eyes for stars of only a narrow range of photospheric
  abundances'.  The importance of `mono-abundance components' arises
  from the fact that in the presence of significant radial migration,
  chemical abundances are the only life-long {\it tags}
  \citep[see][]{Freeman02a} that stars have, which can be used to
  isolate sub-groups independent of presuming a particular dynamical
  history. In a collisionless system, such populations can be modeled
  completely independently; yet they have to `feel' the same
  gravitational potential.
\item the central importance of the (foremost spatial) selection
  function of any stellar sample that enters modeling. Dynamical
  modeling links the stars' kinematics to their spatial
  distribution. Different subsets of Disk stars (differing e.g. by abundance) have
  dramatically different spatial and kinematical distributions. If the
  spatial selection function of any sub-set of stars with measured
  kinematics is not known to better than some accuracy, this will pose
  a fundamental limitation on dynamical inferences, irrespective of
  how large the sample is; with ever larger samples emerging,
  understanding the selection function is increasingly probably to be a
  limiting factor in the analysis.
\end{itemize}

The remainder of the review is structured as follows: in Section 2 we
discuss in more detail the overall characterization of the Milky Way's
disk and detail the open questions of stellar disk formation and
evolution in a cosmological context. In Section 3 we provide an
overview of the existing and emerging stellar Galactic surveys and in
Section 4 we describe how the survey selection function can and should be
rigorously handled in modeling; . In Sections 5 and 6 we present recent
results in dynamical and structural modeling of the Disk, and their
implications for future work. In the closing Section 7, we discuss what
we deem the main practical challenges and promises for this research direction in
the next years.

\section{Galactic Disk studies: an overview}\label{sec:DiskOverview}

`Understanding' the Disk could mean having a comprehensive empirical
characterization for it and exploring which -- possibly competing -- 
theoretical concepts that make predictions for these characteristics
match, or do not. As the Milky Way is only one particular galaxy and
as disk galaxy formation is a complex process that predicts broad
distributions for many properties, it may be useful to consider which
Disk characteristics generically test formation concepts, rather than
simply representing one of many possible disk formation outcomes.

\subsection{Characterizing the current structure of the Disk}\label{CurrentDiskStructure}

In a casual, luminosity or mass-weighted average, the Disk can be
characterized as a highly flattened structure with an (exponential)
radial scale length of $2.5-3\kpc$ and scale height of $\simeq 0.3\kpc$
\citep[e.g.,][]{Kent91a,Lopez02,McMillan11a}, which is kinematically
cold in the sense that the characteristic stellar velocity dispersions
near the Sun of $\sigma_z\simeq \sigma_{\phi}\simeq
\sigma_R/1.5\simeq 25\kms$ are far less than $v_{\mathrm{circ}}\simeq
220\kms$.  Current estimates for the overall structural parameters of the Milky Way
are compiled in Table 1.2 of \citet{binneytremaine}; more specifically
estimates for the mass of the Disk are $\simeq\!
5\times 10^{10}\,\msun$ \citep{Flynn06,McMillan11a},  though the most
recent data sets have not yet been brought to bear on this basic
number. No good estimates for the globally averaged metallicity of the
Disk exist, though $\langle\feh\rangle$ being about at the solar value 
seems likely.

With these bulk properties, the Milky Way and its Disk are very
`typical' in the realm of present-day galaxies: comparable numbers of
stars in in the low-redshift Universe live in galaxies larger and
smaller (more and less metal-rich) than the Milky Way. For its stellar
mass, the structural parameters of the Disk are also not exceptional
\citep[e.g.,][]{vanderkruit11a}. Perhaps the most unusual aspect of
the Milky Way is that its stellar disk is so dominant, with a luminosity
ratio of bulge-to-disk of about 1:5 \cite{Kent91a}: most galaxies
of M$_*>5\times 10^{10}\msun$ are much more bulge dominated
\citep{Kauffmann03a}. 

\begin{figure}
  \centering
  \includegraphics[width=0.85\textwidth,clip=]{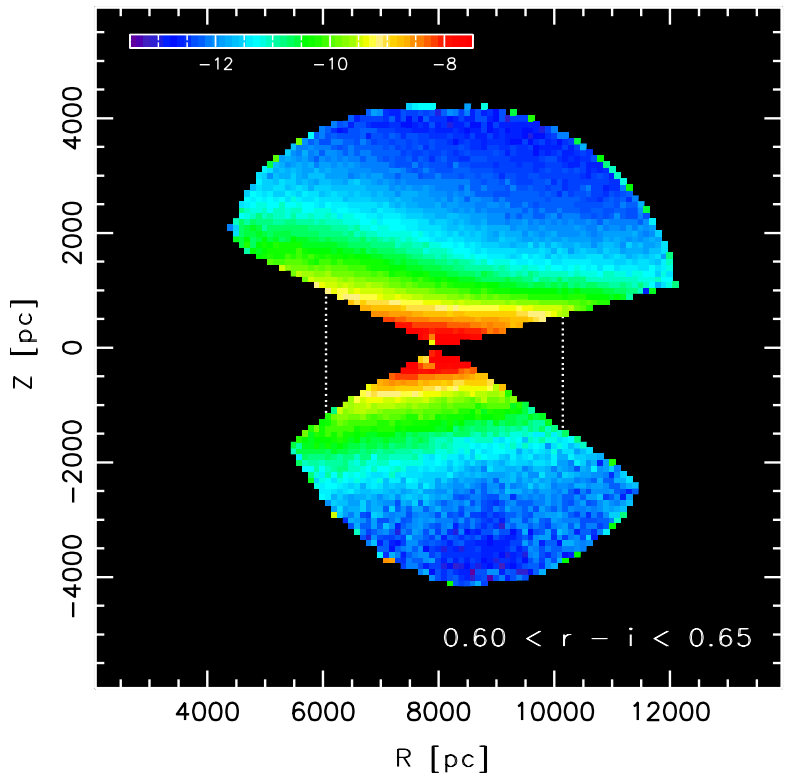}
  \caption{Stellar number density map of the Disk and halo from
    \citet{Juric2008}, for K-stars with colors $0.6\le r-i\le
    0.65$. The map, averaged over the $\phi$-direction was derived
    drawing on SDSS photometry, and applying photometric distance
    estimates that presume (sensibly) that the vast majority of stars
    are on the main sequence, not giants. Note that for these colors
    the main sequence stars will sample all ages fairly, as their MS
    luminosity remains essentially unchanged. With increasing $|z|$
    and $R$, however, the mean metallicity of the stars changes, and
    the stars within that color range represent different masses and
    luminosities.}
  \label{fig:Juric2008}
\end{figure}

But describing the Disk by `characteristic' numbers, as one is often
forced to do in distant galaxies, does not even begin to do justice to
the rich patterns that we see in the Disk: it has been long
established that positions, velocities, chemical abundances, and ages
are very strongly and systematically correlated. This is in the sense
that younger and/or more metal rich stars tend to be on more nearly
circular orbits with lower velocity dispersions. Of course, stellar
populations with lower (vertical) velocity dispersions will form a
thinner disk component. This has led to the approach of defining
sub-components of the Disk on the basis of the spatial distribution,
kinematics, or chemical abundances. Most common has been to describe
the Disk in terms of a dominant thin disk and a thick disk, with
thin--thick disk samples of stars defined spatially, kinematically, or
chemically. While these defining properties are of course related,
they do not isolate identical subsets of stars. Whether it is sensible
to parse the Disk structure into only two distinct components is discussed
below.

Much of what we know about these spatial, kinematical, and chemical
correlations within the Disk has come until very recently from very
local samples of stars, either from studies at $R\simeq R_\odot$ or
from the seminal and pivotal Hipparcos/Geneva-Copenhagen sample of
stars drawn from within $\simeq 100\pc$ \citep{ESA97a,Nordstroem04}.
As dynamics links local and global properties, it is perfectly
possible and legitimate to make inferences about larger volumes
than the survey volume itself; yet,
it is important to keep in mind that the volume-limited
Geneva-Copenhagen sample encompasses a volume that corresponds to
two-millionths of the Disk's half-mass volume.  Only recently have
extensive samples beyond the solar neighborhood with
$p(\vec{x},\vec{v},\X)$ become available.

While a comprehensive empirical description of the Disk (spatial,
kinematical, abundances) in the immediate neighborhood of the Sun has
revealed rich correlations that need explaining, an analogous picture
encompassing a substantive fraction of the disk with direct
observational constraints is only now emerging.

\begin{figure}
  \centering
  \includegraphics[width=0.85\textwidth,clip=]{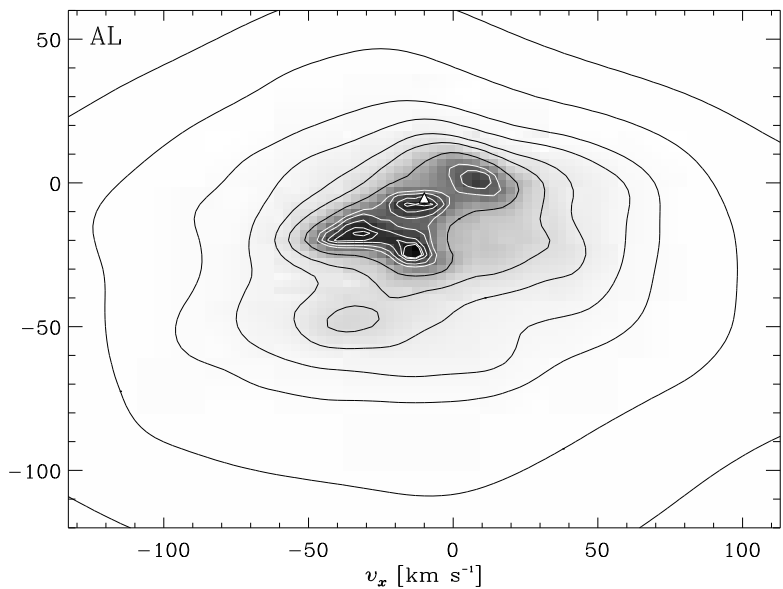}
  \caption{Distribution in $v_x - v_z$-space of very nearby stars in
    the Disk with distances from Hipparcos ($\simeq\!100\pc$;
    \citealt{Dehnen98a}). The sample shows rich sub-structure of
    `moving groups', some of which reflect stars of common birth
    origin, others are the result of resonant orbit trapping.}
  \label{fig:Dehnen98}
\end{figure}

But the Disk is neither perfectly smooth nor perfectly axisymmetric,
as the above description implied. This is most obvious for the
youngest stars that still remember their birthplaces in star clusters
and associations and in spiral arms. But is also true for older
stellar populations. On the one hand, the spiral arms and the Galactic
bar are manifest non-axisymmetric features, with clear signatures also
in the Solar neighborhood
\citep[e.g.,][]{Dehnen98a,Dehnen00a,Fux01a,Quillen03a,Quillen05a}. By
now the strength of the Galactic bar and its pattern speed seem
reasonably well established, both on the basis of photometry
\citep{Binney97a,Bissantz02a} and on the basis of dynamics
\citep{Dehnen99a,Minchev07a}. But we know much less
about spiral structure in the stellar disk. While the location of
the nearby Galactic spiral arms have long been located on the bases of
the dense gas geometry and distance measurements to young stars, the
existence and properties of dynamically-important stellar spiral
structure is completely open: neither has there been a sound measure
of a spiral stellar over-density that should have dynamical effects,
nor has there been direct evidence for any response of the Disk to
stellar spirals. Clarifying the dynamical role of spiral arms in the
Milky Way presumably has to await Gaia.

In addition there is a second aspect of non-axisymmetric
sub-structure, which is known to be present in the Disk, but which is
far from being sensibly characterized: there are groups of chemically
similar stars on similar but very unusual orbits \citep[streams,
  e.g.,][]{Helmi99a,Navarro04a,Klement08a,Klement09a} that point
towards an origin where they were formed in a separate satellite
galaxy and subsequently deposited in the Disk in a (minor) merger. The
process of Disk heating and Disk {\sl augmentation} through minor
mergers has been simulated extensively, both with collisionless and
with hydrodynamical simulations
\citep[e.g.,][]{Velazquez99,Abadi03a,Kazantzidis08a,Moster10a}: these
simulations have shown that galaxy disks can absorb considerably more
satellite infall and debris than initial estimates had suggested
\citep{Toth92a}. However, these simulations also indicated
that---especially for prograde satellite infall---it is not always
easy to recognize satellite debris by its orbit, once it has been
incorporated into the disk.

\subsection{The formation and evolution of the Disk}\label{sec:FormEvDisk}

Explaining the formation and evolution of galaxy disks has a 50-year
history, gaining prominence with the seminal papers by
\citet{Mestel63a}  and
\citet{Eggen62a}. Yet, producing a galaxy through {\it ab initio}
calculations that in its disk properties resembles the Milky Way has remained
challenging to this day.

Exploring the formation of galaxy disks through simplified
(semi-)analytic calculations has yielded seemingly gratifying models,
but at the expense of ignoring `detail' that is known to play a
role. Milestones in this approach were in particular the work by
\citet{Fall80a} that put \citeauthor{Mestel63a}'s idea of gas collapse
under angular momentum conservation into the cosmological context of
an appropriately sized halo that acquired a plausible amount of
angular momentum through interactions with its environment: this
appeared as a cogent explanation for galactic disk sizes. These
concepts were married with the Press \& Schechter
\citep{Press73a,Bond91a} formalism and its extensions by
\citet{Mo98a}, to place disk formation in the context of halos that
grew by hierarchical merging.  This approach performed well in
explaining the overall properties of the present-day galaxy disk
population and also its redshift evolution
\citep[e.g.,][]{Somerville08a}.

Yet, trouble in explaining disks came with the efforts to explain
galaxy disk formation using (hydrodynamical) simulations in a
cosmological context. The first simulations \citep{Katz91a}, which
started from unrealistically symmetrical and quiescent initial
conditions, yielded end-products that resembled observed galaxy
disks. But then the field entered a 15 year period in which almost all
simulations produced galaxies that were far too bulge-dominated and
whose stellar disks were either too small, or too anemic (low mass
fraction), or both. Of course, it was clear from the start that this
was a particularly hard problem to tackle numerically: the initial
volume from which material would come ($\simeq\!500\kpc$) and the
thin-ness of observed disks ($\simeq\!0.25\kpc$) implied very large
dynamic ranges, and the `sub-grid' physics issue of when to form stars
from gas and how this star-formation would feed back on the remaining
gas played a decisive role.

\begin{figure}
  \centering
  \includegraphics[width=0.85\textwidth,clip=]{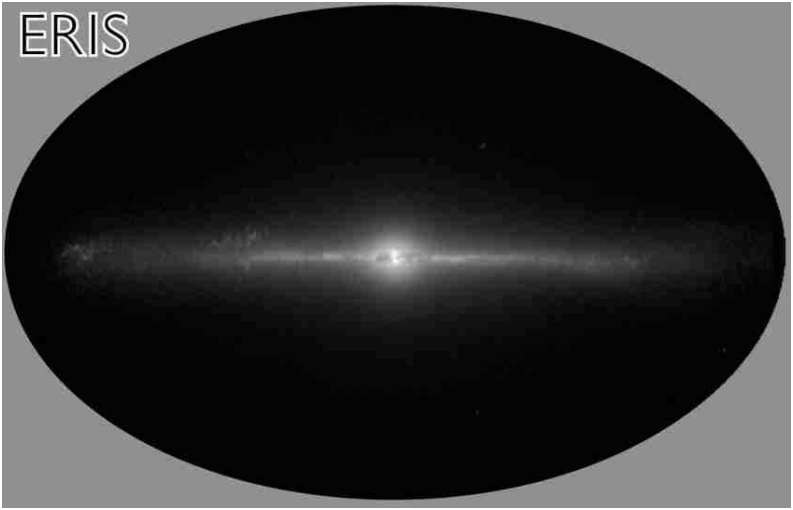}
  \caption{Output of the ERIS disk galaxy formation simulation aimed
    at following the formation history of a Milky Way like galaxy ab
    initio. The Figure shows a projection resembling the 2MASS map of
    the Galaxy, and shows that simulations have now reached a point
    where, with very quiescent merging histories, disk-dominated
    galaxies can result.}
  \label{fig:Guedes}
\end{figure}

As of 2012, various groups
\citep{Agertz11a,Guedes11a,Martig12a,Stinson12a} have succeeded in
running simulations that can result in large, disk-dominated galaxies,
resembling the Milky Way in many properties (cf.
Figure~\ref{fig:Guedes}).  This progress seems to have---in good
part---arisen from the explicit or implicit inclusion of physical
feedback processes that had previously been neglected. In particular,
`radiative feedback' or `early stellar feedback' from massive stars
before they explode as Supernovae seems to have been an important
missing feedback ingredient and must be added in simulations to the
well-established supernovae feedback \citep{Nath09,Hopkins12,Brook12}.
With such --- physically expected --- feedback implemented, disk dominated galaxies can be made in
ab initio simulations that have approximately correct stellar-to-halo
mass fractions for a wide range of mass scales. In some of the other
cases, cf. \citet{Guedes11a}, the radiative feedback is not
implemented directly, but cooling-suppression below $10^4$~K has a
similar effect. It appears that the inclusion of this additional feedback
not only makes disk dominated galaxies a viable simulation outcome, but
also improves the match to the galaxy luminosity function and Tully-Fisher relations.

This all constitutes long-awaited success, but is certainly far from
having any definitive explanation for the formation of any particular
galaxy disk, including ours. In particular, a very uncomfortable
dependence of the simulation outcome on various aspects of the
numerical treatment remains, which has recently been summarized by
\citet{Scannapieco12a}.

In addition to the fully cosmological simulations, a large body of
work has investigated processes relevant to disk evolution from an
analytic or numerical perspective, with an emphasis on chemical
evolution and the formation and evolution of thick-disk
components. The chemical evolution of the Milky Way has been studied
with a variety of approaches
\citep[e.g.,][]{Matteucci89,GilmoreARAA,Chiappini01} mostly with
physically motivated but geometrically simplified models (not
cosmological ab initio simulations): these studies have aimed at
constraining the cosmological infall of 'fresh' gas
\citep[e.g.,][]{Fraternali08,Colavitti08}, explain the origin of
radial abundance gradients \citep[e.g.,][]{Prantzos95}, and explore
the role of 'galactic fountains', i.e. gas blown from the disk,
becoming part of a rotating hot corona, and eventually returning to
the disk \citep[e.g.,][]{Marinacci11}. Recent work by
\citet{Minchev12b} has introduced a new approach to modeling the
evolution of the Disk by combining detailed chemical evolution models
with cosmological $N$-body simulations.

To explain in particular the vertical disk structure
qualitatively different models have been put forward, including
cosmologically-motivated mechanisms where stars from a disrupted
satellite can be directly accreted \citep{Abadi03a}, or the disk can
be heated through minor mergers
\citep{Toth92a,Quinn93a,Kazantzidis08a,Villalobos08a,Moster10a} or
experience a burst of star formation following a gas-rich merger
\citep{Brook04a}. Alternatively, the disk could have been born with
larger velocity dispersion than is typical at $z \approx 0$
\citep{Bournaud09a,Ceverino12a}, or purely internal dynamical
evolution due to radial migration may gradually thicken the disk
\citep{Schoenrich09a,Loebman11a,Minchev12a}. It is likely that a
combination of these mechanisms is responsible for the present-day
structure of the Disk, but what relative
contributions of these effects should be expected 
has yet to be worked out in detail, and none of these have been 
convincingly shown to dominate the evolution of the Disk.

Radial migration is likely to have a large influence on the observable properties 
of the Disk, if its role in shaping the bulk properties of the disk were sub-dominant.
The basic process as described by \citet{SellwoodBinney}
consists of the scattering of stars at the corotation radius of
transient spiral arms; at corotation such scattering changes the
angular momentum, $L_z$, of the orbit ($\simeq$mean radius) without
increasing the orbital random energy. A similar---and potentially more
efficient---process was later shown to happen when the bar and spiral
structure's resonances overlap \citep{Minchev10a,Minchev11}. Such
changes in the mean orbital radii are expected to be of order unity within a few Gyrs,
and to have a profound effect on the
interpretation of the present-day structure of the Disk: \eg
present-day $L_z$ can no longer be used as a close proxy for the birth
$L_z$, even for stars on near-circular orbits.
Quantifying the strength of radial migration in the Milky Way is one
of the most pertinent action items for the next-generation of Milky Way
surveys.

A brief synthesis of the predictions from all these efforts is as
follows:
\begin{itemize}
\item Stellar disks should generically form from the inside out. More
  specifically, it is the low angular momentum gas that settles first
  near the centers of the potential wells forming stars at small
  radii.

\item Disk-dominated galaxies with disk sizes in accord with observations
  can emerge, if, and only if, there is no major merger since $z\simeq\!1$.

\item The luminosity- or mass-weighted radial stellar density profiles at late
  epochs resemble exponentials.

\item Stars that formed at earlier epochs ($z\ge 1$), when the gas
  fraction was far higher, do not form from gas disks that are as
  well-settled and thin as they are at $z\simeq\!0$ with dispersions of
  $\le 10\kms$.

\item Characteristic disk thicknesses or vertical temperatures of
  $400\pc$ and $\sigma_z\simeq\!25\kms$ are plausible.

\item Material infall, leading to fresh gas supply and dynamical disk
  heating, and star formation is not smooth but quite variable, even episodic, with a
  great deal of variation among dark matter halos with the same
  overall properties. It is rare that one particular infall or heating
  event dominates.
 
\item Throughout their formation histories galaxies exhibit
  significant non-axisymmetries, which at epochs later than $z\simeq\!1$
  resemble bars and spiral arms. Through resonant interactions, these
  structures may have an important influence on the evolution of
  stellar disks.
\end{itemize}

\subsection{The Disk and the Galactic gravitational potential}\label{sec:DiskGravPot}

Learning about the orbits of different Disk stellar sub-populations as a
galaxy formation constraint and learning from these stars about the
gravitational potential, $\Phi (\vec{x},t)$, are inexorably linked.
Those orbits are generally described by a {\sl chemo-orbital
  distribution function}, $p\bigl(\vec{J},\vec{\phi},\X, \age| \Phi
(\vx )\bigr )$ that quantifies the probability of being on an orbit labeled by
$\vec{J},\vec{\phi}$ for each subset of stars (characterized
by, e.g., by their ages $\age$ or their chemical abundances $\X =\feh
, \afe, \ldots$). Here, we chose to characterize the orbit (the
argument of the \DF) by $\vec{J}$, actions or integrals of motion
(which depend on both its observable instantaneous phase-space
coordinates $p(\vec{x},\vec{v})$ and on $\Phi (\vec{x},t)$). Each star
then also has an orbital phase (or angle), $\vec{\phi}$, whose
distribution is usually assumed to be uniform in $[0,2\pi]$.

Unless direct accelerations are measured for stars in many parts of
the Galaxy, many degenerate combinations of $\Phi (\vec{x},t)$ and
$p(\vec{J},\vec{\phi},\X, \age)$ exist, unless
astrophysically-plausible constraints and/or assumptions are imposed:
time-independent steady-state solutions, axisymmetric solutions,
uniform phase distributions of stars on the same orbit, etc. This is
the art and craft of stellar dynamical modeling
\citep{binneytremaine,Binney11a}.

Learning about the Galactic potential is one central aspect of such
dynamical modeling.  On scales larger than individual galaxies, the
so-called standard \LCDM\ cosmology has been tremendously successful
in its quantitative predictions. If certain characteristics for dark
energy and dark matter are adopted the large-scale matter and galaxy
distribution can be well explained (including baryon acoustic
oscillations) and be linked to the fluctuations in the cosmic
microwave background. On the scales of galaxies and smaller,
theoretical predictions are more complex, both because all of these
scales are highly non-linear and because the cooling baryons
constitute an important, even dominant mass component. Indeed, the
\LCDM\ paradigm seems to make at least two predictions that are
unsubstantiated by observational evidence, or even in seeming
contradiction. Not only are numerous low-mass dark-matter halos
predicted, almost completely devoid of stars
(\citealt{Kauffmann93a,Klypin99a,Moore99a}), \LCDM\ simulations also
predict that dark matter profiles are cuspy, i.e., have divergent dark
matter densities towards their centers
\citep{Dubinski91a,Navarro96a}. In galaxies as massive as our Milky Way,
baryonic processes could not easily turn the dark matter cusp into a
core, as seems viable in low mass galaxies
\citep[e.g.,][]{Flores93a,Pontzen12a}. Therefore, we should expect for
the Milky Way that more than half of the mass within a sphere of $\simeq
\Ro$ should be dark matter. Yet microlensing towards the Milky Way bulge
\citep{Popowski05a,Hadamache06a,Sumi06a} indicates that in our own
Galaxy most of the in-plane column density is made up of stars
\citep[e.g.,][]{Binney01a}.  Measurements in individual external galaxies 
remain inconclusive, because dynamical tracers only measure the total
mass, but cannot separate the stellar and DM contributions. In the
Milky Way, almost all stellar mass beyond the bulge is in a stellar
disk and hence very flat, while the DM halos emerging from
\LCDM\ simulations are spheroidal or ellipsoidal. So, mapping the
Milky Way's mass near the disk plane as a function of radius through
the vertical kinematics will break the so-called disk---halo degeneracy, when
combined with the rotation curve and
the outer halo mass profile: one can then separate the flat from the
round-ish mass contributions.

Perhaps the most immediate goal of dynamical Disk modeling is to
determine how much DM there is within the Solar radius: as little as
implied by the microlensing results, or as much as predicted by
\LCDM\ cosmology. A second goal for the Galactic potential is to
determine whether the disk-like total mass distribution is as thin as
the stellar counts imply; this is interesting, because the possibility
of a thick dark matter disk has been raised \citep{Read08a}. Finally,
we can combine precise constraints from near the Galactic disk with
constraints from stellar halo streams \citep[e.g.,][]{Koposov10a} to
get the shape of the potential as a function of radius. This can then
be compared to the expectations, e.g., of alternative gravity models,
and provide another, qualitatively new test for the inevitability of some form
of dark matter.

On the other hand, the best possible constraints on $\Phi (\vx )$ are
necessary to derive the (chemo-)orbital distribution of stars (the
`distribution function', \df), as the orbit characteristic, such as
the actions depend of course on both $(\vx , \vv )$ of the stars and
on $\Phi (\vx )$.

\section{Stellar surveys of the Milky Way}\label{sec:DiskMapping}
\subsection{Survey {\it desiderata}}

The ideal survey would result in an all-encompassing catalog of stars
throughout the Disk, listing their 3-D positions and 3-D velocities
($\vx , \vv$), elemental abundances (\abundances), individual masses
($M_*$), ages (\age), binarity, and line-of-sight reddening (A$_V$),
along with the associated uncertainties.  Furthermore, these
uncertainties should be `small',when compared to the scales on
which the multi-dimensional mean number density of stars $n(\vx, \vv,
\X, M_*, \age)$ has structure.  In practice, this is neither
achievable in the foreseeable future, nor is it clear that a `fuller'
sampling of $n(\vx, \vv, \X, M_*, \age)$ is always worth the
additional effort. Such an all-encompassing survey would imply that
the probability of entering the catalog is $p_{\mathrm{complete}}((\vx, \vv,
\X, M_*, \age)\simeq 1$ across the entire relevant
domain\footnote{For the Disk this domain is, e.g., $1\lesssim
  D_{GC}\lesssim20\kpc$, $|\vv|<600\kms$, etc.}  of \vecx, \vecv,
\abundances, $M_*$, and \age.

Any realistic survey is a particular choice of compromise in this
parameter space. Fundamentally, $p_{\mathrm{complete}}$ of any survey, i.e.,
the probability of any given star having ended up in the catalog, is
always limited by quantities in the space of `immediate observables', foremost
by the stars' fluxes (or signal-to-noise) vis-a-vis a survey's flux
limit or image-crowding limit. But these 'immediate observables' 
are usually not the quantities of foremost
astrophysical interest, say $\vx, \vv, \X, m, \age$. While
full completeness is practicable with respect to some quantities,
e.g., the angular survey coverage $(\ell,b)$ and the velocities \vecv ,
it is not in other respects. The effective survey volume will always
be larger for more luminous stars, the survey's distance limit will
always be greater in directions of lower dust extinction at the
observed wavelength; and an all-sky survey at fixed exposure time will
go less deep towards the bulge because of crowding.

So, in general, `completeness' is an unattainable goal. And while
samples that are `complete' in some physical quantity such as volume
or mass are immediately appealing and promise easy analysis, their
actual construction in many surveys comes at the expense of discarding
a sizable (often dominant) fraction of the pertinent catalog entries.
With the right analysis tools, understanding the survey
(in-)completeness, and its mapping into the physical quantities of
interest, is more important than culling `complete' samples. We will
return to this in Section\ref{sec:SelectionFunctions}.  .

\subsection{Observable quantities and physical quantities of interest}\label{sec:ObservableToPhysical}

The physical attributes about a star ($\vx, \vv, \X, m, \age$) that
one would like to have for describing and modeling the Disk are in
general not direct observables.  This starts out with the fact that
the natural coordinates for (\vecx , \vecv ) in the space of
observables are $(\ell,b,D)$ and $(\vlos , \mu_\ell, \mu_b)$, respectively,
with a heliocentric reference system in position and velocity.

In general, Galactic star surveys fall into two seemingly disjoint
categories: imaging and spectroscopy. Imaging surveys get parsed into
catalogs that provide angular positions and fluxes (typically in 2 to
10 passbands) for discrete sources, once photometric solutions
\citep[e.g.,][]{Schlafly12} and astrometric solutions \citep{Pier03a}
have been obtained.  Multi-epoch surveys, or the comparison of different
surveys from different epochs then provide proper motions, and---with
sufficient precision---useful parallaxes
\citep[e.g.,][]{Perryman97,Munn}. Spectroscopic surveys are carried
out with different instrumentation or even as disjoint surveys, usually based
on a pre-existing photometric catalog.  These spectra, usually for
vastly fewer objects, provide $\vlos$ on the one hand and the spectra
that allow estimates of the `stellar parameters', \teff,  \logg, and \X
\citep[e.g.,][]{Nordstroem04,Yanny09}.  But `imaging' and
`spectroscopic' surveys are only seemingly disjoint categories,
because multi-band photometry may be re-interpreted as a (very) low
resolution spectrum and some stellar parameters (e.g., T$_{\mathrm{eff}}$ and
\feh) can be constrained by photometry and/or spectra
\citep[e.g.,][]{Ivezic2008}.

\begin{figure}
  \centering
  \includegraphics[width=0.65\textwidth,angle=90.]{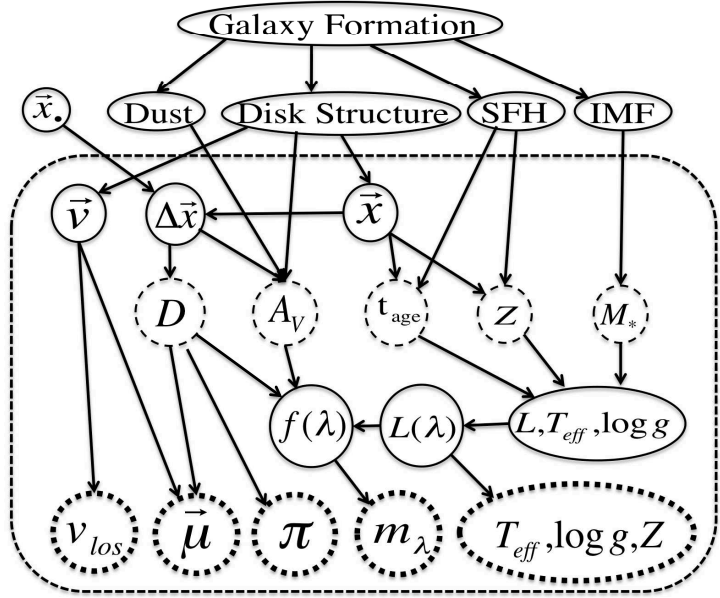}
  \caption{ Constraining Stellar Parameters from Observables in Milky
    Way Surveys: this Figure provides a schematic overview, in the form of a
    simplified graphical model, of the logical dependencies between the
    {\sl stellar observables} in Galactic surveys (thick dotted ovals)
    and the main {\it desiderata} for each star (thin dashed ovals),
    its {\sl stellar parameters} and {\sl distance}, given various
    prior expectations about galaxy formation, star formation and the
    Galactic dust distribution (top).  The basic observables are:
    line-of-sight-velocity, $v_{los}$, proper motions, $\vec{\mu}$,
    parallax $\pi$, multi-band photometry $m_{\lambda_i}$ and
    photospheric parameters derived from spectra ($\Teff$, $\logg$,
    abundances, $Z$); most of them depend on the Sun's position 
    $x_\odot$ through, $\Delta \vec{x}$.
     The main {\it desiderata} are the star's mass
    $M_*$, age \age and abundances $Z$, along with its distance $D$
    from the Sun and the (dust) extinction along the line of sight,
    $A_V$.  The prior probabilities of $M_*$, age \age , $Z$, $D$ and
    \AV are informed by our notions about star formation (the IMF) the
    overall structure of the Galaxy and various constraints on the
    dust distribution.  Overall the goal of most survey analysis is to
    determine the probability of the {\sl stellar observables} for a
    given set of {\sl desiderata}, which requires both isochrones and
    stellar atmospheric models (see \citealt{BurnettBinney10}).  
    In practice, most existing Galactic surveys analyses can mapped
    onto this scheme, with logical
    dependencies often replaced by  assumed logical
    conditions (e.g. `using dereddened fluxes', `presuming the star is
    on the main sequence', etc.).  This graphical model still makes number of
    simplifications and the velocities.}
  \label{fig:GraphicalModel}
\end{figure}

The practical link between `observables' and `quantities of interest'
warrants extensive discussion: along with the survey selection
function (Section~\ref{sec:SelectionFunctions}) it is one of the two
key ingredients for any rigorous survey {\sl analysis}.  For some
quantities, such as $(\ell,b)$ and $\vlos$ this link is quite direct; but
even there a coordinate transformation is required, which involves
$R_\odot$, $v_{\mathrm{LSR}}$ and of course the distances, as most dynamical or
formation models are framed in some rest-frame system with an origin
at the Galactic center.  Similarly, converting estimates of
$\mu_\ell,\mu_b$ to two components of $\vv$ (and $\delta\vv$) requires
knowledge of the distance.  In \figurename~\ref{fig:GraphicalModel},
we show a {\sl graphical model} overview over the task of
`astrophysical parameter determination' in Galactic surveys. In the
subsequent sub-sections, we discuss different aspects of this model.

\subsubsection{Distance estimates}\label{sec:Distances}

Clearly, a `direct' distance estimate, one that is independent of any
intrinsic property of the stars, such as parallax measurements, is to
be preferred. At present, good parallax distance estimates exist for
about 20,000 stars within $\simeq\!100\pc$ from Hipparcos
\citep{Perryman97,ESA97a}. And a successful Gaia mission will extend
this to $\simeq\!10^9$ stars within $\simeq\!10\kpc$
\citep{deBruijne2012}. However, a widely usable catalog with Gaia
parallaxes is still 5 years away as of this writing, and even after
Gaia most stars in the already existing wide-field photometric surveys
(e.g., SDSS or PS1) will not have informative Gaia parallax estimates,
simply because they are too faint.

\begin{figure}
  \centering
  \includegraphics[width=0.60\textwidth,clip=]{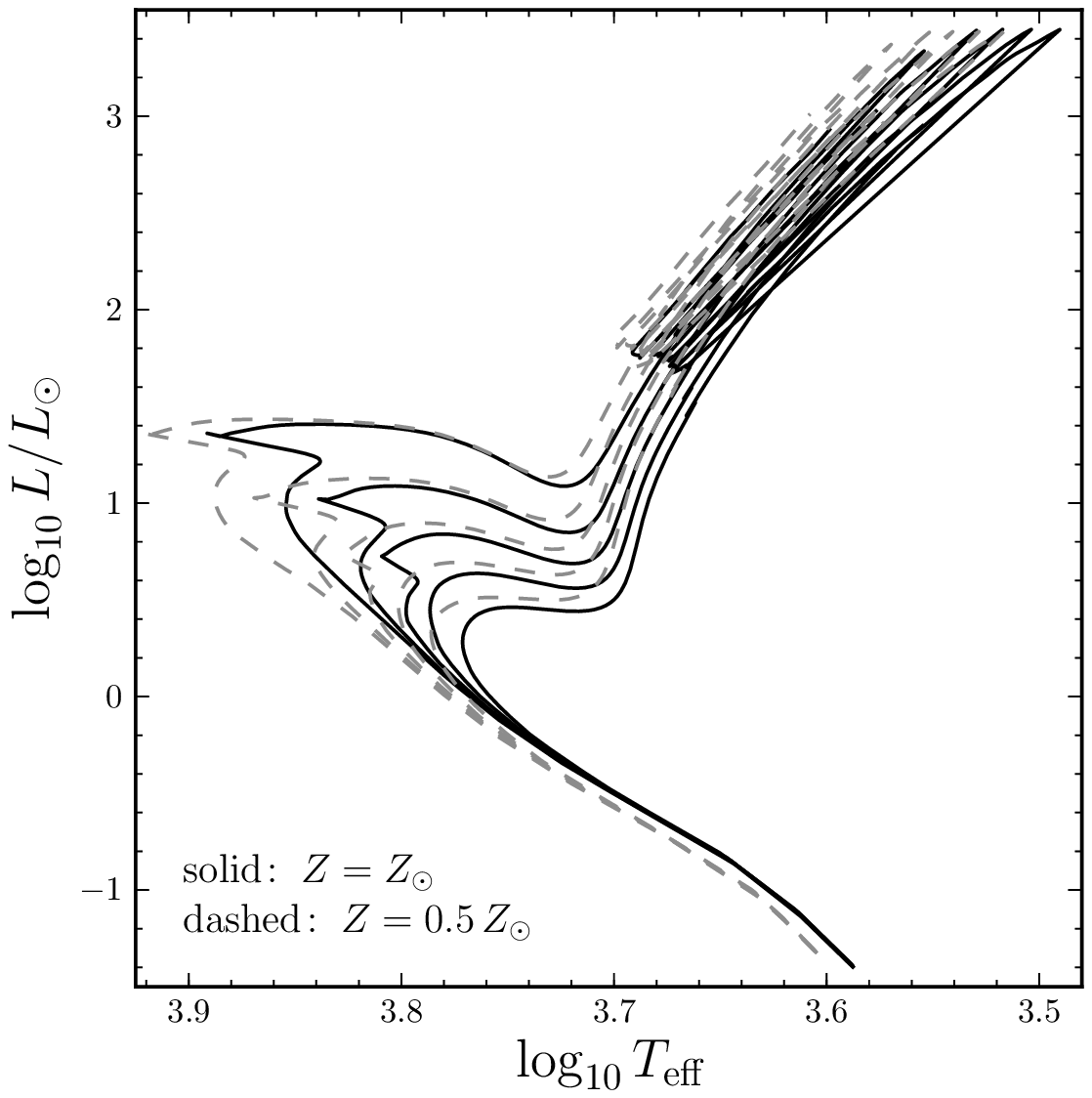}
  \caption{`Physical' color-magnitude diagram, shown by BaSTI
    isochrones for two metallicities, which are spaced $0.2\dex$ in age
    between $10^9$ and $10^{10}$\,yr. This diagram serves as a
    reference for the discussion of how to derive physical parameters
    (distances, ages, abundances) for stars from observables (cf.
    Section~\ref{sec:ObservableToPhysical}): the age-independence of
    the luminosity on the MS makes for robust photometric distance
    estimates; the age and abundance sensitivity of $L$ and $\Teff$
    near the turn-off and on the (sub-)giant branch makes for good age
    determinations, but only if parallax-based distances exist; etc.}
  \label{fig:isochrones}
\end{figure}

Therefore, it is important to discuss the basics and the practice of
(spectro-)photometric distance estimates to stars, which has turned
out to be quite productive with the currently available data
quality. Such distance estimates, all based on the comparison of the
measured flux to an inferred intrinsic luminosity or absolute
magnitude, have a well-established track record in the
Galaxy, when using so-called
standard candles, such as BHB stars \citep{Sirko04,Xue08} or `red
clump stars' in near-IR surveys \citep[e.g.,][]{Alves02}.  But of
course, as the `physical HR diagram' ($L$ vs. $\Teff$) in
\figurename~\ref{fig:isochrones} shows, the
distance-independent observables greatly constrain luminosity and
hence distance for basically all stars. Foremost, these 
distance-independent observables are $\Teff$ or colors
as well \abundances, potentially augmented by spectroscopic $\logg$
estimates.

A general framework for estimating distances in the absence of
parallaxes is given in \citet{BurnettBinney11} or \citet{Breddels10},
and captured by the graphical model in
\figurename~\ref{fig:GraphicalModel}. Basically, the goal is to
determine the distance modulus likelihood
\begin{equation}\label{eq:photdist}
\begin{split}
\ldm =\int\!\!\!\int & p\bigl(\{m_i,\feh\}|
\DM,M_*,\feh_{\mathrm{mod}},\tage\bigr)p_{\mathrm{prior}}\bigl(M_*\bigr)\\ & p_{\mathrm{prior}}\bigl(\feh,\tage|\DM\bigr)dM_*\ d\feh_{\mathrm{mod}}\ d\tage\,,
\end{split}
\end{equation}
where the $m_i$ are the apparent magnitudes of the star in various
passbands (i.e., also the `colors') and $M_*$ is its stellar mass,
which predicts the star's luminosity and colors (at a given \feh\ and
\tage, see \figurename~\ref{fig:GraphicalModel}). Note that $\feh$
appears both as an observational constraint and a model parameter,
which determines isochrone locations. The prior on (\feh,\tage) may depend on
the distance modulus to use the different spatial distribution of
stars of different metallicities (in this case, the position of the
sky would also be included to compute the 3D position).

To get distance estimates, one needs to marginalize over the nuisance
parameters ($M_*$, \feh, and \tage\ in the case above), and one needs to
spell out the external information (i.e. priors) on the relative
probability of these nuisance parameters
\citep[e.g.,][]{BailerJones11}. As \figurename~\ref{fig:isochrones}
makes obvious, the precision of the resulting distance estimates
depends on the evolutionary phase, the quality of observational
information, and any prior information. To give a few examples: for
stars on the lower main sequence, the luminosity is only a function of
\teff\ or color, irrespective of age; hence it can be well determined
($\simeq\!5-10\%$) {\it if} the metallicity is well constrained
($\delta\feh\le 0.2$) and if the probability of the object being a
(sub-)giant is low, either on statistical grounds or through an
estimate of $\logg$. \citet{Juric2008} showed that such precisions can
be reached even if the metallicity constraints only come from
photometry.  On the giant branch, colors reflect metallicity (and also
age) as much as they reflect luminosity; therefore photometric
distances are less precise. Nonetheless, optical colors good to
$\simeq\!0.02\,\mathrm{mag}$ and metallicities with $\delta\feh\le0.2$
can determine giant star distances good to $\simeq\!15\%$, if there is a
prior expectation that the stars are old (Xue \etal\ 2012, in
preparation). The power of spectroscopic data comes in determining
\logg, \teff\ especially in the case of significant reddening, and in
constraining the abundances, especially \feh.  While only
high-dispersion and high-S/N spectra will yield tight $\logg$
estimates, even moderate-resolution and moderate-S/N spectra (as for
the SEGUE and LAMOST surveys; see Section~\ref{sec:SpecSurveys})
suffice to separate giants from dwarfs \citep[e.g.,][]{Yanny09},
thereby discriminating the multiple branches of
$L(\mathrm{color},\feh)$ in the isochrones. While main-sequence
turn-off stars have been widely used to make 3D maps of the Milky Way
(mostly its halo; e.g., \citealt{Belokurov06a,Juric2008,Bell08}), they
are among the stellar types abundantly found in typical surveys the
ones for which precise ($<20\%$) spectrophotometric distances are
hardest to obtain \citep[e.g.,][]{SchoenrichDistances11}. 

The absolute {\it scale} for such spectro-photometric distance
estimates is in most cases tied to open or globular clusters, which are
presumed to be of known distance, age and metallicity
\citep[e.g.,][]{Pont98,An09}. Because the number of open and globular
clusters is relatively small and, in particular, old open clusters are
rare, the sampling in known age and metallicity is sparse and
non-uniform. Therefore, care is necessary to make this work for
Disk stars of any age or metallicity. Interestingly, the panoptic sky surveys
with kinematics offer the prospect of self-calibration of the distance scale
\citep{Schoenrich12a} through requiring dynamical consistency between
radial velocities and proper motions.

\subsubsection{Chemical abundances}\label{sec:ChemicalAbundances}

\newcommand{\mh}{\ensuremath{[\mathrm{M/H}]}}

Chemical abundances are enormously important astrophysically, as they
are witnesses of the Disk's enrichment history and as they are lifelong tags
identifying various stellar sub-populations. Broadly speaking, metals
are produced in stars as by-products of nuclear burning and dispersed
into the interstellar medium through supernova explosions and winds. This leads
to a trend toward higher metallicity as time goes on, with inside-out
formation leading to faster chemical evolution or metal-enrichment in
the inner part of the Disk. While all supernovae produce iron,
$\alpha$-element enrichment occurs primarily through type II
supernovae of massive stars with short lifetimes. Therefore, until
type Ia supernovae, with typical delay times of $2\Gyr$
\citep[e.g.,][]{Matteucci01a,Dahlen08a,Maoz11a}, start occurring, the
early disk's ISM, and stars formed out of it, has \afe\ that is higher
than it is today. For a smooth star formation history, \afe\ decreases
monotonically with time. However, a burst of star formation, e.g.,
following a gas-rich merger, could re-instate a higher \afe, such that
the relation between \afe\ and age is dependent on the star-formation
history \citep[e.g.,][]{Gilmore91a}.

Beyond the astrophysical relevance of abundance information, it is
also of great practical importance in obtaining distances
(Section~\ref{sec:Distances}) and ages (Section~\ref{sec:Ages}).
Among the `abundances' $\X$, the `metallicity' $\feh$ is by far of the
greatest importance for distance estimates. Practical determinations
of $\X$ fall into three broad categories: estimates (usually just
$\feh$, or \mh\ as a representative of the overall metal content) from
either broad- or narrow-band photometry; estimates based on
intermediate-resolution spectra (R=2000 to 3000); or estimates from
high-resolution spectra (R=10,000 and higher) that enable individual
element abundance determinations. We discuss here briefly abundances
obtained in these three different regimes.

\begin{itemize}
\item {Stellar Abundances from Photometry:} That abundances can be
  constrained through photometry---at least for some stellar
  types---has been established for half a century
  \citep{Wallerstein62,Stroemgren66}. The level of overall metal-line
  blanketing or the strength of particular absorption features varies
  with $\feh$ \citep[e.g.,][]{BohmVitense89}, which appreciably
  changes the broad-band fluxes and hence colors. How well this can be
  untangled from changes in \teff\ or other stellar parameters,
  depends of course on the mass and evolutionary phase of the
  star. The most widely used applications are the `color of the red
  giant branch', which is a particularly useful metallicity estimate
  if there are priors on age and distance
 (\eg  \cite{McConnachie10} for the case of M31).  
 In the Galaxy, variants of the `UV
  excess method' \citep{Wallerstein62} have been very successfully
  used to estimate $\feh$ for vast numbers of stars. In particular,
  \citet{Ivezic2008} calibrated the positions of F \& G main sequence
  stars in the SDSS $(u-g)\ - \ (g-r)$ color plane against
  spectroscopic metallicity estimates.  Based on SDSS photometry, they
  showed that metallicity precisions of $\delta\feh\simeq\!0.1$ to $0.3$
  can be reached, depending on the temperature and metallicity of the
  stars: in this way they determined $\feh$ for over 2,000,000 stars.

  Some narrow- or medium-band photometric systems (foremost
  Str\"{o}mgren filters; \citealt{Stroemgren66,Bessell11a}) place
  filters on top of string absorptions features, to constrain stellar
  parameters in particular metallicities can do even better in
  estimating $\feh$. Recently, this has been well illustrated by
  \citet{Arnadottir10a} and \citet{Casagrande11a}, who compared
  Str\"{o}mgren photometry to abundances from high-resolution spectra,
  showing that $\feh$ precisions of $\le 0.1\dex$ are possible in some
  $\Teff$ regimes. \citet{Casagrande11a} also showed that precise
  Str\"{o}mgren photometry can constrain $\afe$ to $\simeq 0.1$ dex. In
  addition, narrow band photometry can provide good metallicity
  estimates ($\delta\feh\simeq 0.2$) for a wider range of stellar
  parameters than SDSS photometry.

  Indeed, the literature is scattered with disputes that a certain
  photometric abundance determination precision is untenable, because
  the same approach yields poor $\feh$ estimates for a sample at
  hand. In many cases, this seeming controversy can be traced back to
  the fact that different approaches work vary radically in their
  accuracy across the $\Teff , \logg $, \age\ parameter plane.

  The abundance information from Gaia will be based on a spectral
  resolution of only R=15 to 80 for most stars; and while these data
  are produced through a dispersive element, they are probably best
  thought of as many-narrow-band photometry; for FGKM stars with
  $g<19$ Gaia data are expected to yield $\delta\feh\lesssim 0.2$
  \citep{Liu12a}.

\item {Stellar Abundances from High-Resolution Spectroscopy:}
  High-resolution spectroscopy (R$\gtrsim 10000$), either in the
  optical or in the near-IR \citep{SteinmetzRAVE,AllendePrieto08}, is
  an indispensable tool to obtain individual element abundances and to
  anchor any abundance scale in physically-motivated models of stellar
  photospheres
  \citep[e.g.,][]{Asplund09,Edvardsson93,Reddy03a,Bensby05a}. The
  abundances are then fit by either modeling the spectrum directly in
  pixel or flux space, or more commonly by parsing the observed and
  the model spectrum into a set of line equivalent widths which are
  compared in a $\chi^2$-sense.  From typical existing spectra, about
  10 to 20 individual abundances are derived, with relative precisions
  of 0.05 to 0.2 \citep[e.g.,][]{Reddy03a,Boeche11}.  High resolution
  spectroscopy has remained the gold standard for investigating
  chemo-dynamical patterns in the Milky Way in detail.

\item {Stellar Abundances from Moderate-Resolution Spectroscopy:} In
  recent years, large data sets at moderate spectral
  resolution\footnote{The RAVE survey, at $R=7500$, is considered
    moderate resolution by some self-respecting stellar
    spectroscopists} have become available, foremost by SDSS/SEGUE
  \citep{Yanny09} in the last years, but in the future also from
  LAMOST data \citep{Deng12}. Such data can provide robust
  metallicities $\feh$, good to $\delta\feh\lesssim 0.2$, and
  constrain \afe\ to$\delta\afe\le 0.15$ \citep{Lee08a,Lee11a}, for F,
  G and K stars. The accuracy of these data been verified against
  high-resolution spectroscopy and through survey spectra in globular
  clusters of known metallicity \citep{Lee08b,Lee11a}. Recently,
  \citet{BovyMAPkinematics} have pursued an alternate approach to
  determine the {\it precision} (not accuracy) if abundance
  determinations $\feh , \afe$, through analyzing the
  abundance-dependent kinematics of Disk stars. They found that for
  plausible kinematical assumptions, the SDSS spectra of G-dwarfs must
  be able to rank stars in $\feh$ and $\afe$ at the $(0.15,0.08)\dex$
  level, respectively.  These precisions should be compared to the
  range of $\feh$ (about $1.5\dex$) and $\afe$ (about $0.45\dex$) of
  abundances found in the Disk, which illustrates that moderate
  resolution spectroscopy is very useful for isolating
  abundance-selected subsamples of stars in the Disk.

\end{itemize}

\subsubsection{Stellar ages}\label{sec:Ages}

Constraints on stellar ages are of course tremendously precious
information for understanding the formation of the Disk, yet are very
hard to obtain in practice. The review by \citet{Soderblom10} provides
an excellent exposition of these issues, and we only summarize a few
salient points here.  For the large samples under discussion here, the
absolute age calibration is not the highest priority, but the aim is
to provide age constraints, even if only relative-age constraints, for
as many stars as possible. In the terminology of \citet{Soderblom10},
it is the `model-based' or `empirical' age determinations that are relevant here:
three categories of them matter most for the current context
($0.5\Gyr\le \age\le13\Gyr$), depending on the type of stars and the
information available.  First, the chromospheric activity or rotation
decay with increasing age of stars, leading to empirical relations
that have been calibrated against star clusters
\citep[e.g.,][]{Baliunas95}, especially for stars younger than a few
Gyr. \citet{Soderblom91} showed that the expected age precision for
FGK stars is about 0.2 dex. Second, stellar seismology, probing the
age-dependent internal structure can constrain ages well; the advent
of superb light-curves from the Kepler mission, has just now enabled
age constraints---in particular for giant stars---across sizable
swaths of the Disk \citep{vanGrootel10a}.

The third, and most widely applicable approach in the Disk context, is
the comparison between isochrones and a star's position in the
observational or physical Hertzsprung-Russell diagram, i.e., the $L$
(or $\logg$) - $\Teff , \feh$ plane (see
\figurename~\ref{fig:isochrones}). Ideally, the set of observational
constraints for a star, $\{\mathrm{data}\}_{\mathrm{obs}}$, would be precise
determinations of $\{\mathrm{data}\}_{\mathrm{obs}} = \{L, \Teff , \feh\}$, or
perhaps precise estimates of $\logg, \Teff , \feh$. But until good
parallaxes to $D\ge 1\kpc$ exist, $L$ is poorly constrained (without
referring to $\Teff$ and $\feh$ etc.) and the error bars on $\logg$
are considerable ($\simeq 0.5\dex$), leaving $\Teff , \feh$ as the
well-determined observables.

All these observable properties of a star depend essentially only on a
few physical model parameters for the star $M_*,\feh,\age$
(\figurename~\ref{fig:GraphicalModel}), and the observables (at a
given age) can be predicted through isochrones
\citep[e.g.,][]{Girardi02,Pietrinferni04}.  For any set of
observational constraints, the probability of a star's age is then
given by a similar expression as that for photometric distances in
\eqnname~(\ref{eq:photdist}) \citep[cf.][]{Takeda07,BurnettBinney10}:
\begin{equation}
\begin{split}
    \label{eq:ages}
\mathcal{L}(\{\mathrm{data}\}_{\mathrm{obs}}| \tage) =\int\!\!\!\int & p\bigl(\{\mathrm{data}\}_{\mathrm{obs}}|
M_*,\feh,\age\bigr)\\ & \,p_p(M_*|\age) p_p(\feh|\tage)\,dM_*d\feh,
\end{split}
\end{equation}
where $p(\{\mathrm{data}\}_{\mathrm{obs}}| M_*,\feh,\tage)$ is the
probability of the data given the model parameters (i.e., the
likelihood); the shape of that distribution is where the uncertainties
on the observables are incorporated.  Further, not all combinations of
$M_*, \feh,\age$ are equally likely, since we have prior information
on $p_p(M_*|\age)$ and $p_p(\feh|\age)$. Those prior expectations come
from our overall picture of Galaxy formation, from our knowledge of
the stellar mass function, or they may reflect the sample properties,
when analyzing any one individual star. To be specific,
$p_p(M_*|\age)$ simply reflects the mass function, truncated at
$M_{*,max}(\age)$, if one accepts that there has been an approximately
universal initial mass function in the Disk.

The integration, or marginalization, over $M_*$ and $\feh$, of course
involves the isochrone-based prediction of observables
$\{\mathrm{params}\}_{\mathrm{ic}}$, as illustrated in
\figurename~\ref{fig:GraphicalModel}, so that
\begin{equation}
    \label{eq:ages2l}
p\bigl(\{\mathrm{data}\}_{\mathrm{obs}}| M_*,\feh,\age\bigr)\rightarrow p\Biggl(\{\mathrm{data}\}_{\mathrm{obs}}| \{\mathrm{params}\}_{\mathrm{ic}}\bigl (M_*,\feh,\age\bigr )\ \Biggr),
\end{equation}
where $\{\mathrm{params}\}_{\mathrm{ic}}$ could for example be $\{L, \Teff ,
\feh\}_{\mathrm{ic}}\bigl (M_*,\feh,\age\bigr )$.

To get the relative probabilities of different presumed ages,
$\mathcal{L}(\{\mathrm{data}\}_{\mathrm{obs}}|\age)$, one simply goes over all
combinations of $M_*$ and $\feh$ at $\age$ and one integrates up how
probable the observations are for each combination of $M_*$ and
$\feh$.  This has been put into practice for various large samples,
where luminosity constraints either come from parallaxes or from
$\logg$ (see \citealt{Nordstroem04,Takeda07,BurnettBinney10}).

The quality of the age constraints depends dramatically both on the
stellar evolutionary phase and on the quality of the observational
constraints.  It is worth looking at a few important regimes, using
the isochrones for two metallicities in
\figurename~\ref{fig:isochrones} as a guide. Note that
stars with ages $>1\Gyr$, which make up over $90\%$ of Disk,
correspond to the last 4 isochrones in this Figure.  

1) For stars on
the lower main sequence (where main sequence lifetimes exceed
$10\Gyr$; $L\lesssim L_\odot$), isochrone fitting provides basically
no age constraints at all; in turn, photometric distance estimates are
most robust there.  For stars on the upper main sequence,
\eqnname~(\ref{eq:ages}) `automatically' provides a simple upper limit
on the age, given by the main sequence lifetime. 

 2) For stars near
the main-sequence turn-off and on the horizontal branch, the
isochrones for a given metallicity are widely spread enabling good age
determinations: \citet{Takeda07} obtained a relative age precision for
their stars with $\logg < 4.2$ of $\simeq\!15\%$.  Similarly, the
parallax-based luminosities of the stars in the Geneva-Copenhagen
Survey (GCS) provide age estimate for stars off the main sequence of
$\simeq\!20\%$. Note that the $\feh$ marginalization in
\eqnname~(\ref{eq:ages}) and the strong metallicity dependence of the
isochrones in \figurename~\ref{fig:isochrones} illustrate
how critical it is to have good metallicity estimates: without very
good metallicities, even perfect Gaia parallaxes (and hence perfect
luminosities) will not yield good age estimates across much of the
color--magnitude diagram.  3) On the red giant branch, there is some
$L-\Teff$ spread in the isochrones of a given metallicity; but for
ages $>1\Gyr$ the metallicity dependence of $\Teff$ is so strong as to
preclude precise age estimates.

Note that the formalism of \eqnname~(\ref{eq:ages}) provides at least
some age constraints even in the absence of good independent $L$
constraints \citep[e.g.,][]{BurnettBinney10}, as the different stellar
phases vary vastly in duration, which enters through the strong $M_*$
dependence in \eqnname~(\ref{eq:ages}).

\subsubsection{Interstellar extinction}\label{sec:Extinction}

The complex dust distribution in the Galaxy is of course very
interesting in itself \citep[e.g.,][]{Jackson08}, delineating spiral
arms, star formation locations, and constraining the Galactic matter
cycle. For the study of the overall stellar distribution of the Milky
Way, it is foremost a nuisance and in some regimes, e.g., at very low
latitudes, a near-fatal obstacle to seeing the entire Galaxy in stars
\citep[e.g.,][]{Nidever12a}.  Unlike in the analysis of galaxy
surveys, where dust extinction is always in the foreground and can be
corrected by some integral measure $A_\lambda(\ell,b)$ \citep[from,
  e.g.,][]{SFD98}, one needs to understand the 3D dust distribution in
the Galaxy, $A_\lambda(\ell,b,D)$.  On the one hand, one needs to
know, and marginalize out, the extinction to each star in a given
sample, foremost to get its intrinsic properties. On the other hand,
for many modeling applications one needs the full 3D extinction
information $A_\lambda(\ell,b,D)$, also in directions where there are
no stars, e.g., in order to determine the `effective survey volume'
(see below): it obviously makes a difference whether there are no
stars in the sample with a given $(\ell,b,D)$ because they are truly
absent or because they have been extinguished below some sample flux
threshold.

Practical ways to constrain the extinction to any given star, given
multi-band photometry or spectra, have recently been worked out by,
e.g., \citet{BailerJones11} and \citet{Majewski11} for different
regimes of $A_v$.  In the absence of parallax distances or spectra for
the stars, estimates of $A_\lambda(\ell,b)$ inevitably involve mapping
the stars' SED back to a plausible unreddened SED, the so-called
stellar locus. This de-reddening constrains the amount of extinction,
but only through the wavelength variation of the extinction, the
reddening.
 
Estimates of $\Teff$ or $L$ from spectra or from well-known parallax
distances, allow of course far tighter constraints, because the
de-reddened colors are constrained a priori and because also the
`extinction' not just the `reddening' appears as constraints.

How to take that star-by-star extinction information, potentially
combine it with maps of dust emissivity
\citep[e.g.,][]{PlanckCollaboration11}, to get a continuous estimate
of $A_\lambda(\ell,b,D)$ has not yet been established. This task is not
straightforward, as emission line maps indicate that the fractal
nature of dust (and hence) extinction distribution continues to very
small scales.  Presumably, some form of interpolation between the
$A_\lambda(\ell,b,D)$ to a set of stars, exploiting the properties of
Gaussian processes, is a sensible way forward.

\subsubsection{Towards an optimal Disk survey analysis}

The task ahead is now to lay the foundation for an optimal survey {\sl
  analysis} that will allow optimal dynamical modeling, by rigorously
constructing the \pdf\ for the physical quantities of interest, $\vx,
\vv, \age , \X, M_*$ from the direct observables.  The preceding
subsections show that many sub-aspects of the overall approach
(\figurename~\ref{fig:GraphicalModel}) have been carried out and
published; and a general framework has been spelled out by
\citet{BurnettBinney10}. What is still missing is a comprehensive
implementation that uses the maximal amount of information and
accounts for all covariances, especially one that puts data sets from
different surveys on the same footing.

\subsection{Existing and current Disk surveys}\label{sec:CurrentDiskSurveys}

There is a number of just-completed, ongoing, and imminent surveys of
the stellar content of our Galaxy, all of which have different
strengths and limitations. Tables 1 and 2 provide a brief overview of
the most pertinent survey efforts. The (ground-based) surveys fall
into two categories: wide-area multi-color imaging surveys and
multi-object, fiber-fed spectroscopy. Broadly speaking, the
ground-based imaging surveys provide the angular distribution of stars
with complete (magnitude-limited) sampling, proper motions at the
$\simeq\!3\,\mathrm{mas\ yr}^{-1}$ level (in conjunction with earlier
imaging epochs), and photometric distances; for stars of certain
temperature ranges (F through K dwarfs) they can also provide
metallicity estimates.

The ground-based spectroscopic surveys have all been in `follow-up'
mode, i.e., they select their spectroscopic targets from one of the
pre-existing photometric surveys, using a set of specific targeting
algorithms. In most cases, the photometric samples are far larger than
the number of spectra can be taken, so target selection is a severe
downsampling, in sky area, in brightness or in color range. The survey
spectra provide foremost radial velocities, along with good stellar
photospheric parameters, including more detailed and robust elemental
abundances.

In this Section we restrict ourselves to surveys that have started
taking science-quality data as of Summer 2012, with the exception of
Gaia.

\begin{deluxetable}{llllll}  
\tabletypesize{\scriptsize}
\tablecolumns{6}
\tablewidth{0pt}
\tablecaption{Stellar Photometric Surveys of the Milky Way}
\tablehead{   
  \colhead{Survey} &
  \colhead{Period} &
  \colhead{Sky Area} &
  \colhead{\# of Filters} &
  \colhead{mag lim.} &
  \colhead{$\delta \feh$}
}

\startdata

2MASS  & 1998 - 2002     & all sky  & 5& H=15& N/A\\
{\tiny \citealt{Skrutskie06}}& &40,000 deg$^2$ & $1.2\mu - 2.2\mu$& & \\  

SDSS I-III  & 2002 - 2012     & North, l$>$20$^\circ$  & 5& g=22& 0.2\\
{\tiny \citealt{Eisenstein11a}} & &15,000 deg$^2$ & $0.4\mu - 0.9\mu$& & \\  

PanSTARRS1 & 2011 - 2013     & $\delta> -20$  & 5& g=22& 0.4\\
{\tiny \citealt{Kaiser02}}& &30,000 deg$^2$ & $0.5\mu - 1.0\mu$& & \\  

VHS & 2010 - 2015     & South  & 5& J=20& N/A\\
{\tiny McMahon \etal, 2012, in prep.}& &20,000 deg$^2$ &$1.2\mu - 2.2\mu$ & & \\  

SkyMapper & 2012 - 2014     & South  & 5& g=21& 0.1\\
{\tiny \citealt{Keller07a}}& &15,000 deg$^2$ & & & 

\enddata

\end{deluxetable}

\begin{deluxetable}{llllllll}  
\tabletypesize{\scriptsize}
\tablecolumns{8}
\tablewidth{0pt}
\tablecaption{Stellar Spectroscopy Surveys of the Milky Way}
\tablehead{   
  \colhead{Survey} &
  \colhead{Period} &
  \colhead{Sky Area} &
  \colhead{\# of Spectra}&
    \colhead{app. mags} &
  \colhead{$\delta$v [km/s]} &
  \colhead{$\delta$[Fe/H]} &
  \colhead{char. distance}
}
\startdata
GCS&1981-2000 &South&16,000&V$\simeq$10?&0.5&indiv&0.003kpc\\
SEGUE I+II & 2004-2009& North, l$>$20$^\circ$& 360,000& g=15-20& 8 & 0.2&2 kpc\\  
RAVE & 2003?2012& South& 370,000+& i=9-12& 3 & 0.2&0.5 kpc\\
APOGEE & 2011-2014& North, l$<$20$^\circ$& 100,000& H$<$13.8& 0.5 & indiv. &10 kpc\\ 
Gaia-ESO & 2012-2015& South&150,000& V$<$18& 0.5& indiv. &4 kpc\\ 
LAMOST & 2012-2018& North& 3,000,000& V$<$18& 10& 0.2&4 kpc\\
Gaia & 2013-2018& all sky& 50,000,000& V$<$16& 10& 0.25&4 kpc\\
\enddata
\end{deluxetable}

\subsubsection{Individual photometric surveys}\label{sec:PhotoSurveys}
\begin{itemize}
\item {\bf 2MASS } \citep{Skrutskie06}: Designed as an `all purpose'
  near-infrared (JHK) imaging survey, it has produced the perhaps most
  striking and clearest view of the Milky Way's stellar distribution
  to date with half a Billion stars, owing to its ability to
  penetrated dust extinction better than optical surveys. Its imaging
  depth is sufficient to see, albeit not necessarily recognize,
  modestly extinct giant stars to distances $>10\kpc$. It has been
  very successfully used to map features in the Milky Way's outskirts, e.g.,
  the Sagittarius stream \citep{Majewski03a} and the Monoceros feature
  in the outer Disk \citep{RochaPinto03a}. It has been the photometric
  basis for several spectroscopic surveys, in the Disk-context most
  notably APOGEE (see below). However, as a stand-alone survey, it has
  done not very much to advance our knowledge of the Disk, owing to
  the limitations of deriving robust distances and abundances from its
  data. But 10 years after its completion, it is still the only
  all-sky survey in the optical/near-IR region of the electromagnetic
  spectrum. In imaging at $\ge 10,000$ square degree coverage, 2MASS
  is now being surpassed (by 4 magnitudes) by the {\sl Vista
    Hemisphere Survey (VHS)} (McMahon \etal, 2012, in preparation).

\item {\bf SDSS} \citep[e.g.,][]{York00,Stoughton02a}: The primary
  science goals of the (imaging \& spectroscopy) Sloan Digital Sky
  Survey were focused on galaxy evolution, large-scale structure and
  quasars, and the 5-band imaging survey `avoided' much of the Milky
  Way by largely restricting itself to $|b| >30^\circ$.  Nonetheless,
  SDSS imaging has had tremendous impact on mapping the Galaxy. Like
  2MASS, its impact has been most dramatic for understanding the
  outskirts of the Milky Way, where its imaging depth (giant stars to
  $100\kpc$, old main-sequence turn-off stars to $25\kpc$), its
  precise colors and its ability to get photometric metallicity
  constraints \citep{Ivezic2008} have been most effective. On this
  basis, SDSS has drawn up a state-of the art picture of the overall
  stellar distribution \citep{Juric2008}, drawn the clearest picture
  of stellar streams in the Milky Way halo \citep{Belokurov06a}, and expanded
  the known realm of low-luminosity galaxies by two orders of
  magnitude \citep[e.g.,][]{Willman05a}. SDSS photometry and proper
  motions ($\delta \vec{\mu}\simeq 3\,\mathrm{mas\ yr}^{-1}$ through
  comparison with USNOB, \citep{Munn} have allowed a kinematic
  exploration of the Disk \citep{Fuchs09a,Bond10a}. However, the
  bright flux limit of SDSS ($g\simeq\!15$) makes an exploration of the
  Solar neighborhood within a few $100\pc$ actually difficult with
  these data.

\item {\bf PanSTARRS1} \citep{Kaiser10}: PanSTARRS~1 (PS1) is carrying
  out a time-domain imaging survey that covers $3/4$ of the sky
  ($\delta>-30^\circ$) in five bands to an imaging depth and
  photometric precision comparable to SDSS
  \citep[e.g.,][]{Schlafly12}.  PS1 has imaging in the $y$ band, but
  not in the $u$ band, which limits its ability to determine
  photometric metallicities. However, it is the first digital
  multi-band survey in the optical to cover much of the Disk,
  including at $b=0$ both the Galactic Center and the Galactic
  Anticenter.

\item {\bf SkyMapper} \citep{Keller07a}: The Southern Sky Survey with
  the SkyMapper telescope in Australia is getting underway in 2012,
  set to cover the entire Southern celestial hemisphere within 5 years
  to a depth approaching that of SDSS. Through the particular choice
  of its five filters, SkyMapper is a survey designed for stellar
  astrophysics, constraining metallicities and surface gravities
  through two blue medium band filters ($u,v$). Together with
  PanSTARRS~1, SkyMapper should provide finally full-sky sky coverage
  in the optical to $g\lesssim 21$.

\end{itemize}

\subsubsection{Spectroscopic surveys}\label{sec:SpecSurveys}
\begin{itemize}
\item {\bf Geneva-Copenhagen Survey (GCS)} \citep{Nordstroem04}: 
This has been the
  first homogeneous spectroscopic survey of the Disk that encompasses
  far more than 1,000 stars. For $\simeq\!13,000$ stars in the Galactic
  neighborhood (within a few $100\pc$) that have Hipparcos parallaxes,
  it obtained Str\"{o}mgren photometry and radial velocities through
  cross-correlation spectrometry, and derived $\Teff$, $\feh$,
  $\logg$, ages, and binarity information from them.  It has been the
  foundation for studying the Galactic region around the Sun for a
  decade.

\item {\bf SEGUE} \citep{Yanny09}: Over the course of its first
  decade, the SDSS survey facility has increasingly shifted its
  emphasis towards more systematically targeting stars, resulting
  eventually in $R\approx\!2000$ spectra from $3800$\AA\ to $9200$\AA\ for
  $\simeq\!350,000$ stars. This survey currently provides the best
  extensive sample of Disk stars beyond the Solar neighborhood with
  good distances ($\simeq\!10\%$) and good abundances ($\feh ,\afe$).
 
\item {\bf RAVE} \citep{SteinmetzRAVE}: RAVE is a multi-fiber
  spectroscopic survey, carried out at the AAO 1.2m Schmidt telescope,
  which, as of 2012, has obtained R$\approx 7000$ spectra in the red
  CaII-triplet region ($8,410$\AA\ $ < \lambda<8,795$\AA) for nearly
  500,000 bright stars ($9<I<13$). RAVE covers the entire Southern
  celestial hemisphere except, regions at low $|b|$ and low $|\ell|$.
   The spectra deliver velocities
  to $\le 2\kms$, $\Teff$ to $\simeq 200\,\mathrm{K}$, $\logg$ to
  $0.3\dex$, and seven individual element abundances to $0.25\dex$
  \citep{Boeche11}. At present, precise distance estimates, even for
  main-sequence stars, are limited by the availability of precise (at
  the $1-2\%$-level) optical colors.
 
\item {\bf APOGEE} \citep{AllendePrieto08}: The APO Galactic Evolution
  Experiment (APOGEE), is the only comprehensive near-IR spectroscopic
  survey of the Galaxy; it started in the Spring of 2011 taking $R\approx 22,500$
  spectra in the wavelength region $1.51\mu m <\lambda< 1.70\mu m$ for
  stars preselected to likely be giants with
  $H<13.8\,\mathrm{mag}$. It aims to obtain spectra for eventually
  100,000 stars that yield velocities to better than $1\kms$,
  individual element abundances and $\logg$; the APOGEE precisions for
  $\X$ and $\logg$ have yet to be verified.  The lower extinction in
  the near-IR $A_H\simeq A_V/6$ enables APOGEE to focus on low
  latitude observations, with the majority of spectra taken with
  $|b|<10^\circ$.

\item {\bf Gaia-ESO} \citep{GilmoreRandich12}: The Gaia-ESO program,
  is a 300-night ESO public survey, which commenced in early 2012, and
  will obtain 100,000 high-resolution spectra with the GIRAFFE and
  UVES spectrographs at the VLT. It will sample all Galactic
  components, and in contrast to all the other surveys, will obtain an
  extensive set of analogous spectra for open clusters with a wide
  rage of properties; this will constitute the consummate calibration
  data set for `field' stars in the Disk.  The stars targeted by
  Gaia-ESO will typically be 100-times fainter than those targeted by
  APOGEE, including a majority of stars on the main sequence; however,
  taking spectra at $\approx\!0.5\mu$m, Gaia-ESO will not penetrate the
  dusty low-latitude parts of the Disk.

\item {\bf LAMOST} \citep{Deng12}: The most extensive ground-based
  spectroscopic survey of the Galaxy currently underway is being
  carried out with the LAMOST telescope. The Galactic Survey, LEGUE,
  has just started towards obtaining moderate-resolution spectra ($R\approx
  2000$) for 2.5 Million stars with $r\le 18$. In the context of the
  Disk, LEGUE is expected to focus on the Milky Way's outer disk,
  carrying out the majority of its low-latitude observations towards
  the Galactic anti-center.
\end{itemize}

\subsubsection{The survey road ahead: Gaia}

Gaia is an astrometric space mission currently scheduled to launch in
September 2013 that will survey the entire sky down to 20th magnitude
in a broadband, white-light filter, $G$. A recent overview of the
spacecraft design and instruments, and the expected astrometric
performance is given in \citet{deBruijne2012}; the expected
performance of Gaia's stellar parameters and extinction measurements
is given in \citet{Liu12a}. For Disk studies in particular, Gaia will
obtain 10\,percent measurements of parallaxes and proper motions out
to about $4\kpc$ for $F$- and $G$-type dwarfs, down to
(non-extinguished) $G\simeq V=15$, for which the Gaia
spectro-photometry also provides line-of-sight velocities good to
$\simeq 5$ to $10\kms$ and \logg\ and \mh\ good to $0.1$ to
$0.2\dex$. Overall, Gaia will observe approximately 400 million stars
with $G_{\mathrm{RVS}} < 17$---where $G_{\mathrm{RVS}}$ is the
integrated flux of the Radial Velocity Spectrometer (RVS)---for which
line-of-sight velocities and stellar parameters can be measured
\citep{Robin12a}.

While high-precision samples from Gaia will provide an enormous
improvement over current data, it is important to realize that most
Gaia projections are in the non-extinguished, non-crowded limit and
Gaia's optical passbands will be severely hampered by the large
extinctions and crowding in the Galactic plane. In practice, this will
limit most Disk tracers to be within a few kpc from the Sun, and Gaia
will in particular have a hard time constraining large-scale Disk
asymmetries that are only apparent when looking beyond the Galactic
center (the `far side' of the Galaxy, $D \gtrsim10\kpc, |\ell| <
45^\circ$). Gaia's lack of detailed abundance information beyond
\mh\ also means that it probably will need to be accompanied by
spectroscopic follow-up to reach its full potential for constraining
Disk formation and evolution. Some follow-up is being planned
\citep[e.g., 4MOST;][]{deJong11a}, but no good studies of the
trade-offs between, for example, sample size and abundance-precision
have been performed to date.

\section{From surveys to modeling: characterizing the survey selection functions}\label{sec:SelectionFunctions}

Spectroscopic surveys of the Milky Way are always affected by various
selection effects, commonly referred to as
`selection biases'\footnote{It could be argued that a 'selection effect' only causes
a 'bias' in the analysis, when not properly accounted for.}. 
In their most benign form, these are due to a set of
objective and repeatable decisions of what to observe (necessitated by
the survey design). Selection biases typically arise in three forms: a)
the survey selection procedure, b) the relation between the survey
stars and the underlying stellar population, and c) the extrapolation
from the observed (spatial) volume and the `global' Milky Way
volume. Different analyses need not be affected by all three of these
biases. 

In many existing Galactic survey analyses emphasis has been 
given---in the initial survey design, the targeting choices and the subsequent
sample culling---to getting as
simple a selection function as possible, \eg \cite{KGTechnique,
Nordstroem04,Fuhrmann11, MoniBidin12a}. This then lessens,  or even obviates
the need to deal with the selection function explicitly in the subsequent 
astrophysical analysis. While this is a consistent approach, it appears 
not viable, or at least far from optimal, for the analyses of the vast data
sets that are emerging from the current 'general purpose' surveys.
In the context of Galactic stellar surveys,
a more general and rigorous way of guarding against these biases 
and correcting for them has been laid our perhaps most explicitly and extensively
in \cite{BovyMAPstructure}, and in the interest of a coherent exposition
we focus on this case.

Therefore, we use the example of the spectroscopic SEGUE G-dwarf
sample used in \cite{BovyMAPstructure}---a magnitude-limited, color-selected sample part of a targeted
survey of $\simeq\!150$ lines of sight at high Galactic latitude
\citep[see][]{Yanny09}. We consider the idealized case that the sample
was created using a single $g-r$ color cut from a sample of
pre-existing photometry and that objects were identically and
uniformly sampled and successfully observed over a magnitude range
$r_{\mathrm{min}} < r < r_{\mathrm{max}}$. We assume that all
selection is performed in dereddened colors and extinction-corrected
magnitudes. At face value, such a sample suffers from the three biases mentioned in
the previous paragraph: a) the survey selection function (\ssf) is
such that only stars in a limited magnitude range are observed
spectroscopically, and this range corresponds to a different distance
range for stars of different metallicities; b) a $g-r$ color cut
selects more abundant, lower-mass stars at lower metallicities, such
that different ranges of the underlying stellar population are sampled
for different metallicities; c) the different distance range for stars
of different metallicities combined with different spatial
distributions means that different fractions of the total volume
occupied by a stellar population are observed.

We assume that a spectroscopic survey is based on a pre-existing
photometric catalog, presumed complete to potential spectroscopic
targets. The survey selection procedure can then ideally be summarized
by a) the cuts on the photometric catalog to produce the potential
spectroscopic targets, b) the sampling method, and c) potential
quality cuts for defining a successfully observed spectrum (for
example, a signal-to-noise ratio cut on the spectrum). We will assume
that the sampling method is such that targets are selected
independently from each other. If targets are not selected
independently from each other (such as, for example, in systematic
sampling techniques where each `N'-th item in an ordered list is
observed), then correcting for the \ssf\ is more complicated. For ease
of use of the \ssf, spectroscopic targets must be sampled
independently from each other.

\begin{figure}
  \centering
  \includegraphics[width=0.48\textwidth]{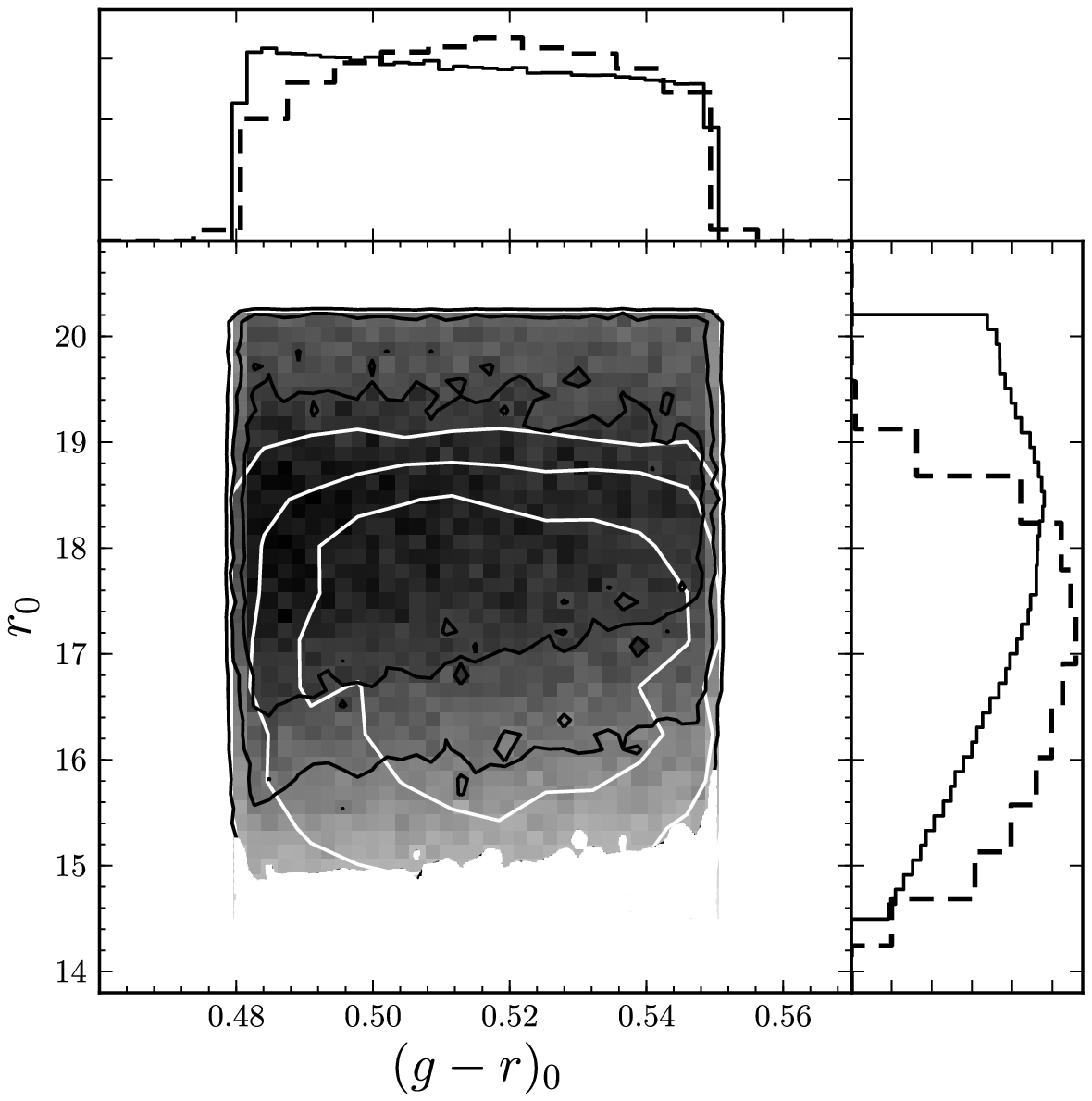}
  \includegraphics[width=0.48\textwidth,clip=]{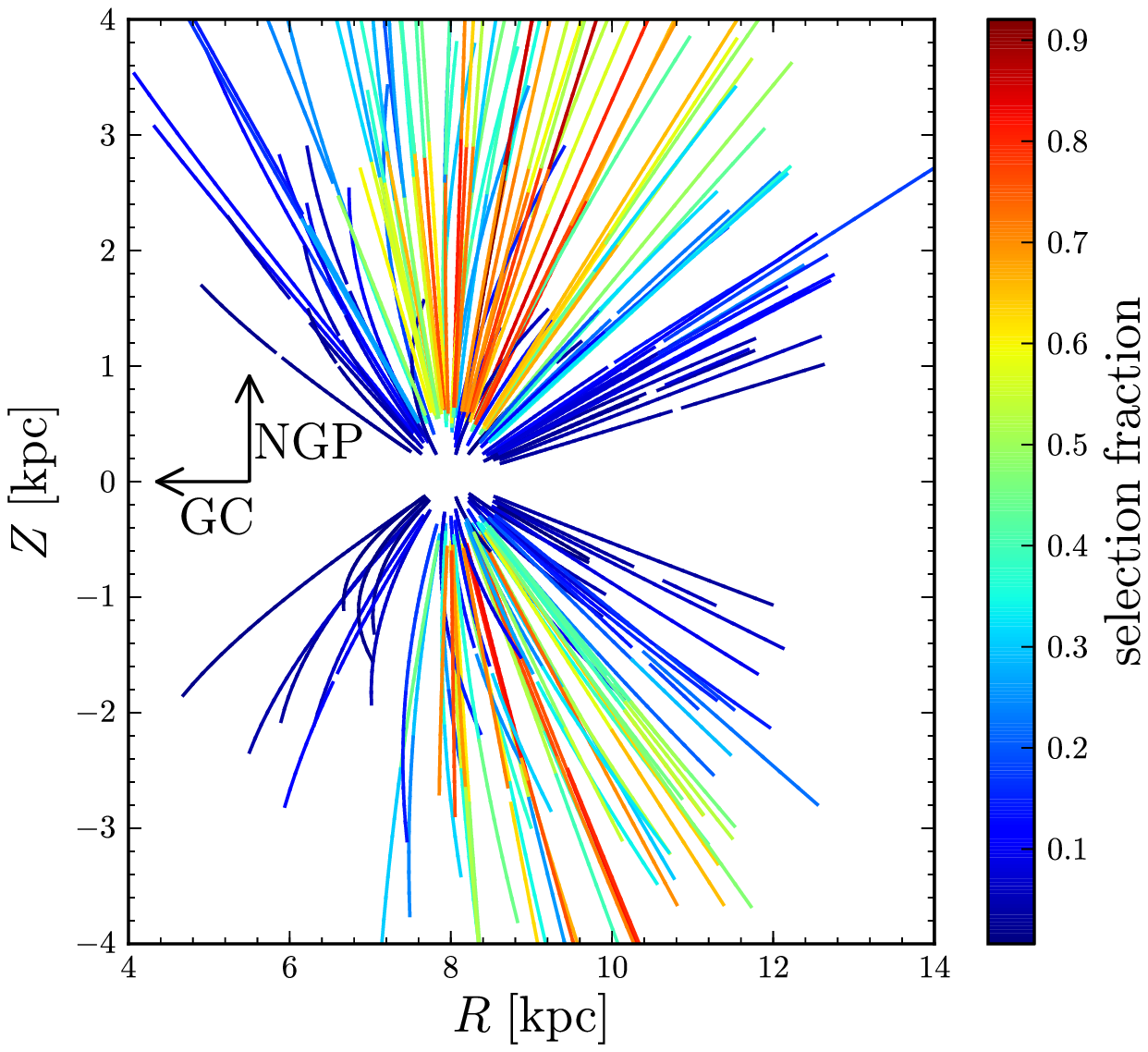}
  \caption{Spectroscopic survey selection functions: this Figure
    illustrates two essential elements of the selection functions in
    spectroscopy surveys that must be accounted for rigorously in
    analyses of spectroscopic surveys of stars in the Galaxy. The
    specific case is taken from \citet{BovyMAPstructure} and discussed
    in Section~\ref{sec:AbundanceDistribution}. The left panel shows in
    grayscale the number density of stars with SDSS photometry
    (presumed to be complete) within the G-dwarf target selection box
    in ($r,g-r$) space; the contours show the distribution of stars
    that resulted in successful spectroscopic catalog entries (the
    spectroscopic completeness), after `bright' and `faint' plates
    were taken (see \citealt{BovyMAPstructure}) obviously the
    distributions differ distinctly, as also the marginalized
    histograms show. The right panel shows the fraction of
    `available' G-dwarf targets that were assigned fibers; clearly
    that fraction varies dramatically with Galactic coordinates.}
  \label{fig:Selection-Function}
\end{figure}
      
The top panel of \figurename~\ref{fig:Selection-Function} shows the
relation between the underlying, complete photometric sample and the
spectroscopic sample for the SEGUE G-dwarf sample. While the sampling
in color is close to unbiased, the sampling in $r$-band magnitude is
strongly biased against faint targets because of the signal-to-noise
ratio cut ($> 15$ in this case). 

The selection function can then be expressed as a function
$S(\selparams)$ of the relevant quantities \selparams\ from the
photometric catalog; this function expresses the relative fraction of
entries in the spectroscopic catalog with respect to the complete
photometric catalog, as a function of \selparams. We will not concern
ourselves here with how this function is derived for the spectroscopic
survey in question; an example is given in Appendix A of
\citet{BovyMAPstructure} and
\figurename~\ref{fig:Selection-Function}. We assume that the \ssf\ is
unbiased in velocity space, that is, that all velocities have equal
probability of being observed, such that the \ssf\ only affects
analyses concerned with the spatial densities of objects. If we then
want to infer the spatial density in $\vec{x} \equiv (R,z,\phi)$ of a
set of spectroscopic objects, we need to constrain the joint
distribution $\lambda(\theta)$ of \selparams, $\vec{x}$, and whichever
other parameters \distparams\ are necessary to relate $\vec{x}$ to
\selparams\ (for example, when photometric distances are used, the
metallicity can be used in addition to purely photometric properties
to calculate $\vec{x}$); we denote all arguments of $\lambda$ as
$\theta \equiv (\selparams,\vec{x},\distparams)$. The joint
distribution can be written as
\begin{equation}
  \lambda(\selparams,\vec{x},\distparams) =
  \rho(\selparams,\distparams|\vec{x})\,\nu_*(\vec{x})\,|J|\,S(\selparams)\,,
\end{equation}
where $\rho(\selparams,\distparams|\vec{x})$ is the distribution of
\selparams\ and \distparams\ as a function of $\vec{x}$, and $|J|$ is
a Jacobian, transforming from the heliocentric frame to the
Galactocentric one. As discussed in \citet{BovyMAPstructure}, the correct
likelihood to fit is
\begin{equation}\label{eq:sflike}
\mathcal{L} = \prod_i \frac{\lambda(\theta_i)}{\int \mathrm{d}\theta\,
  \lambda(\theta)}\,,
\end{equation}
where the product is over all spectroscopic data points. This likelihood
simply states that the observed rate is normalized over the volume in
$\theta$ space that \emph{could have been observed within the survey
  selection constraints as expressed by the \ssf}.

For example, in the SEGUE example discussed above, we may further
assume for simplicity that the distance is obtained simply as
$d(r,g-r)$ (that is, ignoring the metallicity, such that there are no
\distparams) with no uncertainty. and that the distribution of colors
$g-r$ is uniform over the observed color range, then $\lambda(\theta)$
can be written as
\begin{equation}
  \lambda(r,g-r,\vec{x}) = \delta(r-r[\vec{x},g-r,\ell,b])\,\nu_*(\vec{x})\,|J|\,S(r,g-r,\ell,b)\,,
\end{equation}
where $l$ and $b$ are Galactic longitude and latitude, respectively;
$\delta(r-r[\vec{x},g-r,\ell,b])$ is a Dirac delta function that
expresses the photometric distance. The likelihood then reduces to
\begin{equation}
  \mathcal{L} \propto \prod_i
  \frac{\nu_*(\vec{x}_i)}{\int
    \mathrm{d}d\, \mathrm{d}(g-r)\, \mathrm{d}
    l\,\mathrm{d}b\,\nu_*(\vec{x}(r,g-r,\ell,b))\,|J|\,S(r,g-r,\ell,b)}\,,
\end{equation}
where we have assumed that we only want to fit parameters of $\nu_*$
(such that the \ssf\ can be dropped from the numerator). For a survey
of a limited number of lines of sight such as SEGUE, the integral over
$l$ and $b$ can be re-written as a sum over the lines of sight.

Assuming that the \ssf\ does not depend on the velocity, fitting a
joint \DF\ model for the positions and velocity, for example when
fitting dynamical models to data, uses a similar expression, with the
density $\nu_*$ simply getting replaced by the \DF, and the integral
in the denominator in \eqnname~(\ref{eq:sflike}) includes an
additional integration over velocities $\vec{v}$. For example, in the
context of the simplified SEGUE example, we fit a
\df\ $\dff(\vec{x},\vec{v}|\vec{p})$ with parameters $\vec{p}$ using
the likelihood
\begin{equation}
  \mathcal{L}(\vec{p}) = \prod_i
  \frac{\dff(\vec{x}_i,\vec{v}_i|\vec{p})}{\int
    \mathrm{d}r\, \mathrm{d}(g-r)\, \mathrm{d}
    l\,\mathrm{d}b\,\mathrm{d}\vec{v}\,\dff(\vec{x}(r,g-r,\ell,b),\vec{v}|\vec{p})\,|J|\,S(r,g-r,\ell,b)}\,,
\end{equation}

Correcting for selection bias b) requires stellar-population synthesis
models, to connect the number of stars observed in a given color range
to the full underlying stellar population. An example where this
correction is performed by calculating the total stellar mass in a
stellar population given the number of stars in a given color range is
given in Appendix A of \citet{BovyNoThickDisk}. The total stellar mass is
calculated as
\begin{equation}
  M_* = N \langle M \rangle f_M^{-1}\,,
\end{equation}
where $N$ is the number of stars and $\langle M \rangle$ is the
average mass in the observed color range, and $f_M$ is the ratio of
the total mass of a stellar population to the mass in the observed
range. All of these can be easily calculated from stellar-population
synthesis models (see Appendix A of \citealt{BovyNoThickDisk}). 

\begin{figure}
  \centering
  \includegraphics[width=0.45\textwidth,clip=]{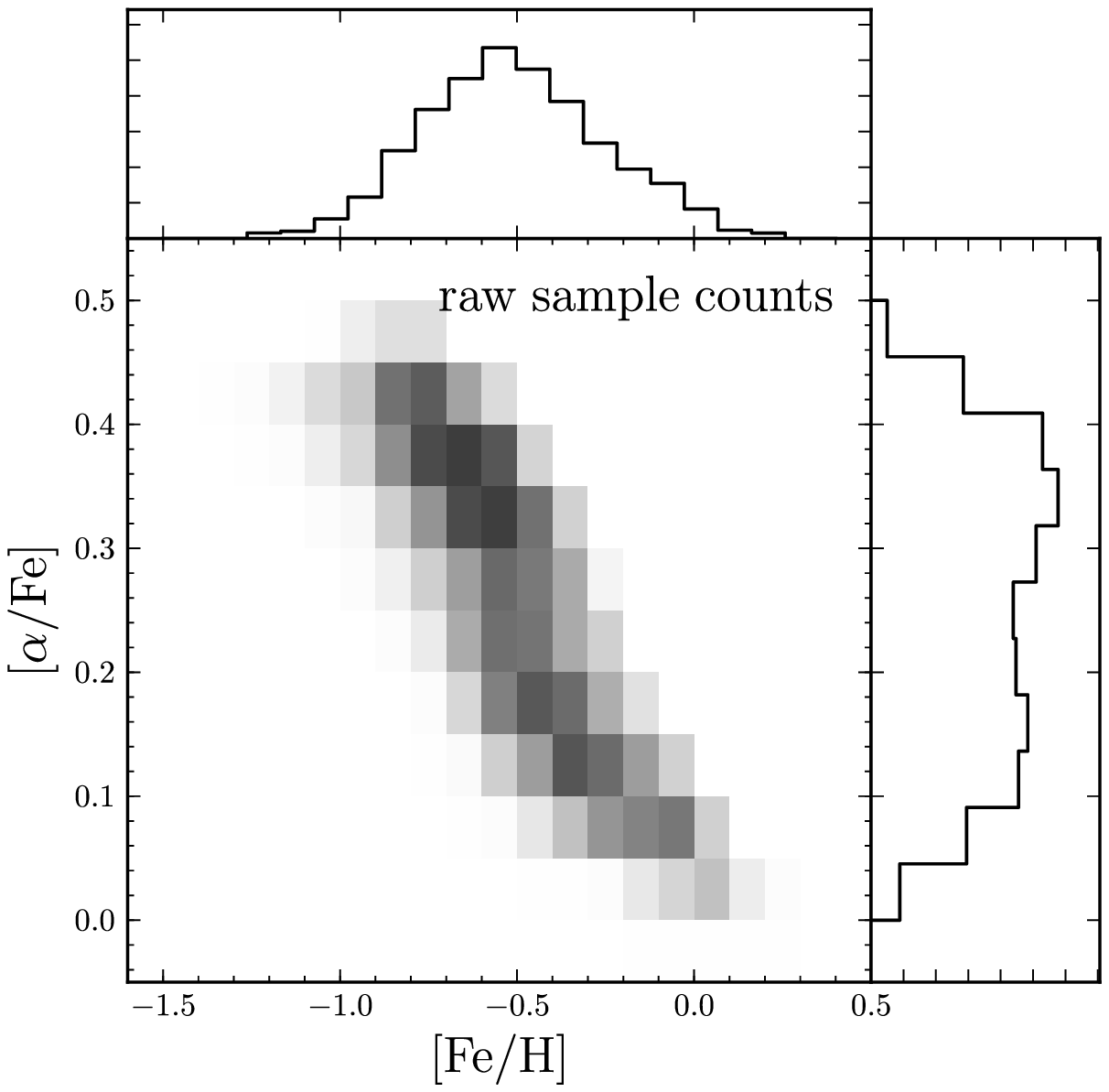}
  \includegraphics[width=0.45\textwidth,clip=]{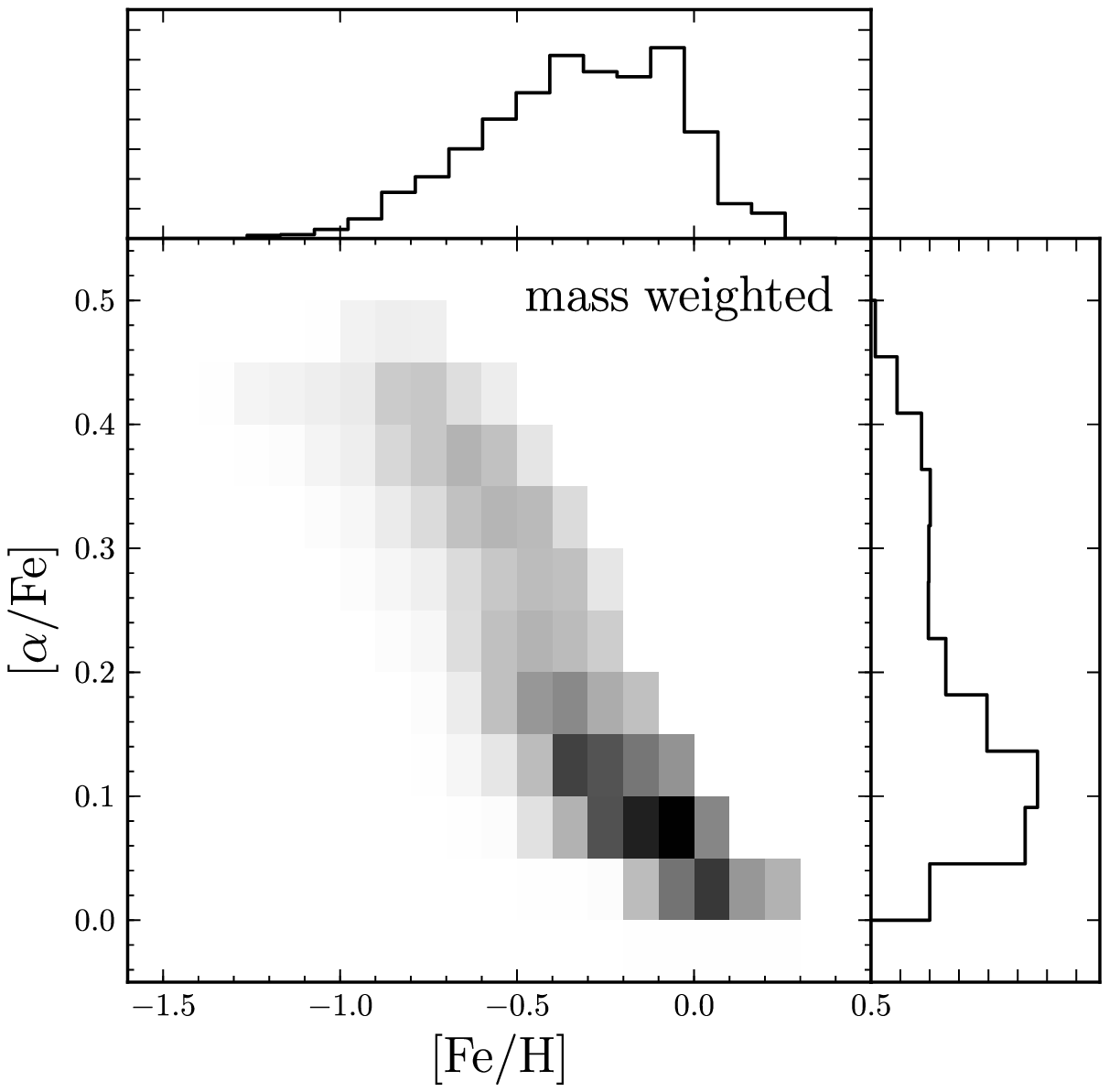}
  \caption{Distribution of stellar element abundances ($\afe$,$\feh$)
    from \citet{BovyNoThickDisk}: the top panel shows the number
    density distribution of stars as they occur within the SDSS/SEGUE
    sample; the bottom panel shows the stellar-mass-weighted,
    $|z|$-integrated distribution, which is quite dramatically
    different, illustrating the importance of incorporating the sample
    selection function and the stellar mass weighting.}
  \label{fig:Mass-Weighting}
\end{figure}

The correction for bias c) above involves extrapolating the densities
of stars in the observed volume to a `global' volume. This is useful
when comparing the total number of stars in the Milky Way in different
components. For example, \citet{BovyNoThickDisk} calculated for
different \map s the total stellar surface density at the solar radius
as the `global' quantity---also correcting biases a) and b)
above---extrapolating from the volume observed by SEGUE (see
below). This extrapolation requires the spatial density of each
component to calculate the fraction of stars in the observed volume
with respect to the `global' volume. Correcting for this bias is then
as simple as multiplying the number of stars in the observed volume by
this factor.

\section{Dynamical modeling of the Disk}\label{sec:DynMod}

\subsection{Goals}\label{sec:GoalsDynMod}

To recall from the Introduction, the goals of dynamical modeling in
the Disk context are two-fold: learn about the gravitational
potential, $p\bigl (\Phi(\vx , t)|
\{\overrightarrow{\mathrm{data}}\}\bigr )$, and learn about the
chemo-orbital distribution of Disk stars,
$p\bigl(\vec{J},\vec{\phi},\X , \age| \Phi (\vx,t )\bigr )$.
The lengthy, explicit notation makes it clear that we need to evaluate
the gravitational potential in light of all the data, and that we need
to marginalize over the different $\Phi(\vx , t )$ to learn about the
orbital distribution. That is, a simultaneous solution for potential
and orbits is needed \citep[e.g.,][]{Binney11a}.  This is a challenge
that much of the existing dynamical analyses of the Disk have
bypassed: when focusing on orbital structure, they have assumed a
`fiducial potential' \citep[e.g.,][]{Helmi99a,Klement08a,Dierickx10},
or by simplifying the treatment of the orbital structure, e.g., by the
use of the Jeans Equation \citep[e.g.,][]{JustJahreiss10} or an {\it
  ad hoc} simplified distribution function. Also, basically all
dynamical models have focused on steady state models, $\Phi(\vx ,
\not t)$, or $\Phi(\vx )$, an assumption that seems wise to retain 
until comprehensive equilibrium modeling has actually been implemented.

The gravitational potential near the Disk has conventionally been
characterized separately by the `rotation curve', $v_c(R)$ and by the
`vertical force' $K_z\equiv \frac{\partial\Phi}{\partial z} (\Ro , z)$
at the Solar radius, which describes the surface mass density near the
mid-plane, $\Sigma (\le z , \Ro )$.  Good basic estimates for both
quantities have been in place for two decades, $v_c(R)\simeq 225\kms$
and $\Sigma (<1.1\kpc , \Ro )\simeq 75\msun\pc{-^2}$, with quoted
uncertainties at the $10\%$ to $15\%$ level. The challenge therefore
is to exploit the emerging wealth of data to take this to the next
precision level, to obtain analogous constraints over a wider portion
of the Disk and to eventually make statements about the
non-(axi-)symmetric aspects of the potential. At the same time, one
would like to eliminate uncertainties in $\Phi (\vx )$ as a factor in
analyzing the stellar orbit (sub-)structure.

These aspects alone imply that a methodology has to be in place to
constrain the potential to $\delta \Phi /\Phi \lesssim 0.1$.  In
addition, one should expect that the ability of data sets to
discriminate models grows as $\sqrt{N_{\mathrm{data}}}$, where
$N_{\mathrm{data}}$ is the number of data points with useful $(\vx,
\vv)$ information. By now, $N_{\mathrm{data}}$ has reached $10^5$,
with Gaia, it will be $>10^{7-8}$: this puts enormous demands on the
`modeling' if the information content of the data is to be
exploited. As even the Jeans Equation in its simplest (1D) form
illustrates, dynamical modeling always involves the link between the
gravitational potential, the kinematics of tracers and the (gradients
of the) spatial density ($\nu$) of tracers,
\[\frac{\partial(\nu_{\mathrm{tracer}}\sigma_{\mathrm{tracer}}^2)}{\partial r} = -\nu_{\mathrm{tracer}} \frac{\partial\Phi}{\partial r}\,;\]
this illustrates the paramount importance of knowing the selection
function of any survey, which enters the determination of
$\nu_{\mathrm{tracer}}$.

It this context it then also becomes no longer tenable to view the
stellar Disk in isolation, as the mass contributions from the dark
matter and the ISM must be taken into account.

\subsection{Approaches}\label{sec:ApproachesDynMod}

To date, the majority of constraints on the Galactic potential from
Disk stars have been derived through the Jeans Equation
(cf. \citealt{binneytremaine}; e.g.,
\citealt{HolmbergFlynn04,BovyTremaine,Zhang12}; see also
Section~\ref{sec:RecentResultsDynMod}). This approach has been
sensible, because for a relatively cold and thin disk, one has a clear
prior on the orbital structure: most orbits should be
`nearly-circular', and the deviations from this can be treated as
vertical or in-plane oscillations around a guiding center.

However, solving the Jeans Equations neither delivers a \DF\ for the
stars, nor can it provide constraints on $\Phi$ \emph{marginalized}
over the possible \DF s; it only can provide a $\Phi$-constraint
\emph{conditioned} on a particular (yet ill-specified) \DF. Hence,
more rigorous modeling approaches that explicitly account for the
orbit-nature of the stars are needed in the Disk context, both given
the quality of the data and the subtlety of the questions one would
like to address. Mostly in the context of modeling external galaxies
or star clusters, three ultimately related approaches have been
developed over the last decades.
\begin{itemize}
\item Methods based on \DF s: Assuming families of analytic \DF s, for
  which $p(\vx , \vv )$ can be predicted {\it in a given potential
    \Pot}\ and compared to the data $\{\vx , \vv \}$. By varying
  \Pot\ and finding the \DF\ whose prediction matches the data in each
  case, one obtains $\Like \bigl ( \{\vx , \vv\}| \Phi(\vx ) \bigr
  )$; this approach also yields the \DF, 
  \df$_{\mathrm{best}}(\Phi_{\mathrm{best}})$.

\item Orbit-based methods: This entails the prediction of $p(\vx ,
  \vv|\mathrm{orbit})$ by explicit calculation of these orbits
  (`Schwarzschild method'), where the best potential is found again by
  varying \Pot\ and optimizing the `weights' of all calculated orbits
  to match the data.  Among the advantages of this approach are that
  no explicit \DF\ needs to be spelled out, which is difficult to do
  in, e.g., non-axisymmetric potentials; among the disadvantages is
  that it is difficult to ensure that the \DF -space has been
  well-sampled by the discrete orbits. This approach has been
  implemented by \citet{Rix97}, \citet{Gebhardt03}, and
  \citet{vdBosch08}.

\item Particle-based methods: Iteratively modifying a self-consistent
  $N$-body model, so that its predictions increasingly better match a
  set of given observations (`made-to-measure'). This approach has the
  advantage that no orbit library needs to be stored and that it makes
  no explicit symmetry assumptions; at least in simple cases, it is
  shown to recover the correct \DF. However, by its particle nature,
  its outcome is only one specific particle sampling of the underlying \DF. This
  approach has been originally devised by \citet{SyerTremaine96}, and
  its practical applications for galaxy modeling have been pursued
  foremost by Gerhard and collaborators \citep[e.g.,][]{Morganti12}).
\end{itemize}

\subsubsection{Data-model comparison in dynamical modeling}\label{sec:DataModComp}

Modeling the Disk also calls for data vs. model comparisons that
differ from those most commonly used in stellar dynamical
modeling. Usually, the data-model comparison is presented by comparing
the surface-brightness, mean-velocity and velocity dispersion profiles with
data that have been binned or never been resolved into individual
stars. The case of the Disk, or most other parts of the Milky Way, is
of course different, in at least two respects: first, the data are
obtained star-by-star, where each star has uncertainty estimates for
each of its phase-space coordinates; second, our position within the
disk means that the direct dynamical observables
$\{\overrightarrow{\mathrm{data}}\}\equiv p(\vlos , \vec{\mu},\ell,b,D)$
map into very different components of $(\vx , \vv )_{\mathrm{GC}}$. Further,
the Sun's motion with respect to the Galactic rest-frame system and
with respect to any sensible local co-rotating reference frame
(`standard of rest') is still under considerable debate
\citep[e.g.,][]{Schoenrich10a,BovyVc}. This enters into the
$(\vlos,\vec{\mu},\ell,b,D)\mapsto (\vx , \vv )_{\mathrm{GC}}$ transformation,
Finally, the size of the uncertainties in $(\vlos,\vec{\mu},\ell,b,D)$
will vary dramatically among sample members, \eg, if some sample stars
have $\vlos$ measurements and others do not.

While this calls for an approach beyond comparing tracer density and
dispersion profiles, a data-model comparison is still straightforward
for any model that predicts $n_*(\vx , \vv)$, or $\nu_*(\vx)$ and
$p(\vv | \vx)$, for a given $\Phi(\vx )$.  Rather than asking
models to match binned moments of the observables, $\nu_*(R,z)$ and
$\sigma_{R/\phi /z}(R,z)$, one simply calculates the likelihood of the
individual data $\Like (\{\mathrm{data}\}^{N_*}_{i=1}|\mathrm{model})$, see
\citet{BovyMAPstructure}.

\subsection{Recent results}\label{sec:RecentResultsDynMod}

In light of the vastly better data, the recent progress in
understanding the dynamical properties of the Milky Way and the local
(few kpc) Galactic potential may seem disappointing: the debate around
the `circular velocity at $\Ro$' has not abated, nor have the
constrains on the stellar and DM mass distribution near $\Ro$
tightened by significant factors.  Before sketching the road towards
comprehensive dynamical modeling (Section~\ref{sec:WitherDisk}), it is
worth looking at some of the recent results to describe the status
quo, first for $\vcirc (R)$ then for K$_z(z|\Ro)$.

\subsubsection{The Disk's `circular velocity'}

Characterizing the Milky Way's central mass distribution by its
circular velocity has tradition and is (operationally) sensible. For
an axisymmetric galaxy, $\vcirc (R)$ is simply $\sqrt{R\,\partial \Phi
  / \partial R (R,z=0)}$. The `rotation curve', especially $\vcirc
(\simeq \Ro )$, can be constrained in many different ways: by
considering the reflex motion of the Galactic Center
\citep[e.g.,][]{Ghez08,Reid04}, by measuring the velocities of
globular clusters or halos stars \citep[e.g.,][]{Sirko04,Deason11},
considering them to be a non-rotating component, by measuring
line-of-sight or 3D velocities to ISM tracers \citep{Fich89,Reid09},
presumed to be on near-circular orbits, by determining an overall mass
model for the Galaxy and its halo and interpolating inferences to
$(\Ro, z=0)$ \citep[e.g.,][]{Xue08,Koposov10a,McMillan11a}, or by
mapping and modeling the stellar velocities across a sufficiently
large portion of the Disk \citep{BovyVc,Schoenrich12b}. Over the last
years, this has led to (published) estimates of $\vcirc (\simeq \Ro
)$, ranging from $215$ to $255\kms$, with little seeming progress
since the IAU recommended a value of $220\kms$, nearly 30 years ago
\citep{Kerr86a}.

This range of values, given the {\it precision} that the data
seemingly can offer, can be traced to the fact a) that the Sun's
motion with respect to $\vcirc (\Ro )$ is uncertain at the
$\simeq\!10\kms$ level, b) that the Galactic potential is not
axisymmetric presumably at the $\simeq\!5\%$ level, and c) that (hence)
cold tracers do not move on circular orbits. Judging from external
galaxies \citep[e.g.,][]{Rix95a}, non-axisymmetries in the
gravitational potential at $\Ro$ in a Milky-Way-like galaxy are
expected to be at the 5 to 10\% level, with causes ranging from
lopsidedness ($m=1$), to spiral arms in the stellar disk {\sl mass}
distribution, to bars in the center, and potentially asymmetric dark
matter halos. This leads to smooth azimuthal variations in $v_\phi$ of
closed orbits, the closest equivalent to $\vcirc$. As of now, the
majority of the analyses have focused on the nearby half of the Disk,
or the local quadrant. This makes analyses susceptible to such
asymmetries and makes it hard to test for them. The observability of
ISM tracers, such as masers \citep{Reid09}, depends potentially
strongly on their orbital phase; therefore the assumption that the
orbital phases are random, underlying most analyses, is poorly
justified.  In principle, stars are better tracers of a `smooth
rotation curve', as their finite kinetic temperature reduces their
response to non-axisymmetric potential perturbations. For them,
however, a model accounting for their velocity dispersion, and
corresponding lag in $\vphi$, the asymmetric drift must be made
\citep[see][]{BovyVc}.

In the context of axisymmetric models, \citet{BovyVc} has thorough
analyzed new stellar radial velocities and approximate distances from
the APOGEE survey, and has extensively explored and accounted for a
wide range of uncertainties and model assumptions. They find a
rotation curve that is very close to flat and has $\vcirc = 218\pm
6\kms$. They show that this result is consistent with all other
existing determinations. This result, however, implies that the Sun's
azimuthal velocity, $\vphi$, is $24\kms$ higher than $\vcirc (\Ro )$.
The main remaining model limitation in \citet{BovyVc}'s approach is
that the possible consequences of non-axisymmetries have not been
thoroughly explored, not even the known ones that must arise from the
Galactic bar.

\subsubsection{The potential perpendicular to the Disk}\label{sec:Kz}

The second focus of dynamical Disk modeling has been the study of the
gravitational potential perpendicular to the Disk at $\simeq
\Ro$. Broadly speaking, $\Kz\equiv | \partial \Phi / \partial z (\Ro
,z)|$ is expected to vary linearly for small $z$ ($\simeq 200$~pc), as
the enclosed (presumably) stellar surface mass density
$\Sigma_*(<|z|)$ grows linearly; for $|z|$ above the dominant layer of
Disk mass, the $K_z$-profile is expected to flatten, as for plane
parallel geometry on has $K_z(z)\rightarrow $ const. for
$\Sigma_*(<|z|)=$ const. Once, or if, a spheroidal dark matter
distribution becomes dominant while $|z|\ll \Ro$ still approximately
holds, one expects again $\Kz\propto z$; for a given radial dark
matter profile, $\rho_{\mathrm{DM}}(\vec{r})$, the relation $\Kz\propto
\rho_{\mathrm{DM}}(\Ro, z=0)\times z$ provides a constraint on the local dark
matter density.

Building on the seminal work of \citet{Kapteyn22a} and
\citet{Oort32a}, who coined the term $K_z$-force, the analysis of
\citet{KGTechnique,KGSurfaceMass} has set the standard for such an
analysis for decades.  They adopted a parameterized 1D vertical force
law of the form
\begin{equation}
  \Kz = 2\pi G \bigl ( \Sigma_{0}\times \frac{z}{\sqrt{z^2+z_h^2}} \bigr ) 
  \ + \ 4\pi G \rho_{\mathrm{DM}}\times z\,,
\end{equation}
where $\Sigma_{0}$ is the integrated disk surface mass density at the
Sun's radius (including the contribution from the cold interstellar
medium). The thickness of the stellar distribution is described by
$z_h$ and the DM density at $(\Ro, 0)$ is described by $\rho_{\mathrm{DM}}$. In
order to break the degeneracy between the two $K_z$-terms that scale
linearly with $z$, it is necessary that $2\Sigma_{0,*}/h_z$ is
substantially larger than $\rho_{\mathrm{DM}}$, which turns out to be the case.

One the one hand, it is therefore necessary to have observational
constraints with $|z|\gg h_z$; on the other hand the present analysis
context is limited by the requirement that a one-dimensional analysis
in cylindrical coordinates makes sense: taken together this leads to a
vertical range over which constraints are needed of
$|z_{\mathrm{max}}|\simeq\!1$ to $4\kpc$.

\citet{KGTechnique} laid out how to link the observables---the
vertical density distribution of tracer particles, $\nu_*(z)$ and
their vertical velocity dispersion profile, $\sigma_z(z)$---to
$\Kz$. Their approach and all subsequent ones have exploited
variants of the Jeans Equation, where---of course---it is
crucial that the radial dependence of all properties is correctly
taken into \citep[e.g.,][]{KGTechnique,BovyTremaine}.  These radial
terms matter least in the case of a perfectly flat rotation curve near
$\Ro$, which fortunately or fortuitously seems to be an excellent
approximation \citep[e.g.,][]{BovyVc}.

Based on an approximately volume-complete sample of K stars towards
the Galactic pole, \citet{KGSurfaceMass} and \citet{KGKz1.1} found at
the time that the best determined quantity was
$\Sigma_{<1.1\kpc}=71\pm6\msun\pc^{-2}$, with a likely baryonic
disk mass of $48\pm8\msun\pc^{-2}$, and no evidence, or at least
no data-driven need, for disk dark matter near the Sun. A number of
conceptually similar analyses have been carried out since
\citep[e.g.,][]{FlynnFuchs94,Siebert2003,HolmbergFlynn04}: these
confirmed the approximate values for $\Sigma_{<1\kpc}$, and
\citet{Siebert2003} obtained a first constraint on the dynamical
thickness of the stellar disk layer.  All of these studies were
inconclusive on estimating local dark matter, as the data could
neither rule out $\rho_{\mathrm{DM}}=0$ nor $\rho_{\mathrm{DM}}\simeq
0.06-0.12\msun\pc^{-3}$, the value expected for the inward
extrapolation of global dark matter halo fits
\citep[e.g.,][]{Xue08,Deason12a}.
 
\begin{figure}
  \centering
  \includegraphics[height=0.3\textheight]{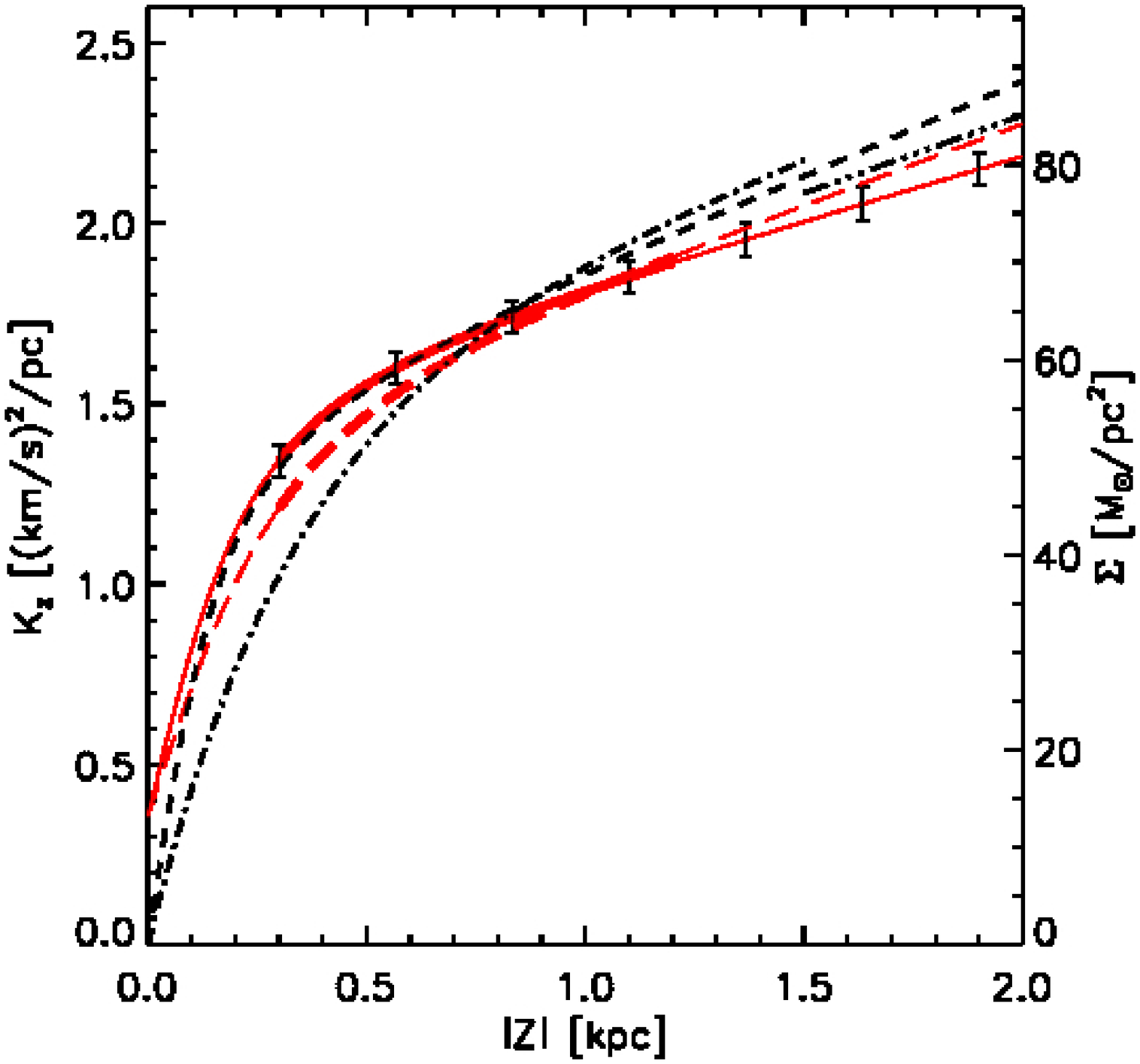}
  \includegraphics[height=0.3\textheight]{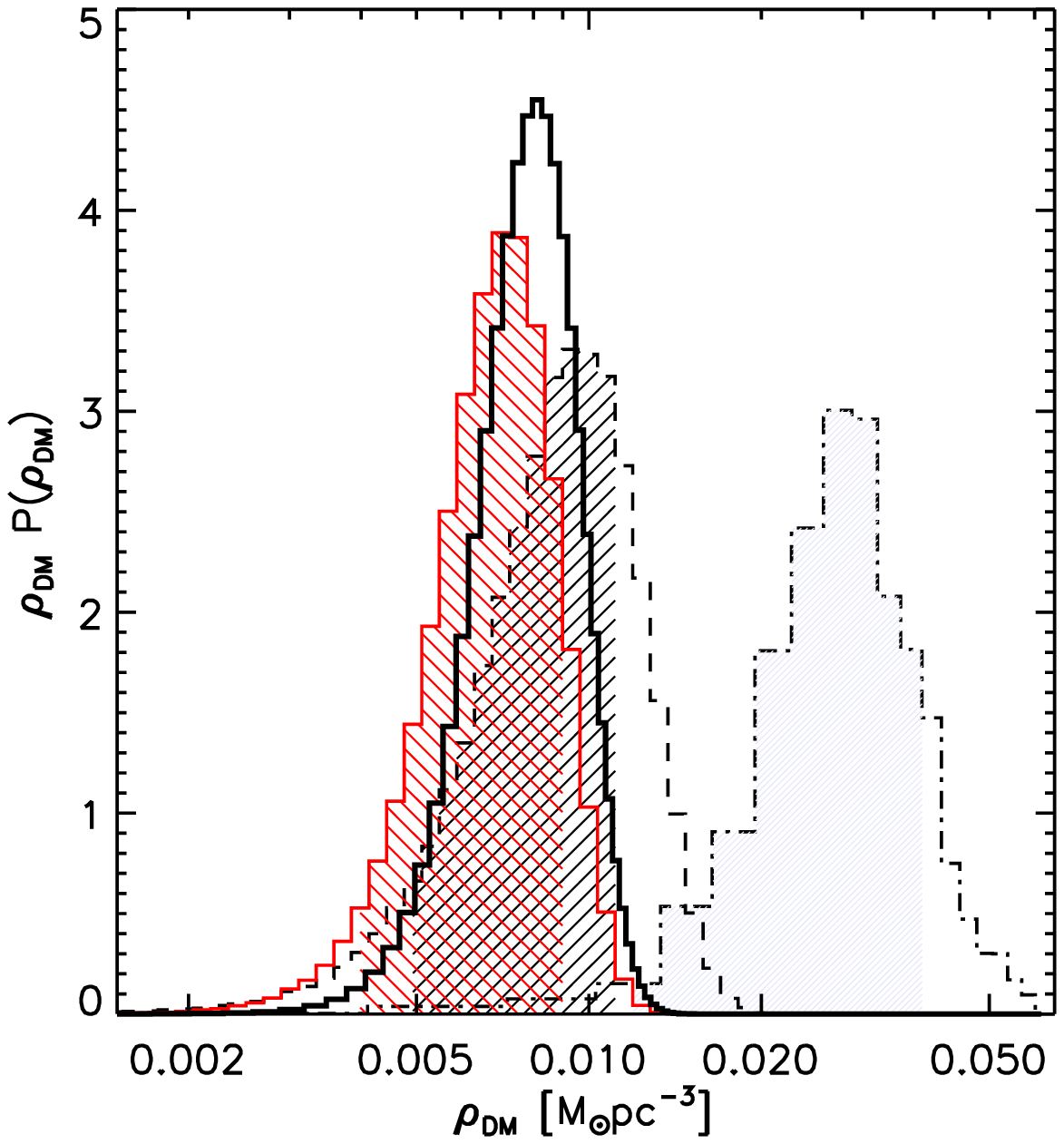}
  \caption{The (vertical) Galactic potential at the Solar radius. This
    Figure, taken from \citet{Zhang12} summarizes various estimates
    for $\Phi(z|R_\odot)$: the left panel shows various estimates of
    $K_z(z)$, where the initial steep rise of $\Kz$ reflects the
    stellar and gaseous disk mass, and the slope beyond
    $\simeq\!1\kpc$ reflects the local dark matter density: the red
    lines show the results from \citet{Zhang12} with the best fitting
    $\rho_{\mathrm{DM}}$ (solid line) and with
    $\rho_{\mathrm{DM}}\equiv 0.008\msun\pc^{-3}$ from
    \citet{BovyTremaine} (dashed line); the fat portions of the red
    lines show the $|z|$-range that are directly constrained by the
    SEGUE data. The dashed line shows the result from \citet{KGKz1.1},
    the gray dash-dotted line for $z<1.5$~kpc the result from
    \citet{HolmbergFlynn04}, and the dash dotted line beyond $z\simeq
    1.5$~kpc the result from \citet{BovyTremaine}.  The right panel
    illustrates recent estimates of the implied local dark matter
    density: \citet{Garbari12a} in blue, \citet{BovyTremaine} in
    black, and \citet{Zhang12} in red; the open histogram shows the
    joint probability of these estimates.}
  \label{fig:Zhang}
\end{figure}
     
This has recently changed, as three studies have claimed significantly
non-zero $\rho_{\mathrm{DM}}$ estimates from $K_z$-type
experiments. \citet{Garbari12a} reanalyzed literature data, but
properly marginalized over a number of assumptions and parameters in
previous analyses, and inferred
$\rho_{\mathrm{DM}}=0.025^{+0.013}_{-0.014}\msun\pc^{-3}$, potentially
indicating a flattened halo.  \citet{BovyTremaine} re-analyzed a set
of $\simeq\!400$ stars with distances and kinematics assembled by
\citet{MoniBidin12a}, and obtained an estimate of
$0.008\pm0.003\msun\pc^{-3}$ 
(where the uncertainties also incorporate a systematic component); 
despite the modestly-sized sample
this local dark matter Milky Way was enabled by the large vertical extent of the tracer
stars ($\simeq\!4\kpc$), sampling high above the disk.
 
All of these analyses had to carefully model the metallicity
(selection) distribution of the stars, as the vertical density profile
and effective $\sigma_z$ dispersion depend sensitively on the
abundance mix of the sample stars (see
Section~\ref{sec:MonoAbundanceProps}).  \citet{Zhang12} tackled this
limitation, by analyzing the $\Kz$-force problem considering and
fitting abundance-selected, nearly isothermal sub-samples
separately. Using $\simeq\!9000$ K-type dwarfs from SDSS/SEGUE, they
found $0.0065\pm0.0025\msun\pc^{-3}$, which in conjunction with
\citet{BovyTremaine} yields
$\rho_{\mathrm{DM}}=0.0075\pm0.0021\msun\pc^{-3}$, or $0.28\pm
0.08\,$GeV\,cm$^{-3}$.
 
They also obtained $\Sigma_{<1.1\kpc}=68\pm6\msun\pc^{-2}$,
consistent with \citet{KGKz1.1} and other previous results. It may
seem startling that the error bars have not become substantially
smaller. However, recent determinations have fitted directly for many
more aspects of the model, rather than simply assuming a prior value,
which broadens their confidence limits, even with larger samples. For
example, \citet{Zhang12} determined not only the tracer scale-heights
consistently, but also fitted the Disk's effective mass scale height,
finding $100\pc < z_h < 350\pc$. 
 
To date, the dark matter constraints from $\Kz$ analyses corroborate
other evidence, but are not yet better than other approaches. This
should to change in the near future, as we outline below.
 
\section{New ways of looking at the Milky Way's stellar Disk}\label{sec:DiskResults}

\subsection{Mono-abundance sub-populations}
We now lay out in more detail why we believe that dissecting the Disk
in terms of `mono-abundance stellar populations' (\map
s), i.e. in term of stellar subcomponents with very similar abundances
(\eg $\feh$ and $\afe$), is a productive way forward, both for studying galaxy evolution and
for dynamical modeling.  We do so by synthesizing a number of recent
results. We also show that for the most part this new (or, newly
implemented) way of looking at the Disk is largely consistent with a
range of earlier results when properly compared.

Work over the last 30 years has shown that the Disk is complex, with
stars of different ages and abundances showing a different dynamical
structure \citep[e.g.,][]{GilmoreARAA}). To simplify the problem of
understanding the Disk, it is sensible to dissect it into different
components. A dissection into a {\sl thin} and a {\it thick} disk,
either by spatial or by kinematical criteria seems obvious. But the
hierarchical assembly and secular evolution processes erase, or at
least diffuse, dynamical memory with time
(\citealt{Wielen,SellwoodBinney,Kormendy04,Schoenrich08a}; see Section~\ref{sec:WitherDisk}).
Further, any (sub-)sample selection based on spatial or kinematic
criteria feeds back into the inferred structure of that component in
very complex ways.  This leaves `age' and `chemical abundances' $\X$
as lifelong tags to mark sub-populations \citep{BlandHermes}, and disk
sub-components have long been defined that way \citep{Fuhrmann11}.
Because good age determinations, spanning 1 to 12 Gyrs, are at present
only available for tiny volumes ($\simeq\!10^{-3}\kpc^3$; where
parallaxes exist), this leaves `mono-abundances' criteria as
the sub-population marker of choice for the near future. Of course,
abundances and ages are linked through chemical enrichment, making
`mono-abundances' a sensible, albeit qualitative, proxy for mono-age
populations, an issue which we explore in
Section~\ref{sec:WitherDisk}. The concept of \map s is different from
a mere abundance-based thin-thick disk distinction on the basis of
abundances: it presumes nothing about the spatial or kinematic
properties of stars at a given $\X$, nor does it presume that there is
a small number of distinct components.

As we will see, the consideration of \map s allows quite
direct inferences about galaxy formation, but it has also two
advantages for dynamical modeling: first, in dynamical modeling it is
convenient to have tracer populations that have simple orbit \DF\
properties, which \map s turn out to have.  Second, all \map s live in
the same gravitational potential, and hence provide opportunities to
cross-check dynamical inferences (cf. Section~\ref{sec:Kz}).

\subsection{Properties of the Disk's mono-abundance sub-populations}\label{sec:MonoAbundanceProps}

\begin{figure}
  \centering
    \includegraphics[width=0.99\textwidth]{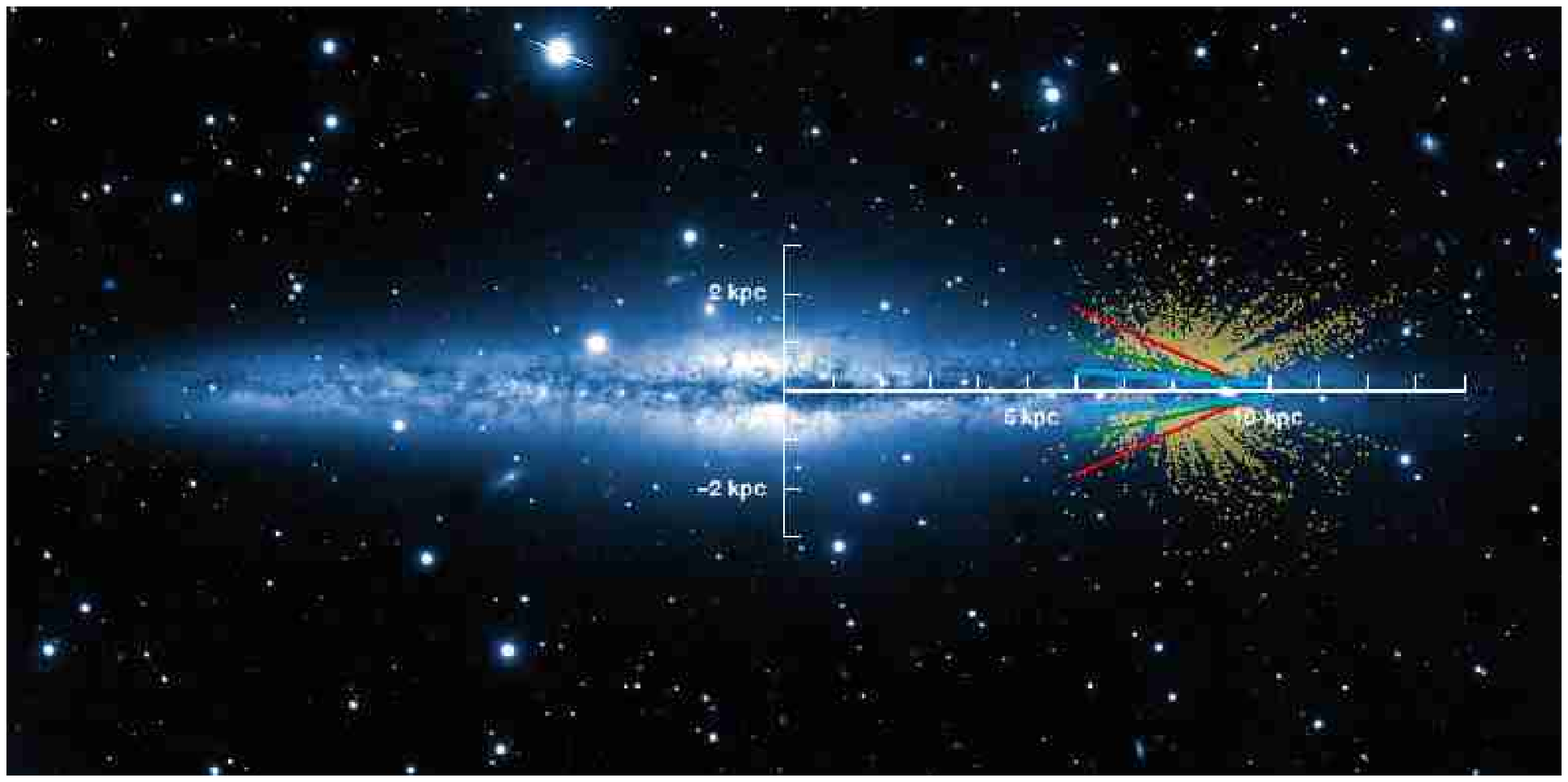}
  \caption{The geometry of `mono-abundance populations' (MAPs), as
    derived from SDSS/SEGUE data in \citet{BovyMAPstructure}. The
    Figure shows lines of constant stellar number density (red, green
    and blue) for \map s of decreasing chemical age (i.e. decreasing
    $\afe$ and increasing $\feh$, with color coding analogous to
    Fig.~\ref{fig:MAPspatial}), illustrating the sequence of \map s from
    `old, thick, centrally concentrated' to `younger, thin, radially
    extended'. The Figure also puts the SDSS/SEGUE survey geometry
    into perspective of the overall Galaxy (here represented by an
    image of NGC~891).}
  \label{fig:MAPstructure}
\end{figure}

Following Bovy \etal\ (2012b,c,d) we now lay out the results of
analyzing \map s in the Disk, drawing on spectra from
SDSS/SEGUE\footnote{Analogous analyses could and should be carried out
  with other specspectroscopictroscopic surveys, such as RAVE, APOGEE,
  Gaia-ESO, etc.}. Again, qualitatively we are asking {\sl "What would
  the spatial, kinematical, and dynamical structure of the Disk look
  like, if we had eyes only for stars of a particular abundance?"} In
the present context `abundances' refer to $\feh$ and $\afe$ only, both
because such data are available from SDSS, and because they describe
the bulk of the variation of individual abundances
\citep[e.g.,][]{Ting12a}. While the \citet{BovyMAPstructure} analysis
goes well beyond the solar neighborhood (so thoroughly probed by the
the GCS, but it still covers only $6\kpc <R < 10\kpc$ and $0.3\kpc
<|z|<2\kpc$ (see \figurename~\ref{fig:MAPstructure}), not the entire
Disk.  In practical terms, \map s mean ensembles of stars whose
abundances are within $\simeq 0.15\dex$ in $\feh$ and
$\simeq\!0.08\dex$ in $\afe$ \citep{BovyMAPkinematics}, leading to
about 50 different \map s in \citet{BovyMAPstructure}.  Given the
total sample size of $\simeq\!20,000$ G-type dwarfs in SEGUE, typical
\map\ sample sizes in \citet{BovyMAPstructure} were a few hundred.

As laid out in Section~\ref{sec:DiskOverview}, $\feh$ and $\afe$
abundances depend on both the degree of chemical enrichment and the
speed with which it occurred. In general, less $\afe$-enhanced and
more metal-rich star probably have formed later
\citep[e.g.,][]{Schoenrich09a}. In particular for metallicity, the
formation radius also plays an important role, as most galaxies show
an outward decay in their mean metallicity (see
Section~\ref{sec:WitherDisk}). For language convenience, however, we
will simply refer to $\afe$-enhanced stars as `$\alpha$-old', or
`chemically-old'.

\subsubsection{The spatial structure of mono-abundance populations}

As a main result from the \citet{BovyMAPstructure} analysis, the
spatial structure of the Disk's \map s turned out to be remarkably
simple over the observed range: fitting a number-density model that is
a simple exponential in both the $z$ and the $R$ directions matches
the data well. If the model complexity is enhanced to include two
vertical scale-heights (a `thin' and `thick' component) the data do
not point towards two scale heights of significant mass fractions and
significantly different scale heights for any \map.  Therefore, each
\map\ can be characterized by its scale length $\re$, scale height
$\hz$ and number-density normalization.

\begin{figure}
  \centering
  \includegraphics[width=0.45\textwidth,clip]{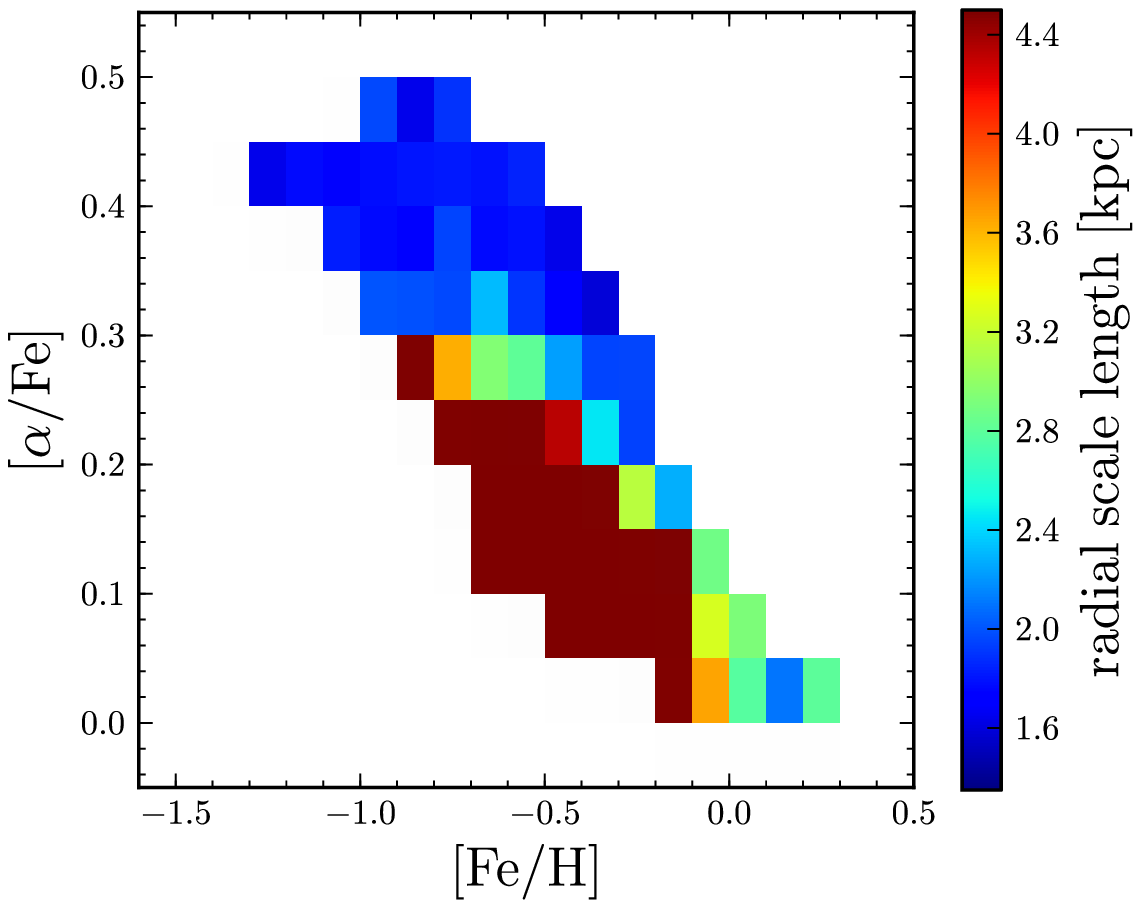}
  \includegraphics[width=0.45\textwidth,clip=]{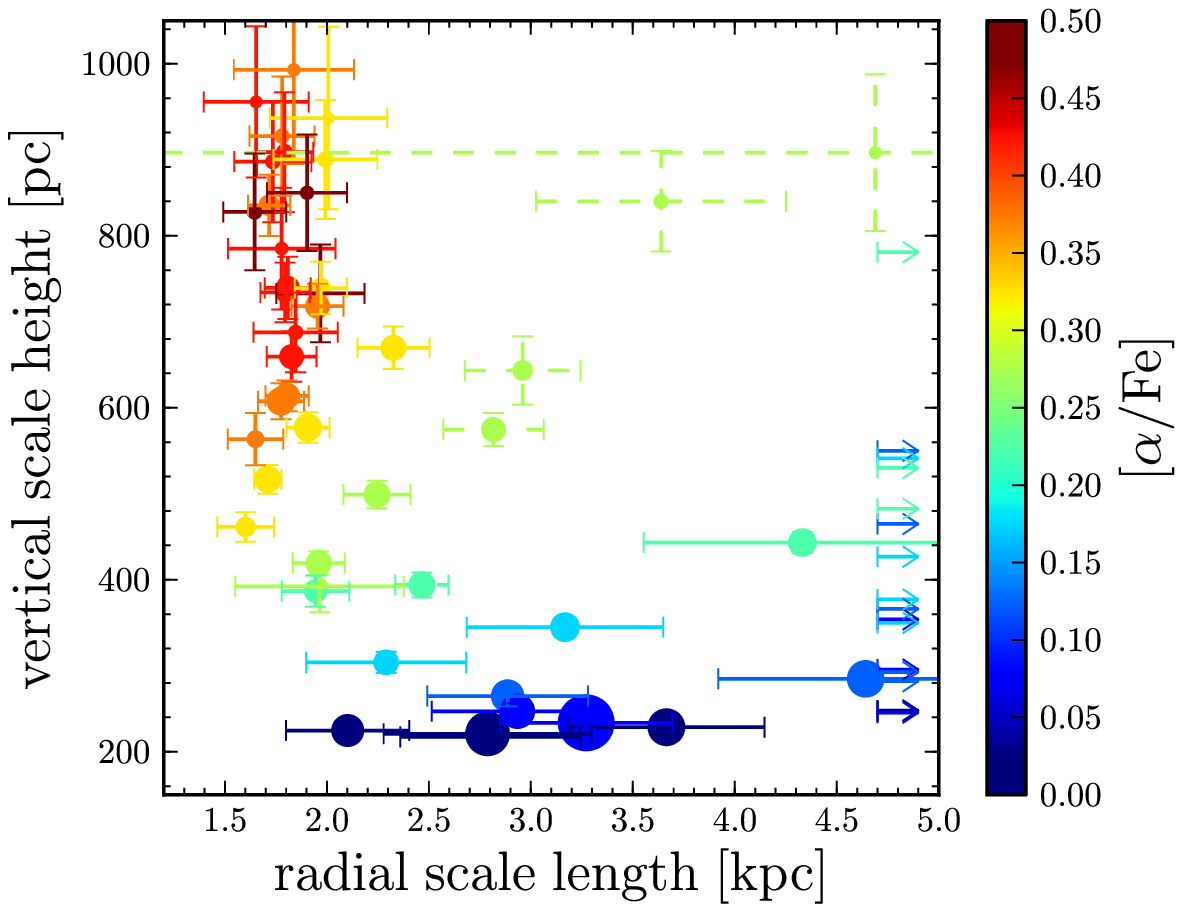}\\ 
  \includegraphics[width=0.45\textwidth,clip=]{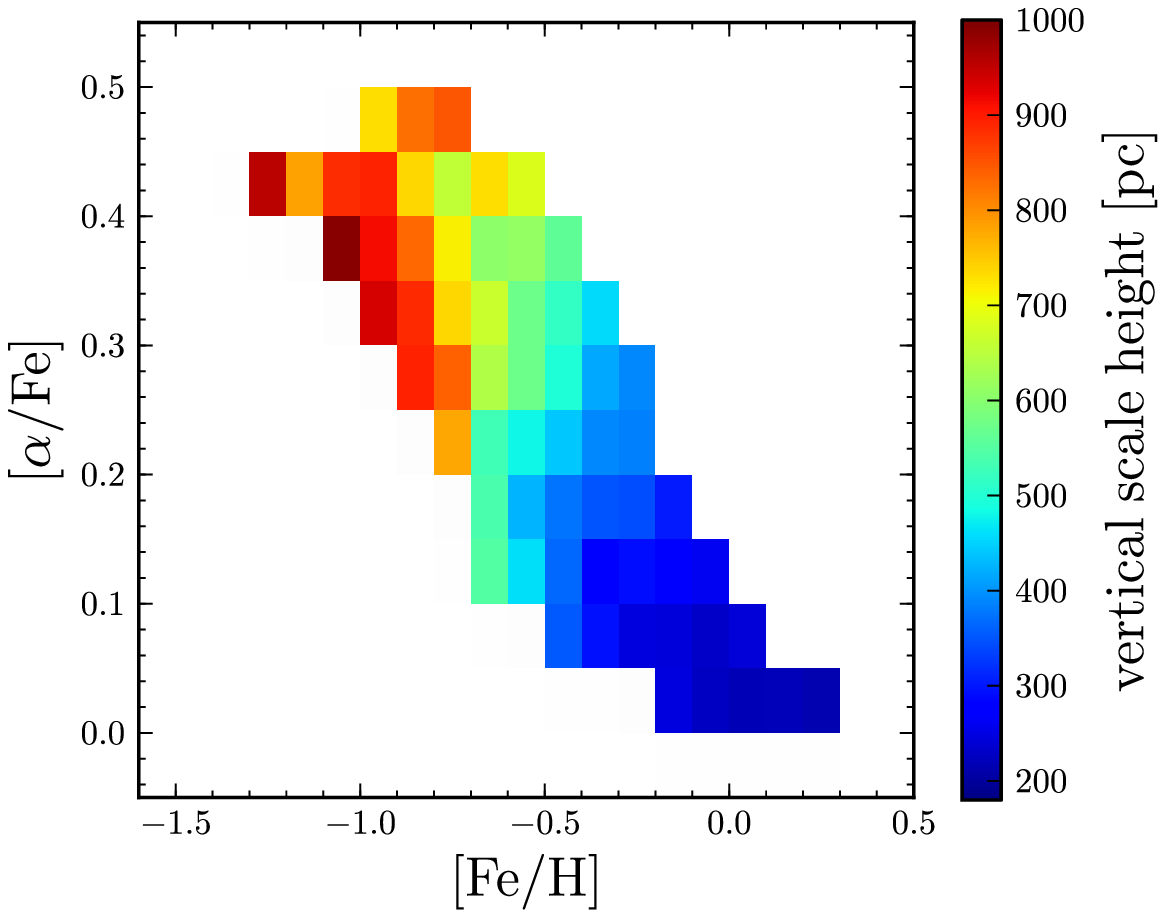}
  \includegraphics[width=0.45\textwidth,clip=]{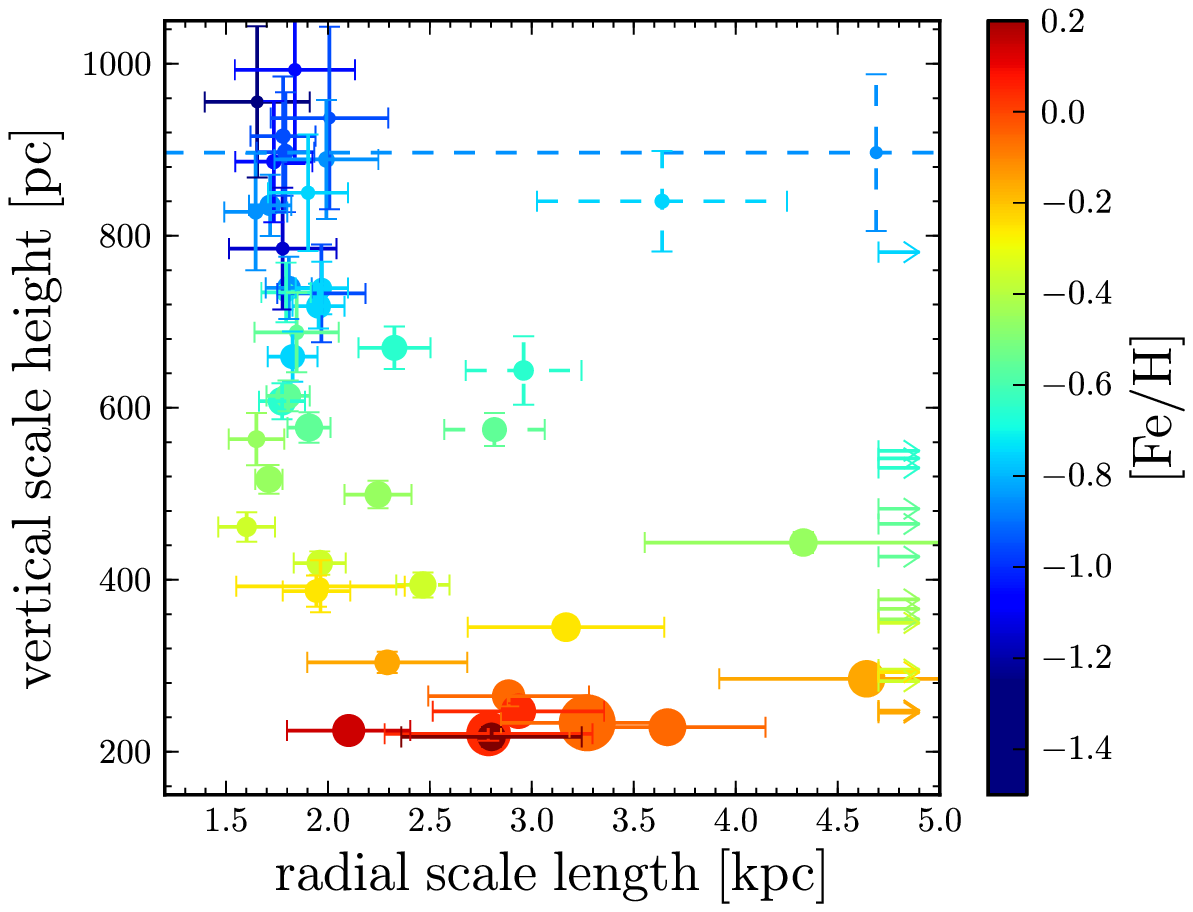}
  \caption{Spatial structure of the `mono-abundance populations' (\map
    s) in the Disk from \citet{BovyMAPstructure}.  Each of the
    components, characterized by ($\feh , \afe$), can be
    well-described by a simple exponential density profile in both the
    radial direction and vertical direction. The left two panels show
    the radial and vertical scales as a function of $\feh , \afe$; the
    right two panels show \hR\ vs. \hz, as a function of $\afe$ and
    $\feh$, respectively. This Figure illustrate that from the
    chemically-old to chemically-younger \map s the structure changes
    systematical from `old, thick, radially concentrated' to `younger,
    thin, radially extended', but in a way that cannot be fully
    captured by $\afe$ or $\feh$ alone
    (cf. \citealt{BovyMAPstructure}).}
  \label{fig:MAPspatial}
\end{figure}

The complexity of the Disk and the power of \map s comes in when
considering the $\feh$ - $\afe$-dependence of these structural
parameters: $\re$ and $\hz$ vary systematically with abundances, in a
way that depends both on $\feh$ and on $\afe$, as shown in
\figurename~\ref{fig:MAPspatial}.  Broadly speaking, there is a
simple trend: chemically-older \map s form thicker disk components
with shorter radial scale-lengths.  Among the \map s, the
scale-heights range from $\simeq\!200\pc$, the classical `thin--disk'
regime to $\gtrsim\!\kpc$, the classical `thick-disk' regime.  Note
that components with $\hz\ge\!200\pc$ should still be well sampled by
SDSS/SEGUE ($|z|\gtrsim\!300\pc$), while very thin, young components
with $\hz\simeq\!100\pc$ could be missed.  Similarly, the radial scale
lengths of these \map s vary widely: from $\hR\lesssim 2\kpc$ for the
chemically-old \map s, to an essentially flat radial profile,
$\hR\gtrsim\!5\kpc$ for the chemically-young \map s (with solar
$\afe$) of low $\feh$; note that for thin components and large $\hR$
the SEGUE survey geometry is problematic (confined
mostly to $l>30^\circ$), as it
samples only volumes will above the plane at $R\neq \Ro$.

This analysis shows empirically that the Disk contains a continuum of
(\map) stellar components that form a sequence from {\sl
  chemically-old, metal-poor, thick, and radially concentrated} to
{\sl chemically-young, metal-rich, thin, and radially extended}.  We
compare these results with cosmological simulations in
Section~\ref{sec:WitherDisk}; but even at face value, these results
point quite directly towards inside-out growth of the Disk.

On second look, the distribution of the \map s' structural parameters
has some subtleties: e.g, for a given \afe, the more metal-poor \map s
have longer scale lengths: there is an outward metallicity gradient in
the disk.  Finally, it is worth noting that $\hR$ and $\hz$ depend on
abundances in a way that cannot be captured by any one-dimensional
description of the abundances, i.e., by $\feh$, \afe, or any
combination of them alone. This may not be surprising, as birth epoch
and birth radius should be reflected in different ways in the two
abundance coordinates.
 
This view of the Disk appears to be on first sight in stark contrast
to purely geometric thin-thick Disk decompositions
\citep[e.g.,][]{Juric2008}. However, the \citet{BovyMAPstructure}
analysis is the only case to date of a large-scale Disk structure analysis where
the disk components have been selected solely by a
structure-independent property: $\feh-\afe$. In all other analyses
beyond the Solar neighborhood, the disk components have been defined
by their geometry, leaving inevitably some level of circularity.
 
\subsubsection{The kinematical structure of mono-abundance populations}

\begin{figure}
  \centering
  \includegraphics[width=0.45\textwidth,clip=]{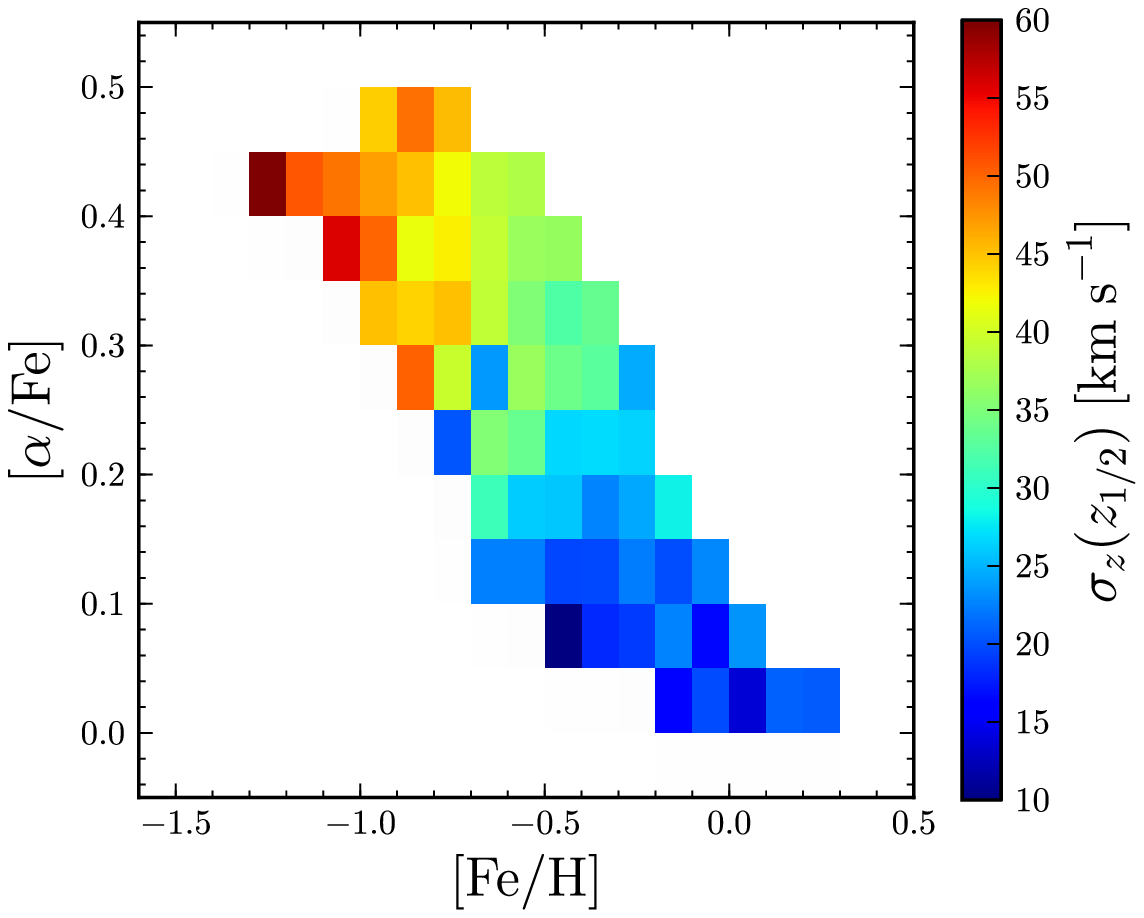}
  \includegraphics[width=0.45\textwidth,clip=]{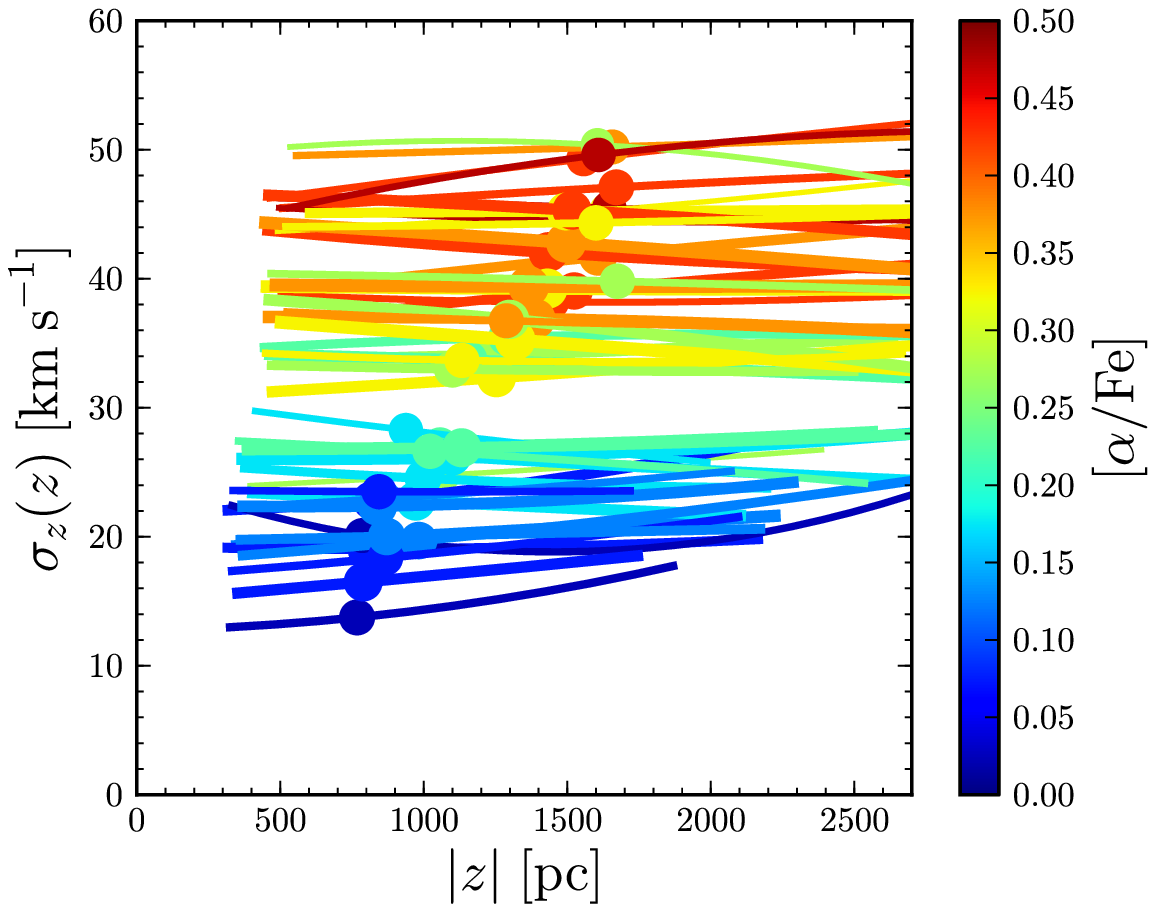}
  \caption{Vertical kinematics of \map s from
    \citet{BovyMAPkinematics}. The right panel shows the vertical
    dependence of the vertical velocity dispersion of \map s: all \map
    s exhibit an isothermal vertical profile at an individual level of
    a few $\kms\kpc^{-1}$; jointly they are consistent with being
    isothermal to $0.3\kms\kpc^{-1}$. The left panel shows that the
    vertical velocity dispersion increases when moving from
    metal-rich, solar \afe\ to more metal-poor and/or
    $\alpha$-enhanced \map s, similar to the \hz\ behavior in
    \figurename~\ref{fig:MAPspatial}.}
  \label{fig:Vertical-Kinematics}
\end{figure}
  
Analogously, the kinematical structure of individual \map s turns out
to be simple, in some sense a simple as can be
\citep{BovyMAPkinematics}: for each MAP, the velocity dispersions
$\sigmaz$ and $\sigmar$ are vertically approximately isothermal:
$\sigma_{z,R}(z|R,\X )\simeq$ const.  In particular for $\sigmaz$,
which \citet{BovyMAPkinematics} have investigated in detail, the degree
of isothermally is remarkable (see
\figurename~\ref{fig:Vertical-Kinematics}): the mean gradient is only
$|\partial \sigma_z /\partial z| = 0.2\pm0.3\kms\kpc^{-1}$.

As a function of Galactocentric radius, the dispersions show a slow
decrease, $\sigma_{z,R}(R|z,\X)\propto \exp{\bigl (-(R-\Ro )/7\kpc
  \bigr )}$. These kinematics warrant careful dynamical modeling, but
it seems qualitatively plausible that the redial dependence simply
reflects the decrease in the restoring force to the disk plane
\citep{BovyMAPkinematics}.

Also analogous to the spatial structure, the characteristic velocity
dispersion of the \map s shows a distinct pattern as a function of
$\feh$ and $\afe$ (\figurename~\ref{fig:Vertical-Kinematics}): the
chemically older and thicker components have higher dispersions, as
anticipated from $\hz$ and dynamics \citep{BovyMAPkinematics}.

As mentioned already in Section~\ref{sec:ChemicalAbundances},
\citet{BovyMAPkinematics} used the isothermality of \map s to test the
SEGUE abundance precision. This is possible, because the trends of
thicker and kinematically-hotter components for more metal-poor,
$\afe$-enhanced \map s would lead to an increase in vertical velocity
dispersion as a function of height when substantial abundance errors
lead to abundance mixing in \map s. \citet{BovyMAPkinematics} show
that the degree of isothermality observed for \map s requires an
abundance precision of $\simeq\!0.15\dex$ in \feh\ and $\simeq\!0.07\dex$
in \afe, close to the stated SEGUE-pipeline precision.

While the radial velocity dispersion seems to vary rather analogously
to the vertical dispersion among the \map s, there may be qualitative
differences in the orbit structures of \map s (especially at the
extremes of $\feh - \afe$ space), as pointed out by
\citet{Liu12}. Overall, the kinematics of all \map s are dominated by
orbits with relatively high angular momentum
\citep{Dierickx10,Wilson11,Liu12}, making a denotation of {\sl disk},
rather than {\sl halo} sensible.

In summary, each \map\ of the Disk has a very simple spatial and
kinematical structure, their properties vary widely: the SDSS analysis
yields components with $\sigma_z$ from $15$ to $50\kms$, $\hz$ from
$150\pc$ to $900\pc$, and $\hR$ from $\simeq\!1.5\kpc$ to an essentially
flat radial profile at $\Ro$.

\subsubsection{The overall structure of the Disk}

Deconstructing the overall structure of the Disk into a large set of
\map s is a different approach to slicing the Disk than is commonly
done.  It behoves one then to explore what this implies for the
overall structure of the Disk, when viewed as a superposition of \map
s.

This requires some additional considerations in order to put the
different \map s on the same footing.  At first, such components in
any survey such as SEGUE are defined at first in terms of the volume
corrected number-densities of the sample members.  However, any
operative sample definition, such as SEGUE's color cut
$0.48<(g-r)<0.55$, means that those stars stand in for differing
fractions of the stellar mass of their underlying stellar population,
depending on their $\feh$ and \age (see
Section~\ref{sec:SelectionFunctions}) By marginalizing over plausible
age distributions, one can convert the $z$-integrated
surface density of target stars into a surface mass density of stellar
mass \citep{BovyNoThickDisk,Schlesinger12}.

\begin{figure}
  \centering
  \includegraphics[width=0.85\textwidth,clip=]{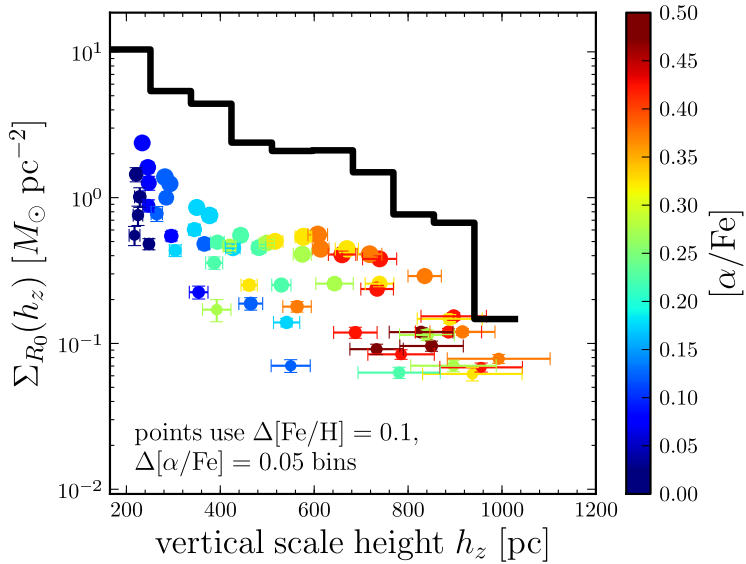}
  \caption{The Milky Way has no distinct thick disk. This Figure, from
    \citet{BovyNoThickDisk}, shows the stellar surface mass density
    contributions of individual \map s (colored symbols), each of which
    has associated a unique vertical scale height. The black histogram
    shows the total stellar surface mass density contributions from
    \map s with scale height $\hz$, which is exponentially decaying
    towards higher $\hz$, and shows no hint of thin-thick disk
    bimodality.}
  \label{fig:No-Thick-Disk}
\end{figure}

Along these lines, \citet{BovyNoThickDisk} worked out what the
surface-mass density distribution of each \map\ is at the Solar
radius, $\Sigma_{\Ro} (\feh , \afe )$. As each \map\ has a unique
scale height $\hz$, or kinematical temperature $\sigma_{z,R}$,
associated with it, one can then answer: {\sl How much of the total
  stellar surface mass density at $\Ro$ comes from stars with a
  scale-height $\hz$ (or with a certain $\sigmaz$)?  } If there was a
distinct `thin' and `thick' disk, one would expect a bi-modal
distribution in $\Sigma_{\Ro} (\hz )$, with $\simeq\!85\%$ of it in a
thin disk peak ($\hz \simeq\!150$ to $250\pc$) and $\simeq\!15\%$in a
thick disk (with $\hz\simeq\!700\pc$) for a canonical disk decomposition
\citep[e.g.,][]{Juric2008}. However,
\figurename~~\ref{fig:No-Thick-Disk} shows a different picture. There
is a continuous distribution of $\Sigma_{\Ro} (\hz )$, with no sign of
bi-modality, a distribution that is quite well-approximated by
$\Sigma_{\Ro} (\hz )\propto \exp{\bigl( \hz/280\pc\bigr )}$.  The
distribution of $\Sigma_{\Ro} (\sigmaz )$ shows analogous behavior.

This shows directly that while the Disk has a wide range of scale
heights and temperatures, thinking of only two distinct thin and thick
disk components is not consistent with the data: the Milky Way has no
{\sl distinct} thick disk \citep{BovyNoThickDisk}.  A few things are
important to keep in mind: First, this result was able to emerge
because selecting \map s only by their abundances, and being able to
associate a unique scale-height to each star, simply on the basis of
its $ \X$, has enabled a very different look at the Disk. Second, the
demonstrated $\X$ precision of SDSS/SEGUE argues strongly against the
smoothly decreasing $\Sigma_{\Ro} (\hz )$ distribution as merely a
consequence of poor abundance determinations (see also
\citealt{BovyMAPstructure}).  Third, this way of looking at the Disk,
as made up from a continuum of components of {\sl many} different
scale heights (or temperatures), does not change the integrated disk
properties (mass-weighted $\sigmaz$ or $\hz$), compared to the
traditional thin-thick disk dichotomy; it just shows that there is no
structural or kinematical dichotomy\footnote{Perhaps the clearest arguments for a thin-thick disk dichotomy come from
  bimodality in $\X$, which are discussed below.}.
  
\begin{figure}
  \centering
  \includegraphics[width=0.85\textwidth]{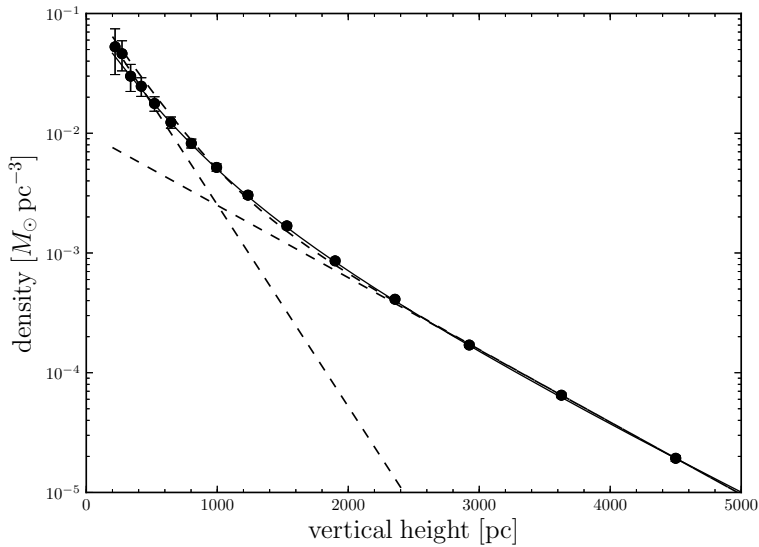}
  \caption{Overall vertical density profile in the Solar neighborhood
    implied by the \map\ decomposition. The full line is the
    mass-weighted density profile obtained by summing the
    contributions from the various \map s (\figurename
    s~\ref{fig:MAPspatial} and \ref{fig:No-Thick-Disk}). The points
    are a noisy sampling of this density and the straight dashed lines
    are exponential fits to the low- and high-height `data
    points'. This Figure can be directly compared to Figure 6 of
    \citet{Gilmore83a}. The sum of the two exponentials is the curved
    dashed line, which is barely distinguishable from the full line,
    showing that the Disk's overall vertical density profile implied
    by \map s can still be represented by a two-exponential fitting
    function.}
  \label{fig:GilmoreReidComparison}
\end{figure}

To test for consistency of this picture with previous work, we can
`re-assemble' our decomposition of the disk into \map s into the
overall spatial, kinematical, and elemental-abundance structure of the
Disk. \citet{Gilmore83a} determined the overall vertical structure of
the Disk in the solar neighborhood out to several kpc and found that
it can be represented as the sum of two exponential distributions: a
`thin disk' with a scale height of $300\pc$ and a `thick disk' with a
scale height of $1350\pc$. Later observations have confirmed this
measurement \citep[e.g.,][]{Juric2008}. We can compare the
\map\ measurements of the Disk's structure to these studies by
synthesizing the overall structure of the Disk---i.e., \emph{not}
split by elemental abundance---by summing the (mass-weighted)
contributions from the \map s. In
\figurename~\ref{fig:GilmoreReidComparison}, we show the overall
(mass-weighted) vertical density profile in the Solar neighborhood
implied by the \map\ measurements. This Figure shows that the overall
density can be well described by the sum of two exponential
distributions, a `thin disk' at low heights and a `thick disk' that
starts dominating at heights $>1\kpc$, even though the distribution is
made up of dozens of \map s with the scale height distribution of
\figurename~\ref{fig:No-Thick-Disk}. Thus, the \map\ decomposition
does not conflict with the measurement of \citet{Gilmore83a}, but the
decomposition into \map s based on elemental abundances has allowed a
qualitatively somewhat different description of the overall Disk
structure to be found; the description in terms of two exponentials
can certainly be views as a convenient and well-working fitting
function to the mass-weighted structure of the Disk.

\begin{figure}
  \centering
  \includegraphics[width=0.48\textwidth,clip=]{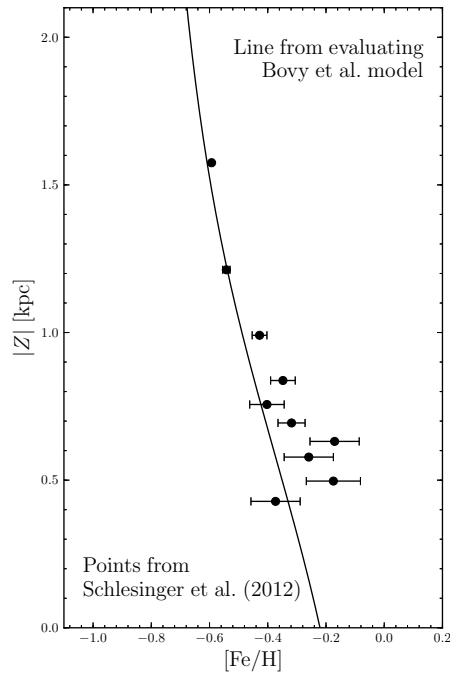}
  \caption{Comparison of the metallicity gradients implied by the \map\
    way of looking at the Disk with a direct estimate from
    \citet{Schlesinger12}.}
  \label{fig:SchlesingerComparison}
\end{figure}

Analogously, in the \map\ way of looking at the Disk, any vertical or
radial gradients in abundances or kinematical properties arise foremost
from the changing contributions of different \map s at a given
$(R,z)$. Of course, given a set of \map s, one can construct spatial
gradients in the population mean of, e.g., metallicity or velocity
dispersion. We illustrate for the concrete case of the SEGUE-derived
\map s and for $z$-gradients that this leads to predictions that are
consistent with direct measurements. As an example,
\figurename~\ref{fig:SchlesingerComparison} shows that the predicted
gradients from the Bovy \etal\ 2012b,c,d \map s and the direct
abundance gradients from \citet{Schlesinger12}. Similarly, integrating
over all abundances and deriving $\sigmaz (z)$ leads to a nearly
linear rise of the dispersion with height above the plane.

\begin{figure}
  \centering
  \includegraphics[width=0.85\textwidth]{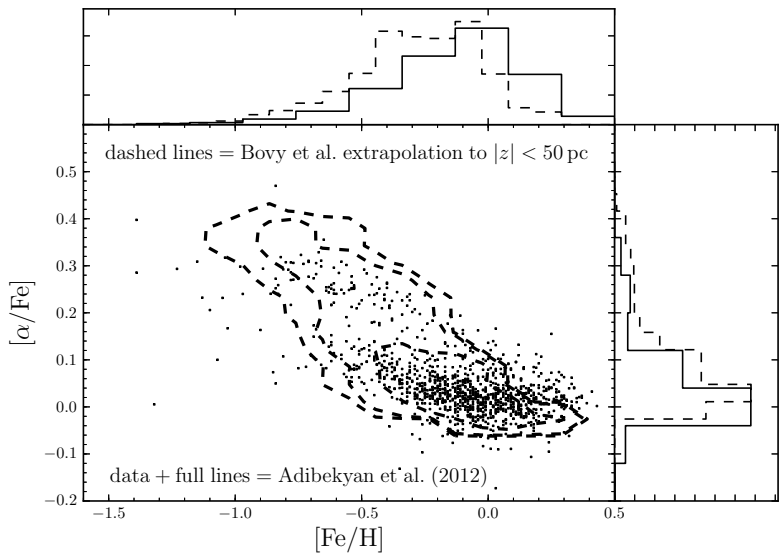}
  \caption{Abundance distribution of stars within $\le 100$~pc based
    on high-dispersion spectroscopy (based on data form
    \citet{Adibekyan12a}). This Figure shows that the distribution of
    $\afe$-abundances appears to be bi-modal, albeit not as two
    disjoint distribution as claimed by \citet{Fuhrmann11}; see
    discussion in Section~\ref{sec:AbundanceDistribution}.}
  \label{fig:AdibekyanComparison}
\end{figure}

We can also use the \map\ model for the Disk to predict the \feh --
\afe\ distribution at $|z| \approx 0$. The \map\ abundance
distribution extrapolated to $|z| < 50\pc$ is shown as contours in
\figurename~\ref{fig:AdibekyanComparison}, where it is compared to the
observed distribution from \citet{Adibekyan12a}. The \afe\ in this
Figure is the combination measured by
SEGUE---$0.5\,[\mathrm{Mg/Fe}]+0.3\,[\mathrm{Ti/Fe}]+0.1\,[\mathrm{Ca/Fe}]+0.1\,[\mathrm{Si/Fe}]$---and
we have subtracted 0.06 from the model's \afe\ to put the \afe\ on
approximately the same scale. The \map\ model, constrained at $|z| >
300\pc$, does a remarkably good job of producing the main features of
the $|z| < 50\pc$ abundance distribution, except for appearing too
metal-poor, which is probably due to an imperfect description of the
more metal-rich, $|z| < 300\pc$ stellar content.

The dissection and subsequent global synthesis of the Disk through
\map s is ultimately an empirical approach that is distinct from
synthetic models such as the Besancon model \citep{Robin03a}, Trilegal
\citep{Girardi05a}, or the model of \citet{JustJahreiss10}, which we
do not discuss in detail here. These models build a global Disk model
using a mix of stellar populations of different spatial, kinematical,
and chemical properties through the application of stellar-population
synthesis models and observational relations and correlations between
the model variables, such as the age--velocity relation and
metallicity gradients. The strength of these synthetic Galaxy models lies in their
ability to summarize our current understanding of the Disk into a
consistent---if not always dynamically
consistent---stellar-populations-based model that can be used to
simulate the expected content of new surveys
\citep[e.g.,][]{Robin12a}, test Disk analyses on realistic mock data,
and provide fiducial or background models for the interpretation of
new data.

\subsection{The abundance distribution of the Disk stars}\label{sec:AbundanceDistribution}

Beyond a few 100 pc, the abundance distribution of the Disk has only
been studied very recently with sizable samples \citep{Boeche11}.
However, in the immediate Solar neighborhood a number of studies,
usually based on the Hipparcos catalog, have revealed quite striking
abundance patterns in their samples (especially
\citealt{Bensby03,Nordstroem04,Bensby05a,Feltzing08,Navarro11,Fuhrmann11}).  To be
explicit, we take the {\it chemical abundance distribution} to be the
mass-weighted probability distribution of chemical abundances,
$p_{\mathrm{mass}}(\X|R,z)$ at $(R,z)$. It is worthwhile stating this explicitly,
because the published sample abundance distributions often differ (for
good reasons) drastically from this $p_{\mathrm{mass}}$ distribution, making a
direct comparison difficult if not impossible.  In particular, many
studies have explicitly included kinematical pre-selections of the
target stars \citep[e.g.,][]{Bensby03,Feltzing08,Bensby11}; others, e.g. \citet{Navarro11},
are foremost literature compilations without quantitative accounting
for the selection function.

Rightfully prominent among these abundance distribution studies are those of Fuhrmann
\citep{Fuhrmann98,Fuhrmann04,Fuhrmann08,Fuhrmann11}, which assembled
an approximately volume-limited sample of about 300 stars within
$25\pc$ of the Sun (i.e., a volume of $6\times 10^{-5}\kpc^3$), with
precise individual element abundances from high-resolution
spectroscopy.  Fuhrmann finds a distribution that is strikingly
bi-modal in the $\feh - \afe$ plane. Recent studies, initially geared
at exoplanet searches, have yielded similar information for stars at
typical distances of $\simeq\!100\pc$ \citep{Adibekyan12a}: they, too,
find a bi-modal distribution in the $\feh - \afe$ plane, though they
do not confirm an actual `gap' in the abundance plane, as
\figurename~\ref{fig:AdibekyanComparison} shows.
So, detailed abundance studies based on high-resolution spectra show
consistently, that when nearby samples are split by their motions into
those with kinematics typical `thin' and of `thick' disk stars, their abundance
patterns are distinctly and clearly different (see also \citealt{Bensby03,Nordstroem04,Bensby05a,Feltzing08}).

The questions remains to which extent this abundance bimodality in
itself argues that there is (at least chemically) a distinct thick
disk component (cf. Fuhrmann's work), attributable to an early (thick
disk) star-formation epoch in the Galaxy, and a significantly later
thin disk formation epoch, with a star-formation hiatus in between.
Because in contrast to that view, \citep{Schoenrich08a} have argued that ($\afe$-)abundance
bimodality, though no gap, could arise from a perfectly `smooth'
star-formation and enrichment history.  This is because the
$\afe$-enhancement fades rather rapidly, once SN~Ia enrichment becomes
important.

At the moment, excellent abundance information over very small volumes
tell a somewhat different story about an abundance dichotomy in the
stellar Disk population than the less precise information on $1\kpc$
scales. The next years, providing accurate abundances for large
samples over kpc-scales from APOGEE and Gaia-ESO should sort this out.

In summary, the authors' view---certainly guided and perhaps tainted
by their own work---points towards a structural description of the
Disk in terms of an approximately continuous distribution of 
stellar scale-heights or velocity
dispersions, which appears inconsistent with a structural
thin-thick Disk dichotomy or clear bi-modality.  The recent, vastly
larger data sets with abundances seem to imply that a thin-thick disk
description is too simplistic in, rather than being `wrong'. Yet, as
of this writing, there is no universal acceptance in the community of
this view, for understandable reasons: on the one hand, there is a
large number of previous results that seem very well explained by a
thin-thick disk dichotomy; while for some of these results,
consistency with the continuously varying \map~ picture has been
demonstrated (see above and cf
\figurename~\ref{fig:GilmoreReidComparison}, and
\figurename~\ref{fig:SchlesingerComparison}), other previous results
still await their comparison with the continuous \map~ picture.  On
the other hand, the element abundance structure in the Solar
neighborhood ($<100$~pc) does show clear evidence for abundance-bimodality,
with some data sets pointing towards an actual dichotomy. These
have yet to be fully confronted or reconciled with the continuous \map~
picture.

An currently open question is whether this \map\ view of the galactic disk provides
useful and new constraints on the extensive body of chemical evolution models
of the disk that has been developed over the last two decades
\citep[e.g.,][]{Matteucci89,Prantzos95,Chiappini01,Fraternali08,Marinacci11}.
 An exploration of this issues is beyond the scope of this review.

Taken together, viewing the Disk as a superposition of \map s appears
as a promising framework for studies ahead, irrespective of the 
question of a thin-thick dichotomy. In the final Section of
this Review, we will sketch two aspects where \map s are useful, which
will bring us to the initial issues, the implications of Disk studies
for dynamics and galaxy formation.

\section{Whither Disk studies?}\label{sec:WitherDisk}

In the closing Section of this review, we now return to the two broad
issues laid out in the introduction: First, what steps still need to
be taken to understand the 3D gravitational potential near the Disk as
well as we can, and to draw optimal inferences about the dark matter
distribution in the Galaxy.  Second, how can the empirical description
of the Disk be linked to galaxy formation mechanisms? We start by
describing emerging approaches towards comprehensive dynamical
modeling, followed by illustrative initial comparisons of the Disk
structure to cosmological simulations in the frame-work of MAPs. We
end with our best guess of what the conceptual and practical modeling
and interpretation challenges are in the Gaia era.

\subsection{Towards implementing comprehensive dynamical modeling}\label{sec:ComprehensiveApproachDynMod}

Approaches to stringent dynamical modeling of the Disk, which are
going beyond the modeling frameworks described, have also greatly
progressed foremost owing to J. Binney and P. McMillan.  Yet, all the
elements still have not yet quite come together to address: what are
the best quantitative constraints on the Galactic potential and on the
Disk's \DF one can derive from the available, or soon available, data?
Neither is there a stringent tool for experiment forecasting' to
address what pieces of observational data are most informative on a
given dynamical questions.

These questions require models that are both {\sl global}, {\sl
  dynamically consistent}, and can predict the likelihood of diverse
data sets. The following section reflects not only the authors'
thoughts and attempts for a roadmap towards such a modeling machinery,
but extensively draws on the work and plans laid out eloquently by
Binney and McMillan
(\citealt{Binney10a},\citealt{Binney11a},\citealt{Binney12a},
\citealt{McMillan12a}):

\begin{itemize}

\item Given the complexity of the problem, fully dynamically
  self-consistent models should be implemented in the steady-state,
  axisymmetric regime first, before proceeding to non-axisymmetric,
  time-dependent Disk modeling.  That alone would provide enormous
  progress over the status quo. The limitations of such modeling can
  be straightforwardly tested against mock-data from numerical disk
  simulations, which have now reached high-enough resolution($\ge 300$
  Million particles, \eg, \citealt{DOnghia12a}).

\item Given the vast number of discrete observational constraints,
  model predictions need to be continuous in $\vx,\vv$, to directly
  calculate the likelihood for (discrete) data.  This argues for a
  `distribution function based' approach, rather than a discrete
  particle or orbit based representation of the \DF.

\item To make the joint optimization, or sampling, of \Pot\ and
  \DF\ in light of large data sets practical, both \Pot\ and
  \DF\ should be describable by a modest number of parameters. For the
  gravitational potential this leaves two options: either describe the
  potential by a number of discrete, traditional components, such as
  the bulge,bar,halo,thin disk, thick disk and gaseous disk
  \citep[e.g.,][]{McMillan11a}. As the mono-abundance analysis of Bovy
  \etal\ (2012b,c,d) indicates that a sub-division into a few discrete
  Disk components is not sensible, a description of \Pot\ in terms of
  two parameterized `components', a spheroidal (not spherical) one and
  a flat, disk-like component, with flexibility in their $r$, and
  $R,z$ profiles, seems like a more sensible starting point.

\item The family of distribution functions advocated by
  \citet{Binney11a}, which is cast in terms of actions appears as a
  conceptually attractive and now also practical approach.
  Specifically, a \DF cast in terms of the radial, azimuthal and
  vertical actions $\vec{J}\equiv \bigl (J_R,J_\phi\equiv L_z,
  J_z\bigr )$ of the form
\begin{equation}
\begin{split}
f(J_R,L_z, J_z| \lambda_{\vec{J}}) = & \tf (L_z) \Biggl [\frac{\kappa(L_z)}{C_R(L_z)}\exp\Bigl (-\frac{\kappa(L_z)J_R}{C_R(L_z)}\Bigr )\Biggr ]\\ & \ \ \ \ \times \Biggl [\frac{\kappa(L_z)}{D_R(L_z)}\exp\Bigl (-\frac{\nu(L_z)J_z}{D_R(L_z)}\Bigr )\Biggr ]
\end{split}
\end{equation}
seems like a sensible choice \citep{Binney12a}, where $\lambda_J$ is a
set of parameters to describe the \DF . Elements of $\lambda_J$ in
this case are $\tf L_z)$, $C_R(L_z)$, and $D_R(L_z)$ \citep{Ting12b}.
Overall $\tf (L_z)$ sets the radial profile of the disk, given \Pot;
$C_R(L_z)$ and $D_R(L_z)$ set the radius-dependent disk `temperature'
in the radial and vertical direction. \citet{Ting12b} showed that
these \DF\ families can reproduce the radial and vertical properties of
mono-abundance Disk populations well, demonstrating that suitable
simply-parameterized families of \DF s are available.

While action-based \DF s are easy and elegant to write down, community
acceptance seems to have been hampered by their reputation to be hard
(or slow) to calculate, beyond the `azimuthal action', $J_\phi\equiv
L_z$, which is the angular momentum.  Recent progress in calculating
and testing the accuracy of approximate actions
\citep{Solway12a,Binney12b,Sanders12a} should help overcome this
issue.

\item `Modeling' requires efficient computation of $(\vx , \vv )
  \leftrightarrow (J_R,L_z, J_z)$, given \Pot. Clearly, $(\vx ,
  \vv| \Phi(\vx) ) \rightarrow (J_R,L_z, J_z)$ is easier to
  calculate, as this does not involve explicit treatment of the
  orbital {\sl angles} that complement the {\sl actions}; in
  steady-state modeling, the {\sl angles} are assumed to be distributed
  uniformly. This argues for modeling approaches that only go
  computationally from configuration to action space.

\item Yet, models need to be evaluated against data in the
  `configuration space' of observable data, i.e., we have to determine
  $\Like \bigl ( \{ \mathrm{data}\}|\lambda_{\vec{J}},
  \Potlambda)\bigr )$, with $ \{
  \mathrm{data}\}=\{p(\vx,\vv,\X,\tage)\}_i$. This will then inform us about
  the potential after marginalizing over the \DF\ parameters, $\Like
  \bigl ( \{ \mathrm{data}\}| \Potlambda \bigr )$; or about the \DF\ after
  marginalizing over \Pot, $\Like \bigl ( \{ \mathrm{data}\}|
  \lambda_{\vec{J}}\bigr )$.

\item To calculate meaningful likelihoods of the data for different
  $\lambda_{\vec{J}}$ given $\Potlambda$, it is necessary to interpret
  the \DF\ as a probability distribution. This is where in this
  context the spatial sample selection functions come in
  \citep[e.g.,][]{McMillan12a}. One approach to incorporating it
  \citep{Ting12b}) is to `normalize' the \DF\ over the observable
  volume, $f(\lambda_{\vec{J}})\rightarrow c_{\mathrm{selection}}\cdot
  f(\lambda_{\vec{J}})$, where

\begin{equation} 
c_{\mathrm{selection}}^{-1}\equiv \int d\vx d\vv\  p_{\mathrm{selection}}(\vx) \cdot 
       f\bigl (\vec{J}( \vx, \vv)| \lambda_{\vec{J}},\Potlambda \bigr ).
\end{equation} 

\item Of course, the `data', $\{p(\vx,\vv,\X,\tage)\}_i$ are not
  precise points in $(\vx,\vv)$, but have uncertainties, or may even
  be missing in some dimensions.  In practice, this implies yet
  another marginalization over the data's uncertainties, i.e., an
  integral over $\delta\vx,\delta\vv,\delta\X,\delta\tage$. How to do
  this computationally efficiently and well enough, is yet to be
  clarified.

\item All of the observables of course depend on the Sun's position
  and motion, where the aspects of the observed velocities that simply
  are the Sun's reflex motion are a notorious source of uncertainty.
  As recent work has showed, the `local standard of rest' is still
  under extensive debate : as \citet{BovyVc} showed, the Sun's motion
  should also be an explicit model parameter.

\item Any modeling should exploit the measured chemical abundances as
  `integrals of motion' that separate sub-populations. In essence, the
  above procedure should hold for any \map, with a distinct \DF\ for
  each, but of course the same \Pot\ for all. By adding the likelihood
  contributions from all the \map's, inferences about \Pot\ should be
  straightforward.

\item Finally, the dynamics of the Disk do not care about the
  particulars of the various surveys described in
  Section~\ref{sec:CurrentDiskSurveys}, which offer observational
  constraints in different $(\ell,b,D)$ regimes. Hence models need to be
  fit simultaneously to data from different surveys, which has
  happened far too little to date. Again, this should be
  straightforward, as any given model (\DF, \Pot) can predict the
  likelihoods for any survey, and data (log-)likelihoods simply have
  to be added.  The practical difficulties lie in the tedious task of
  compiling the various sample selection functions.

\end{itemize}

As of now, only initial demonstrations of theses approaches exist,
retrieving information from pseudo-data: \citet{McMillan12a} showed
how well the distribution functions of ensembles of 5000 stars could
be retrieved, if a priori disjoint thin and thick disk components were
presumed. \citet{Ting12b} explored how well the parameters of a
simplified 3D model for the Galactic potential could be retrieved,
with mock-survey data that resemble SDSS/SEGUE G-dwarfs; they could
show that the {\sl shape} of the potential could be constrained, but
that such constraints are much harder to get than constraints on, say,
$\vcirc$.

Overall, the above roadmap shows that basically all elements are in
place to do such comprehensive modeling. This suggest that much more
can and will be learned about the Galactic potential, drawing on
existing data well before the first extensive Gaia releases, even if
those stay on their 2012 schedule.  This should provide us with a much
firmer picture of \Pot\ for the Disk and with a solid base-line
distribution function description of the Disk.

\subsubsection{Chemo-dynamical substructure in the Disk}\label{DiskSubstructure}

A \DF\ description of the Disk and its MAPs will also provide a much
more sensible basis to characterize deviations from it as {\it
  ``sub-structure"}. Note that sub-structure can be both `clumps' in
action-space as well as in angle space, an aspect that has yet to be
explored.

\subsection{Mono-abundance populations in a galaxy formation context}\label{sec:MAP_in_GalForm}

A second avenue for the near future is to place the emerging empirical
results about the Disk in the context of galaxy formation. Here we
outline what we deem to be a promising approach, looking at \map s in
cosmological disk formation simulations.  As laid out in
Section~\ref{sec:DiskResults}, an empirical picture seems to emerge in
which the Disk is `seen' to exhibit an age-sequence from `thick'
(large $\hz$) and centrally-concentrated (small $\hR$) \map s to thin
and radially-extended ones, {\it if} $\afe$ and $\feh$ can indeed
provide an approximate relative age ranking of \map s in the Disk.
  
In two ways, this picture qualitatively resonates with well
established concepts of galaxy formation.  One the one hand, disk
galaxy formation should proceed from the inside out
\citep[e.g.,][]{Mo98a}.  On the other hand, there are several reasons
why older components of the Disk should be thicker (or have higher
velocity dispersion): they may have been born with higher dispersion
\citep{Bournaud09a,ForsterSchreiber11a}, they may have been heated by
subsequent tidal interactions or satellite infall
\citep[e.g.,][]{Quinn93a}; outward radial migration appears to conserve approximately 
the vertical action (not the vertical energy, cf. \citet{Schoenrich09a}), 
which will slightly thicken the disk portions that have migrated outward 
{\it and} slightly decrease their vertical velocity  
\citep{Solway12a,Minchev12a}.

\begin{figure}
  \centering \includegraphics[width=0.85\textwidth]{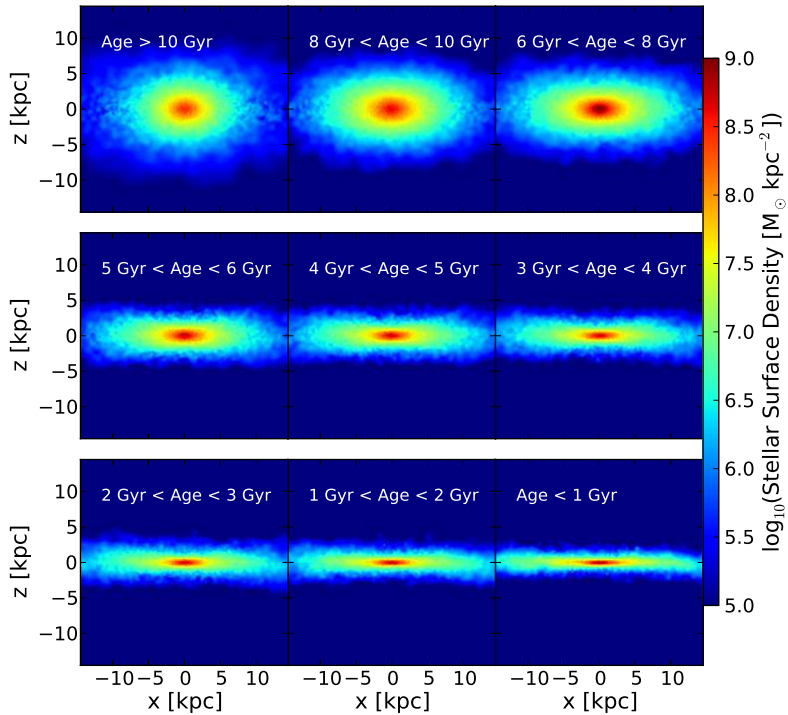}
  \caption{Present-day structure of `mono-age populations' in a
    cosmological formation simulation than led to a Milky-Way-like galaxy
    from \citet{Stinson12a}, where each panel shows a present-day, edge-on
    of stellar subsets in the simulations, sorted by their age: there is a clear trend of `old, thick,
    centrally concentrated' to `young, thin, extended'.}
  \label{fig:Stinson2012a}
\end{figure}

\begin{figure}
  \centering
  \includegraphics[width=0.85\textwidth]{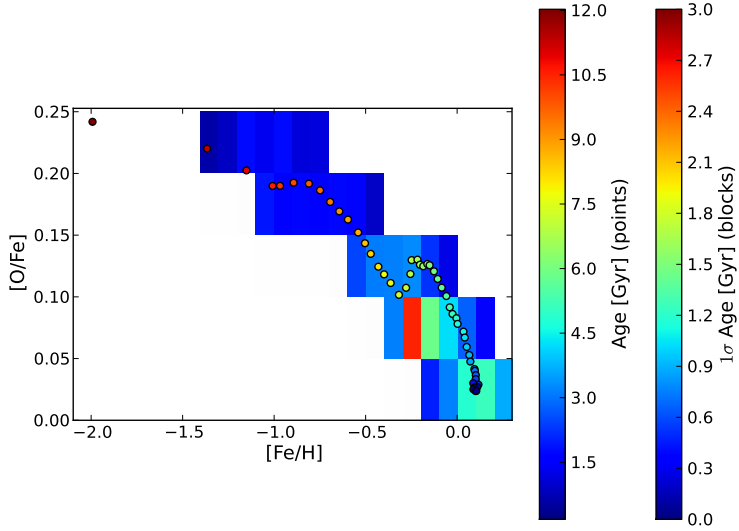}
  \caption{How well do \map s reflect mono-age populations in
    simulations. This Figure from \citet{Stinson12a}, whose
    simulations track chemical enrichment, shows the age dispersion of
    stars within each (simulated) \map\ bin. For most \map s the age
    dispersion is less than $1\Gyr$, implying that \map s are reasonably
    good approximations to mono-age populations.}
  \label{fig:Stinson2012b}
\end{figure}

To push a comparison with expectations in the cosmological context
further, predictions of galaxy formation simulations for \map s are
needed. As described in Section~\ref{sec:DiskOverview}, disk galaxy
formation simulations have made enormous strides in producing outcomes
that resemble Milky Way like galaxies
\citep{Guedes11a,Martig12a,Stinson12a}, with dominant flat disks and
only a modest fraction of the stars in a central bulge.  Some of these
simulations also treat the chemical enrichment self-consistently
\citep[e.g.,][]{Stinson12a} and these simulations provide an initial,
at least qualitative, comparison of \map s between cosmological
simulations and direct Milky Way observations. Recently, \citet{Stinson12a}
has carried out such a comparison with the patterns found by
\citet{BovyMAPstructure}. As illustrated in
Figure~\ref{fig:Stinson2012a}, they found remarkable agreement.  This
Figure shows edge-on views of the present-day stellar density
distribution in the simulations, sorted according to the
$[\mathrm{O/Fe}]$ (as proxy for $\afe$). As observed in the Disk,
there is a sequence of centrally-compact and thick configurations at
the $[\mathrm{O/Fe}]$-enhanced end to radially-extended, thin
distributions at Solar $[\mathrm{O/Fe}]$ values.  Quantifying this
picture by plotting $\hz$ vs. $\hR$ for \map s in the simulations,
confirms that this simulation shows the same behavior as the
data. This agreement lends encouragement to query the simulations
about the extent to which mono-abundance populations can serve as
proxy for mono-age populations. Figure~\ref{fig:Stinson2012b} shows
the mean age and the age dispersion as a function of $\feh$ and $\afe$
in the simulations. Remarkably, in most mono-abundance bins
of these simulations, the age
dispersion is $\lesssim 1\Gyr$, justifying the use of 2D-abundances as
age proxies.

Of course, at present the above is only an anecdotal comparison with
one simulation; albeit a state-of-the-art simulation that was `tuned'
to match the stellar-mass / halo-mass relation of \citet{Moster12a} at
redshift zero, and that produces a large, fairly massive disk and a
small bulge, with the spatial and temporal abundance pattern only a
consequence of it.  Clearly, this kind of data model comparison
warrant much further exploration, as we lay out in the final
Subsection.

Some readers may pine for `tests' of galaxy evolution mechanism that
do not depend on comparison with numerical simulations. But such tests
may be far and few between, as the present-day dynamical structure
mostly tells us about the present day. This can be illustrated by the
quest to find tests for the efficacy of radial migration. The
Hermes/GALAH experiment \citep{Freeman10a} has set out to provide such
a clean test: the ambition is to identify, through `abundance
fingerprinting', stars that were formed basically at the same time
with the same orbital actions and phases. If we can look at their
orbits (say angular momenta) now, and if we know their ages, then one
gets a direct estimate of the importance of radial migration, though
still no clear hint about the cause of that migration (bars, spiral
arms, satellites, etc.).

\subsection{Peripheral Disk issues}

The Galactic stellar Disk, as defined here, does of course not live in
isolation, but it interfaces with other Galactic `components' at its
extremes: with the bar and (pseudo-)bulge at small radii; with the
slightly warped, complex, and almost messy outer reaches of the Disk
($>12\kpc$); and with the Galactic stellar halo, which the data
suggest is truly a `distinct' component from the bulk of the Disk
\citep{Majewski93a,Ivezic2008}. We review only briefly the status and
prospects in these areas.

\begin{itemize}
\item {\bf The innermost Disk:} Massive late-type galaxies like ours
  often have bars at the center \citep{Barazza08} and a thick central
  portion that is described as a pseudo-bulge
  \citep[e.g.,][]{Kormendy04}; the Milky Way fits that pattern, as it
  has a large-scale stellar bar
  \citep{Binney91,Blitz91a,Weinberg92,Binney97a}. This bar must have
  formed from stars that were part of the pre-existing Disk, and clear
  separation between the components is perhaps moot
  \citep[\eg,][]{Shen10a}. In practice, the present end of the bar is
  a sensible dividing line, presumably at co-rotation $2.5\pm0.5\kpc$
  \citep{Binney91}. As the patterns speeds of bars change secularly
  \citep{Debattista00}, it is unclear---and should be
  clarified---whether this radius implies any changes in the abundance
  or age structure.

\item {\bf The Outer Fringes of the Disk:} The stellar disk of our
  neighbor galaxy, M31, is manifestly frayed in its very outer parts
  \citep{Ferguson02,Ibata05}, possibly because much of the material is
  tidally disrupted satellite debris or possibly because disk material
  has been grossly perturbed by such infall events. There is ample
  evidence that the outer parts of the Galactic Disk are just as
  messy: the structure variously known as the `anticenter ring',
  `Monoceros ring', `Canis Major feature' \citep{Yanny03, Martin04,
    deJong10} reflects the fact that there are far more stars at
  $R_{\mathrm{GC}}=15$ to $20\kpc$ \citep{Conn12} with $|z|>1$~kpc than a
  simple double exponential disk model suggests.  This `feature' of
  the outer Disk has elicited extensive discussion, as two very
  different (and in the pure form probably simplistic) explanations
  have been advanced to explain it: either purely as deposited stellar
  debris from a satellite that had merged on a low-latitude,
  low-eccentricity prograde orbit
  \citep[e.g.,][]{Penarrubia05}, or simply as a combination
  of a disk warp and flare \citep[e.g.,][]{Momany06}. Because the
  abundance distribution and kinematics of possible `Monoceros ring'
  member stars are not grossly different from the expectations for an
  outer disk, it will take comprehensive area coverage (\eg, from
  PS1), good distances from Gaia, and extensive spectroscopy (1000's
  of stars) yielding velocities and abundances to sort out to which
  extent these are stars that have been `dragged into' the Disk
  vs. `kicked out of' the Disk.
\end{itemize}

\subsection{Some specific tasks for the next years}

Having described the various elements of gathering, analyzing and
(dynamically and cosmologically) modeling stellar Disk surveys, we now
try to cast answering some of introductory questions in this review
into possible projects that appear feasible and
necessary for the next years:

\subsubsection{What factors limit dynamical Disk modeling?}

With the modeling machinery of
Section~\ref{sec:ComprehensiveApproachDynMod} in place, at least in
principle, all elements have come together to answer this question.
The most straightforward approach to doing this is to create mock data
sets from $N$-body simulations of disks that are self-consistent by
construction, and feed them though the modeling machinery to see how
well \Pot\ and \DF\ can be recovered.  Recently, non-cosmological
simulations of disk galaxies have reached particle sizes of
$\simeq\!300$ Million particles \citep{DOnghia12a}, where mock-data
sets that match the current generation of survey data (RAVE, SDSS,
APOGEE, etc.) can be drawn directly without supersampling.  Then
selection functions, distance estimates with either random or
systematic errors, and (appropriate or inappropriate) de-reddening
estimates can be applied, where the reddening model would be a version
of the currently emerging 3D extinction maps for the Milky Way
(cf. Section~\ref{sec:Extinction}). Deriving probability distribution
functions on \Pot\ in light of a suite of mock data sets, would reveal
the factors that limit the accuracy of the potential inference.
Similar, or even the same, mock-data sets can explore to which extent
non-axisymmetry of disk galaxies such as the Milky Way, in particular
spiral arms and a central bar, would affect the \Pot\ inferences made
under axisymmetric assumptions. In turn, with such analyses one can
explore what it would take to, \eg, measure the strength of the spiral
arms dynamically.  What the mock data, at least in the very high
resolution, hence `non-cosmological' simulations, do not easily supply
is an `abundance tag' that allows to define MAPs. Hence, the best way
to test how much is gained in {\it dynamical} analyses by splitting
the tracer samples into MAPs still needs to be devised.  Current
experiments along those lines \citep{Ting12b} indicate that the sheer
sample size plays a sub-dominant role in constraining the potential in
the current regime ($\ge 10^{4-5}$~stars).  Such mock data sets will
also be crucial in finding out whether vast numbers of stars with
partial phase-space information (\eg, without $v_{\mathrm{los}}$) have
comparable information content than far smaller sets with complete
phase-space information.

\subsubsection{How to get to the best chemo-orbital distribution function?}

In the context of comprehensive dynamical modeling
(Section~\ref{sec:ComprehensiveApproachDynMod}), an estimate of the
\DF\ of course results `automatically'. Yet, the issues of getting an
optimal estimate for \DF\ are different. On the one hand, slight
systematic errors in \Pot\ probably lead to a (smooth) \DF\ that looks
very similar to the correct one; so to get a sense of whether the
vertical action distribution of MAPs varies little with radius (as
expected for asymptotically-efficient radial migration), an
approximate potential probably suffices.  On the other hand, it is
perhaps \DF\ sub-structure or fine-scale structure that is most
interesting. In the limit of a `cold stellar stream', i.e., stars of
the same actions that differ only in phase \citep[\eg,][]{Koposov10a},
the most plausible potential may be the one that makes the \DF\ of
that stream most like a $\delta$-function, even if that potential is
not necessarily the most likely in light of the overall set of
tracers.  Further, sample size, precision of the individual $(\vx ,
\vv )$ estimates, and chemical abundances matter far more than for
merely estimating \Pot. Sample size matters, because diagnostically
precious sub-populations (\eg, streams) may only make up a very small
fraction of the Disk mass; $(\vx , \vv )$-precision matters, because
`cold' substructures, those very compact in \DF\ space, are of
preeminent interest, and $\X$ matters, because the \DF\ has to be
strictly separable in chemical abundances, which resemble effective
separating integrals of motion.  Therefore, even with only an initial
estimate of \DF, a full exploration of \map\ \DF s, with the largest
possible samples and the most precise phase-space coordinate estimates
will be the next step. This is clearly also a direction where dramatic
progress using Gaia data will be very straightforward.

\subsubsection{The next steps mapping the gravitational potential}

Initial attempts have been made to bring the diverse constraints on
the Galactic gravitational potential on the same footing
\citep{CatenaUllio10,McMillan11a,BovyTremaine}, the outer halo
constraints, the rotation curve, and the local dark matter estimates
from $\Kz$. But the step of measuring $\Kz$ as a function of
$R_{\mathrm{GC}}$, which is feasible even with the existing data from
$6\kpc$ to $12\kpc$, is currently missing. Getting the
dynamically-measured disk mass scale length (presumably over 2 to 3
scale lengths, i.e., an order of magnitude in $\Sigma(R)$) will break
the `disk--halo degeneracy' and allow us to take strides towards
better quantifying the amount of dark matter within $\Ro$. In
constraining the rotation curve, the emerging stellar surveys will
eventually allow for testing less constrained models, and the
combination of stellar kinematics with good distances and the emerging
results from the Galactic maser survey \citep{Brunnthaler11} will be
the most powerful constraints on global asymmetries of the Galactic
potential, such a lopsidedness. This is because masers are the most
feasible tracer of the rotation curve on the far side of the
Galaxy. Untangling the disk and halo contributions to the potential at
$R\le 10\kpc$ will then permit tighter constraints on the shape of the
dark matter and---ultimately---a test for the existence of dark matter
(as opposed to alternative gravity laws) on the bases of the shape of
the acceleration/potential map alone; such analyses probably require
Gaia data, foremost to get very good proper motions to
$D\simeq\!10\kpc$.

\subsubsection{Closing remarks}
 
In summary, an enormous amount of practical work needs to be carried
out and conceptual modeling issues need to be sorted out, even before
Gaia data are available. It seems crucial to do this work, even if
some aspects (but clearly not all---abundances, radial velocities of
faint stars) will be dramatically superseded by Gaia. This is because
the currently emerging data will be key in making sure we are asking
the right questions of the Gaia data, and think through what the most
critical `complementary data' to ESA's next flagship mission are. In
return, the field promises a information revolution that is probably
unmatched in the field of galaxy studies in this decade, ALMA and JWST
notwithstanding.

But the attentive and persistent reader will have noticed that we have
not fully closed the circle on using the Galactic Disk to test
`mechanisms of disk galaxy formation'.  While
Section~\ref{sec:MAP_in_GalForm} has sketched some specific examples
of how to test aspects of disk galaxy formation by comparing ab initio
simulations to the data, it has not tackled head-on some of the broad
initial questions (Section~\ref{sec:FormEvDisk}): To which extent
where stars born (vertically) hot or subsequently heated? Is there
testable evidence that the feedback implemented in the simulations
actually took place?  Can we recognize stellar satellite debris well
enough to quantify the fraction of Disk stars with external origin?

That means that also much conceptual work needs to be done. It may
serve as a useful compass to thoroughly think about the following
scenario: if one had the perfectly `analyzed and modeled' Gaia data (and
those from all other surveys) at hand, what are the crispest
inferences one could draw about how our Galaxy and other disk galaxies,
have formed?

{\bf Acknowledgements} It is a pleasure to thank Vardan Adibekyan, Dan
Foreman-Mackey, Ralph Sch\"onrich and David W. Hogg for helpful comments and
assistance. H.-W.R. and J.B. were partially supported by SFB 881 funded
by the German Research Foundation DFG. J.B. was supported by NASA
through Hubble Fellowship grant HST-HF-51285.01 from the Space
Telescope Science Institute, which is operated by the Association of
Universities for Research in Astronomy, Incorporated, under NASA
contract NAS5-26555. J.B. is grateful to the Max-Planck Institut
f\"ur Astronomie for its hospitality during part of the period during
which this research was performed.

\end{document}